\documentclass[prd,preprint,floatfix,nofootinbib,tightenlines,preprintnumbers,superscriptaddress]{revtex4-2}

\usepackage[english]{babel}
\usepackage{amsmath,amssymb,bm,slashed}
\usepackage{graphicx}
\usepackage[sort&compress]{natbib}
\usepackage[usenames,dvipsnames,svgnames]{xcolor}
\usepackage[normalem]{ulem}
\usepackage{hyperref}
\usepackage[capitalize]{cleveref}
\crefname{paragraph}{Paragraph}{Paragraphs}
\definecolor{red}{rgb}{1.0, 0, 0}
\definecolor{green}{rgb}{0.0,0.7,0.2}

\usepackage[section]{placeins}

\usepackage{pifont}

\hypersetup{
  colorlinks,
  linktoc=page,
  linkcolor=blue,
  citecolor=ForestGreen
}

\allowdisplaybreaks

\definecolor{baic}{rgb}{0.00392156862745098, 0.45098039215686275, 0.6980392156862745}
\definecolor{bpic}{rgb}{0.8705882352941177, 0.5607843137254902, 0.0196078431372549}
\definecolor{paic}{rgb}{0.00784313725490196, 0.6196078431372549, 0.45098039215686275}
\definecolor{ppic}{rgb}{0.8352941176470589, 0.3686274509803922, 0.0}
\definecolor{indv}{rgb}{0.8, 0.47058823529411764, 0.7372549019607844}
\definecolor{naive}{rgb}{0.5803921568627451, 0.5803921568627451, 0.5803921568627451}
\definecolor{indv2}{rgb}{0.71, 0.71, 0.71}
\definecolor{indv_2state}{rgb}{0.984313725490196, 0.6862745098039216, 0.8941176470588236}

\makeatletter\AtBeginDocument{\let\@elt\relax}\makeatother
\usepackage{extarrows}
\usepackage{float}
\floatstyle{plaintop}
\usepackage[caption=false]{subfig} 
\usepackage{array}
\usepackage{multirow}

\hypersetup{
  colorlinks,
  linktoc=page,
  linkcolor=blue,
  citecolor=ForestGreen
}

\allowdisplaybreaks

\bibpunct{[}{]}{,}{n}{}{,}

\newcommand{\ev}[1]{\ensuremath{\left\langle #1 %
                     \right\rangle}} 
\newcommand{\tr}{\text{tr}}

\newcommand{\diag}{\text{diag}}

\renewcommand{\vec}[1]{{\mathbf{#1}}}

\providecommand{\para}[1]{\left(#1\right)}
\providecommand{\bracket}[1]{\left[#1\right]}
\providecommand{\abs}[1]{\left\lvert#1\right\rvert}
\providecommand{\norm}[1]{\left\lVert#1\right\rVert}
\newcommand{\beq}{\begin{equation}}
\newcommand{\eeq}{\end{equation}}

\newcommand{\pr}{\ensuremath{\mbox{pr}}}

\newcommand{\avg}[1]{\ensuremath{\left\langle #1 \right\rangle}}

\newcommand{\yd}{\{y\}}

\DeclareMathOperator{\aug}{aug}
\DeclareMathOperator{\BC}{BC}
\DeclareMathOperator{\perf}{perf}
\DeclareMathOperator{\C}{C}
\DeclareMathOperator{\K}{K}

\DeclareMathOperator{\T}{T}
\DeclareMathOperator{\off}{O}
\DeclareMathOperator{\argmin}{argmin}

\DeclareMathOperator{\MA}{MA}
\DeclareMathOperator{\MLE}{MLE}
\DeclareMathOperator{\PM}{PM}
\DeclareMathOperator{\SL}{SL}

\DeclareMathOperator{\CV}{CV}
\DeclareMathOperator{\KL}{KL}

\DeclareMathOperator{\plug}{plug-in}
\DeclareMathOperator{\marg}{marg}
\DeclareMathOperator{\post}{post-avg}
\DeclareMathOperator{\pred}{post-pred}

\DeclareMathOperator{\AIC}{AIC}
\DeclareMathOperator{\ABIC}{ABIC}
\DeclareMathOperator{\BAIC}{BAIC}
\DeclareMathOperator{\BIC}{BIC}
\DeclareMathOperator{\BPIC}{BPIC}
\DeclareMathOperator{\BTIC}{BTIC}
\DeclareMathOperator{\DIC}{DIC}
\DeclareMathOperator{\IC}{IC}
\DeclareMathOperator{\PAIC}{PAIC}
\DeclareMathOperator{\PPIC}{PPIC}
\DeclareMathOperator{\TIC}{TIC}

\DeclareMathOperator{\esssup}{ess\:sup}

\usepackage{amsthm}

\begin{document}

\title{Improved information criteria for Bayesian model averaging in lattice field theory}

\author{Ethan T.~Neil}              \email[Email: ]{ethan.neil@colorado.edu}
\affiliation{Department of Physics, University of Colorado, Boulder, CO 80309, USA}

\author{Jacob W.~Sitison}          \email[Email: ]{jacob.sitison@colorado.edu}
\affiliation{Department of Physics, University of Colorado, Boulder, CO 80309, USA}

\date{\today} 

\begin{abstract}
Bayesian model averaging is a practical method for dealing with uncertainty due to model specification.  Use of this technique requires the estimation of model probability weights.  In this work, we revisit the derivation of estimators for these model weights.  Use of the Kullback-Leibler divergence as a starting point leads naturally to a number of alternative information criteria suitable for Bayesian model weight estimation.  We explore three such criteria, known to the statistics literature before, in detail: a Bayesian analogue of the Akaike information criterion which we call the BAIC, the Bayesian predictive information criterion (BPIC), and the posterior predictive information criterion (PPIC). We compare the use of these information criteria in numerical analysis problems common in lattice field theory calculations.  We find that the PPIC has the most appealing theoretical properties and can give the best performance in terms of model-averaging uncertainty, particularly in the presence of noisy data, while the BAIC is a simple and reliable alternative.
\end{abstract}


\maketitle
\newpage

\tableofcontents

\newpage

\section{Introduction}
\label{sec:intro}

In many data analysis applications, particularly in lattice gauge theory, model uncertainty is a common challenge.  Model uncertainty arises when multiple candidate model descriptions exist for a given set of observations, with the desired analysis results dependent on which model is used.  A simple solution to this problem is model selection, i.e., choosing a single model from the available candidates based on some (generally data-driven) criteria.  Model selection is appealing due to its relative simplicity: once a model is chosen using some procedure, inference on parameters or prediction of future observations can be done using standard statistical methods within the chosen model.

However, this approach is not always optimal, especially when the primary goal of analysis is parameter inference and model selection is only an intermediate step.  By choosing a single ``best'' model, model selection neglects the effects of model uncertainty compared to other sources of error such as parameter uncertainty from a regression procedure (e.g., least squares) \cite{Wasserman2000,Parkinson2013,Fragoso2018}. As a result, model selection can lead to overly confident results based on limited statistical information.

To incorporate model uncertainty into statistical analyses, a natural alternative to model selection is model averaging.  With model averaging, quantities of interest are determined for each model in a space of candidates,  and a final estimate is made by taking a weighted average over the model-dependent estimates. The weights correspond to how likely each respective model is to describe the observed data.  Combining models in this way accounts for the model uncertainties in the overall statistical uncertainty of the analysis.  Moreover, the probabilistic weighting of models can yield smaller uncertainties compared to overly conservative procedures such as taking the full difference between plausible model variations as a systematic error, without introducing asymptotic bias.

Bayesian inference gives a natural framework in which to carry out the procedure of model averaging.  Specifically, Bayes' theorem gives a way to construct a posterior distribution over the combined model-parameter product space and allows analysts to incorporate whatever prior information is available.  Bayesian model averaging has been well-known in the statistics literature for some time \cite{Leamer1978,Racine1986,Madigan1994,Kass1995,Raftery2003}. The central problem in applying Bayesian model averaging is the estimation of model probability weights, which is generally formulated in terms of quantities known as ``information criteria'' (ICs).  The most well-known information criterion is the Akaike information criterion (AIC) \cite{akaike1973information,akaike1974new,Akaike1998}, which by construction is inherently a frequentist estimator (although a close analogue, the ``Bayesian AIC'', may be derived in a Bayesian context as we will show).

In this paper, we explore several ICs that may be used to determine model weights for Bayesian model averaging.   As a unifying concept, and inspired by the work of Zhou \cite{ZhouThesis,zhou2020posterior}, we focus on the derivation of information criteria based on the Kullback-Leibler (K-L) divergence, which can be thought of as an information-theoretic starting point for evaluating model probability.  Variations on the explicit definition of the K-L divergence in the case of parametric models (which are of primary interest for model averaging) are shown to lead to different ICs. 

This work is motivated specifically by a need for improved statistical methods in lattice field theory.   Bayesian model averaging is well-suited for lattice applications because of the notorious stability issues of the functional forms that arise in lattice application (e.g., two-point correlators modeled with an infinite tower of exponentials). As a result, statistical analyses of lattice data typically require model and/or data set truncation, which introduces systematic errors to a typical model selection procedure. By averaging over models and data subsets (which we will see is equivalent to the general model averaging framework), this systematic error is accounted for. Furthermore, the firm physical foundation of lattice field theory complements the use of Bayesian inference by giving well-motivated families of models to consider.  Other explorations of model averaging in a lattice field theory or effective field theory context include \cite{Chen:2004gp,Davies:2008sw, Schindler:2008fh,Durr:2013goa, Berkowitz:2017gql, Chang:2018uxx, Rinaldi:2019thf, Miller:2020xhy,Borsanyi:2020mff,Phillips:2020dmw,Connell:2021qcd}; the AIC was first used in the context of lattice field theory analysis in \cite{BMW:2014pzb}.  Our current work inherits directly from \cite{Jay:2020jkz}, which rigorously studied model averaging for lattice field theory in a Bayesian context.

The remainder of the paper is structured as follows.  In the next subsection \cref{subsec:summary}, we give an overview of our key results.  We then review some general results important for model averaging in \cref{sec:prelim}, including the Bayesian framework for model averaging developed in \cite{Jay:2020jkz} and some general concepts from mathematical statistics; as part of this discussion, we establish how bias on information criteria influences bias of parameter estimates.  In \cref{sec:kl-div}, we define the K-L divergence and give several distinct variations for parametric distributions; here we also introduce the information criteria that result from these variations. We specialize our discussion of Bayesian model averaging to least-squares regression in \cref{sec:least_squares} and derive formulas to approximate the model probability weights from the aforementioned information criteria. In \cref{sec:data_subset_select}, we reformulate the data subset selection problem as one of model variation and derive the corresponding expressions for the information criteria in this case. \cref{sec:num-tests} gives three numerical examples to demonstrate the performance of each information criterion in model averaging; these include linear least squares applied to a fixed data set (\cref{subsec:ex1}), a nonlinear toy problem that resembles fitting a two-point correlation function to demonstrate the effectiveness of model averaging in lieu of manual data subset selection (\cref{subsec:ex2}), and finally a similar two-point nucleon correlator example on a set of real lattice QCD data (\cref{subsec:ex3}). \cref{sec:conclusion} summarizes our findings and gives some concluding remarks.  Appendix~\ref{sec:app_last_paper} connects the theoretical details of \cite{Jay:2020jkz} to our updated view in terms of the K-L divergence, and provides some additional discussion.   In Appendix~\ref{sec:asymptotic-appendix}, we discuss the asymptotic equivalence of the various information criteria in the limit of infinite data.  A bound on the asymptotic bias of model averaging is derived in detail in Appendix~\ref{sec:bias-appendix}.  Another information criteria known as the posterior averaging information criterion (PAIC) was proposed by Zhou \cite{ZhouThesis,zhou2020posterior} to generalize and improve the performance of the BPIC; however, using the same integral approximation as the BPIC and PPIC gives a lower order (in the inverse sample size $N^{-1}$) approximation to the PAIC and hence worse performance in practice. Therefore, the PAIC is not discussed as thoroughly as the other ICs, but the relevant formulas are given in Appendix~\ref{sec:paic_formulas}. Appendix~\ref{sec:app_laplace_method} gives a brief derivation of the asymptotic approximation known as Laplace's method used in \cref{sec:least_squares} as well as some Gaussian integrals used in \cref{sec:data_subset_select}. Appendix~\ref{sec:app-subset-alt} contains an alternative derivation of the data subset selection criteria introduced in \cref{sec:data_subset_select}.  Finally, some of the relevant derivative tensors used in the calculations are given in Appendix~\ref{sec:derivatives}.

\subsection{Summary of key results }
\label{subsec:summary}

Since we derive a number of technical results in this paper in some detail, we include here an overview of some of our key findings.  Our primary focus is on information criteria (ICs), which quantify the (logarithmic) probability of a given model; for a review of the basic formalism of model averaging including rigorous definition of ICs, see \cref{subsec:review} below.  

Included in our work are two important clarifications.  First, we study the effect of bias in ICs on bias in parameter estimates; our result \cref{eqn:asymptotic_bias} establishes that unbiased ICs are important for obtaining unbiased model-averaged parameter estimates.  Second, we clarify some key points in how the AIC arises in a Bayesian context compared to the earlier work in \cite{Jay:2020jkz}; see \cref{subsec:kl-div-plug} and \cref{sec:app_last_paper}.  We call our revised formula the ``Bayesian AIC'', or BAIC, see \cref{eqn:atic-gen,eqn:baic,eqn:baic_subset}; the differences between AIC, $\ABIC_{\CV}$ (our name for the formula defined in \cite{Jay:2020jkz}) and BAIC are subtle and irrelevant in the limit that the priors do not influence the results.

A central aspect of our work is to approach the problem of data modeling using the Kullback-Leibler (K-L) divergence as a foundation.  Given a ``true model'' $M_{\T}$, the K-L divergence (or relative entropy) between the true model and a candidate model $M_\mu$ is (notation is defined more fully in \cref{sec:kl-div}):
\begin{equation}
\KL(M_\mu) = E_z[\log \pr_{M_{\T}}(z)] - E_z[\log \pr_{M_\mu}(z)].
\end{equation}
where future data $z$ are drawn from the true likelihood $\pr_{M_{\T}}(z)$.  Broadly speaking, minimization of this divergence will select the model $M_\mu$ that most closely resembles the true model $M_{\T}$.  However, this definition is non-parametric, and therefore ambiguous when the model likelihood $\pr_{M_\mu}(z|\mathbf{a})$ depends on some fit parameters $\mathbf{a}$.  

This ambiguity allows us to define several variations on $\KL(M_\mu)$, which then lead naturally to different ICs.  Each variation can be thought of as representing a  choice for how to obtain the non-parametric model predictive distribution $\pr_{M_\mu}(z)$ by starting from a parametric model. For example, we may adopt a ``plug-in'' estimate using the best-fit parameter value $\vec{a}^*$, leading to a Bayesian version of the well-known Akaike information criterion (AIC) as described in \cref{subsec:kl-div-plug}:
\begin{equation*}
E_z[\log \pr_{M_\mu}(z)] \sim E_z[ \log \pr_{M_\mu}(z|\vec{a}^*)] \simeq -\frac{\BAIC}{2N},
\end{equation*}
where $N$ is the data sample size, ``$\sim$'' indicates a choice of construction for the non-parametric model predictive distribution, and ``$\simeq$" indicates that the IC on the right-hand side may be used to compute an unbiased sample estimate of the term on the left-hand side.   The plug-in approach is simple, but unnatural from a Bayesian point of view since it focuses on a single best-fit parameter \textit{value} instead of a posterior \textit{distribution}.

We also explore two alternatives that are more manifestly Bayesian and lead to two other ICs.  Specifically, we study the ``Bayesian predictive information criterion" (BPIC, \cref{eqn:bpic_unsimp}) and the ``posterior predictive information criterion" (PPIC, \cref{eqn:ppic-unsimp}):
\begin{align*}
E_z[\log \pr_{M_\mu}(z)] \sim E_z[E_{\vec{a}| \yd} [\log \pr_{M_\mu}(z|\vec{a})]] &\simeq -\frac{\BPIC}{2N}, \\
E_z[\log \pr_{M_\mu}(z)] \sim E_z[\log E_{\vec{a}| \yd} [\pr_{M_\mu}(z|\vec{a})]] &\simeq -\frac{\PPIC}{2N},
\end{align*}
These two ICs are not the unique constructions that may be used to estimate the corresponding expectation values; in \cref{sec:kl-div} we discuss a wide variety of other ICs that have appeared in the statistics literature before.  We emphasize that none of these ICs are new, although we believe that our approach to deriving them in a unified way from variations on the K-L divergence is novel.  

Although these two constructions look superficially similar, we will find that the form of the PPIC makes it uniquely sensitive to fluctuations within a given data sample, and therefore more robust in the presence of noise.  On the other hand, the BPIC is somewhat more aggressive in selecting models with fewer parameters.  This can lead to lower variances at the cost of higher bias at finite sample size due to the bias-variance tradeoff (see discussion in \cref{subsec:asymptotic-bias} and explicit demonstration of this effect in our numerical results in \cref{sec:num-tests}).

In \cref{sec:asymptotic-appendix} we demonstrate that the BPIC and PPIC are asymptotically equivalent to the BAIC, so that in the limit of large sample size $N$ all three will give identical results; in this sense, the BPIC and PPIC may be viewed as finite-sample size modifications of the BAIC.  Including the possibility of data subset selection (see \cref{sec:data_subset_select}), our simplified approximate formulas for the case of least-squares fitting (see \cref{sec:least_squares}) are \cref{eqn:baic_subset}, \cref{eqn:bpic_subset}, and \cref{eqn:ppic_subset}; we reproduce these here for convenience:
\begin{align}
\BAIC_{\mu,P}=&\hat{\chi}^2(\vec{a}^*)+2k+2d_{\C},\\
\BPIC_{\mu,P}\approx&\hat{\chi}^2(\vec{a}^*)-\frac{1}{2}\tilde{H}_{ba}(\Sigma^*)_{ab}+\frac{1}{2}\tilde{g}_dT_{cba}(\Sigma_2^*)_{abcd}+3k+3d_{\C},\\
\nonumber \PPIC_{\mu,P}\approx&\hat{\chi}^2(\vec{a}^*)+2k+d_{\C} + Nd_{\C} \log\left(1 + \frac{1}{N} \right)\\
&-2\sum_{i=1}^{N}\log\left[1+\frac{1}{2}\left(\frac{1}{4}(g_i)_b(g_i)_a-\frac{1}{2}(H_i)_{ba}\right)(\Sigma^*)_{ab}+\frac{1}{4}(g_i)_dT_{cba}(\Sigma_2^*)_{abcd}\right],
\end{align}
where $k$ is the number of model parameters, $d_{\C}$ is the number of cut data points, and the other symbols (defined in \cref{sec:least_squares}) represent various derivatives of  $\chi^2$ functions.  For use in model averaging, all of these ICs should also include a model prior probability term $-2\log \pr(M_\mu)$ when it is non-constant, see \cref{subsec:review}.

The above formulas for BPIC and PPIC are approximate, based on expansion of integrals in the inverse sample size $1/N$, as discussed in \cref{sec:least_squares}.  For the BPIC and PPIC, we recommend the use of these formulas combined with an ``optimal truncation'' prescription, explained in \cref{subsec:super}, based on the theory of superasymptotics.  In the numerical tests we have performed, optimal truncation improves the agreement of these formulas with direct numerical evaluation of the associated integral formulas.  Optimal truncation has the additional practical benefit of ensuring that the sum appearing in the PPIC only includes logarithms with positive argument, so that the formula is always well-defined.

In order to understand the practical performance of these ICs, we carry out numerical tests on both synthetic data and on real lattice QCD data.  As we will see, based on both theoretical considerations and numerical performance, the PPIC is generally the most attractive information criterion for Bayesian model averaging.  The BAIC, although its performance in terms of uncertainty is somewhat worse in certain tests, is by far the simplest IC and often gives indistinguishable results from the more complex and expensive to calculate PPIC.  Based on these results, we recommend the PPIC as the primary information criteria for Bayesian model averaging in all cases, with BAIC as a backup option in use cases where computing the PPIC is impractical.  The BPIC is more aggressive in penalizing model complexity, particularly in the context of data subset selection, which can lead to statistically significant biases at finite sample size as seen in our numerical examples in \cref{sec:num-tests}.  As a result, we do not generally recommend the use of BPIC in practice.

\section{Model averaging and bias correction}
\label{sec:prelim}

In this section, we review some preliminary material necessary to understand and motivate the content of subsequent sections, including basic concepts of model averaging and statistical bias.  We then examine the relationship between bias in information criteria and bias in model-averaged statistical estimates, finding that the use of asymptotically unbiased ICs is key.

\subsection{Bayesian model averaging procedure}
\label{subsec:review}

Bayesian model averaging is a tool that allows for quantitative treatment of uncertainty due to model choice, in situations where many candidate models can plausibly describe a given data set. Such problems occur commonly in lattice field theory.  Even in situations where physical arguments strongly motivate the use of a single theory to describe the data, often the theory is an effective field theory that must be truncated, and uncertainty in the order of truncation is equivalent to model selection uncertainty. For a detailed discussion of Bayesian model averaging with derivations of the basic formulas appearing here, see \cite{Jay:2020jkz}.

Suppose we are interested in determining expectation values of functions of some model parameters $\vec{a}$, marginalized over a set of models $\{M\}$ from a set of data $\yd$. The key idea behind Bayesian model averaging is that we can obtain these expectation values as a weighted average over models,
\begin{align}
\label{eqn:model_avg} \avg{f(\mathbf{a})} =& \sum_{\mu} f(\mathbf{a}_{\mu}^*) \pr (M_{\mu} | \yd),\\
\sigma_{f(\mathbf{a})}^2=&\avg{f(\mathbf{a})^2}-\avg{f(\mathbf{a})}^2\\
\label{eqn:model_avg_var}=&\sum_{\mu}\sigma_{f(\mathbf{a}_{\mu})}^2\pr(M_{\mu}|\yd)+\sum_{\mu}f(\vec{a}_{\mu}^*)^2\pr(M_{\mu}|\yd)-\left(\sum_{\mu}f(\mathbf{a}_{\mu}^*)\pr(M_{\mu}|\yd)\right)^2,
\end{align}
where $\vec{a}_{\mu}^*$ denotes the best-fit parameters for the model $M_{\mu}$, and the probabilities $\pr (M_{\mu} |\yd)$ (the ``model weights'') represent the probability of each model given the data.  The quantity $\sigma_{f(\mathbf{a})}^2$ is the estimated variance of the expectation value $\avg{f(\mathbf{a})}$, and includes contributions both from statistical error within each model (the first term) as well as a ``systematic error'' contribution due to variation of the individual model estimates (second and third terms); see \cite{Jay:2020jkz} for further discussion.
The central problem in computing expectation values is thus to determine the model weights $\pr(M_{\mu}|\yd)$. From Bayes' theorem,
\begin{align}\label{eqn:bayes-model-wt}
\pr(M_{\mu}|\yd)=\frac{\pr(\yd|M_{\mu})\pr(M_{\mu})}{\pr(\yd)},
\end{align}
where $\pr(M_{\mu})$ is the model prior probability.  We will only consider cases where the data $\yd$ is fixed for all candidate models $M_{\mu}$, so $\pr(\yd)$ will henceforth be omitted when irrelevant.  The sum of model weights over the space of all models is normalized to 1,
\begin{align} \label{eqn:model_normalize}
\sum_{\mu}\pr(M_{\mu}|\yd) = 1.
\end{align}

At this point, we remark on the connection of model weights to the common idea of an ``information criterion'' (IC). Most ICs are defined explicitly in terms of a likelihood function $\pr(\yd|M_{\mu})$, so that a generic information criterion is
\beq \label{eqn:IC}
\IC_{\mu} \equiv -2 \log \pr(\yd | M_{\mu}).
\eeq
By Bayes' theorem, we may define a similar concept of information criterion for use in model averaging simply by including the model prior probabilities,
\beq \label{eqn:IC_MA}
\IC_{\MA,\mu} \equiv -2 \log \pr(M_{\mu} | \yd) = -2\log\pr(M_{\mu})+\IC_{\mu},
\eeq
where the subscript ``MA'' denotes model averaging.  We will generally work with the former version of the ICs in the text below, to avoid repeatedly writing the factor $-2 \log \pr(M_\mu)$ that is shared between all of them.

We note in passing that any constant terms (i.e., identical for all models considered) in the definition of an IC can be safely ignored, since they will cancel when the normalization condition \cref{eqn:model_normalize} is applied.  This applies to the factor $-2 \log \pr(M_\mu)$ in the case of a flat model prior, i.e., if equal prior probability is assigned to all models $M_\mu$ then this term becomes a constant and drops out.  We also exploit this observation to define an equivalent formula for the unnormalized model weight,
\beq \label{eq:model_weight_remove_min}
\pr(M_\mu | \yd) \propto \exp \left[ -(\IC_{\MA,\mu} - \min_\nu \IC_{\MA, \nu}) / 2 \right].
\eeq
This form of the model weight formula is less prone to numerical instability when working at fixed floating-point precision, and is used in practice in our numerical implementations.

\subsection{Bias correction and model averaging}
\label{subsec:asymptotic-bias}

There are a staggering number of information criteria present in the statistics literature.  To motivate a specific subset of ICs to study, we first introduce the concept of bias for statistical estimators.  Roughly speaking, bias measures the difference between an estimator and the true population value that the estimator is intended to reflect.  There are many possible sources of bias in any statistical study; we will focus here on estimator bias, arising specifically from the choice of sample estimator and not from other systematic effects.

Suppose that $\yd$ is a random sample of size $N$ drawn independently from an unknown true distribution with probability density function $\pr_{M_{\T}}(z)$ (i.e., $\yd$ are iid samples). Consider a sample estimator $X(\yd)$ for some property $\xi$ of the true underlying probability distribution $\pr_{M_{\T}}(z)$ from which $N$ independent data samples are drawn.  The bias of $X(\yd)$ is defined as \cite{larsen2005introduction} %
\begin{equation} \label{eqn:bias-finite-sample}
b_y[X(\yd)] \equiv E_y[X(\yd) - \xi] = E_y[X(\yd)] - \xi,
\end{equation}
where $E_y$ denotes expectation with respect to the population distribution (i.e., the limit of an infinite number of trials, carried out at fixed sample size $N$.)  An unbiased estimator satisfies $b_y[X(\yd)] = 0$. The quantity
\begin{equation} \label{eqn:bias-asymptotic}
b_z[X(z)] = \lim_{N \rightarrow \infty} b_y[X(\yd)],
\end{equation}
is known as the asymptotic bias of the sequence of estimators $\{X(\yd)\}_{N\in\mathbb{N}}$.

Obviously, it is ideal if one can find a sample estimator that is unbiased at finite $N$.  However, it is not always practical to calculate (and hence correct for) the bias of a given estimator \emph{a priori}.  Instead, one can settle for removal of only the asymptotic bias.  In the context of lattice simulations, where lattice ``data'' are generated through a Monte Carlo process, the sample size $N$ tends to be quite large and it can always (in principle) be extended in order to approach the $N \rightarrow \infty$ limit.  For this reason, we insist on asymptotic unbiasedness as a primary quality of interest in lattice applications; this guarantees at least that any estimator bias will vanish in the large-$N$ limit.  For lattice applications where the goal is typically inference of some physical parameters in a well-motivated theoretical model, this requirement ensures that parameter estimates will converge to the correct answers as $N \rightarrow \infty$. 

It is important to emphasize that this goal (removal of asymptotic bias) is not universal across all fields of research. For example, in machine learning the model space is much less well-understood, and the primary goal is generally out-of-sample prediction rather than parameter inference.  As a result, machine learning applications are often better served by joint optimization of bias and variance; for an accessible review of this so-called ``bias-variance tradeoff," see \cite{Mehta2019}.  As will be demonstrated in \cref{sec:num-tests}, the use of model averaging itself represents a form of bias-variance tradeoff: inference with a single fixed model will typically have lower variance than a model average but at the risk of asymptotic bias if the model is wrong.

It is important to place the idea of bias properly in the present context of model selection.  Suppose that within the space of models $\{M_\mu\}$, there is one model $M_{\T}$  that corresponds to the true distribution $\pr_{M_{\T}}(z)$  (assuming that any model parameters are set to their correct asymptotic values $\vec{a}_{\T}^*$). Assuming that $M_{\T}$ is in the space of candidate models $\{M_\mu\}$,\footnote{The assumption that there is only one model $M_{\T}$ in the space of candidates is for simplicity. For example, say the true distribution is nested within two candidate models $M_{\T,1}$ and $M_{\T,2}$. In this case, $\lim_{N\rightarrow\infty}(\pr(M_{\T,1} | \yd)+\pr(M_{\T,2} | \yd))=\pr(M_{\T,1} | z)+\pr(M_{\T,2} | z)=1$, $\lim_{N\rightarrow\infty}\vec{a}_{\T,1}^*=\lim_{N\rightarrow\infty}\vec{a}_{\T,2}^*=\vec{a}_{\T}^*$,
and the model averaged results \cref{eqn:model_avg,eqn:model_avg_var} will be the same as if there were only one true model in the space of candidates.} asymptotically we should find that
\begin{align}
\label{eqn:asymptotic_model_weights} \lim_{N\rightarrow\infty}\pr(M_{\mu} | \yd)=\pr(M_{\mu} | z)=\begin{cases}1,&\mu=\T,\\0,&\mu\neq\T.\end{cases}
\end{align}

As discussed above, our primary goal is to remove asymptotic bias from model averaged estimates. If we assume that the model parameter estimation procedure is consistent\footnote{Informally, consistency here means that the parameter estimates converge in probability to their true, asymptotic values; see \cref{sec:bias-appendix} for a formal definition.} (this is true for, say, least-squares regression), then the asymptotic bias of \cref{eqn:model_avg} is bounded by
\begin{align}\label{eqn:asymptotic_bias}
\abs{b_z[\avg{f(\mathbf{a})}]}&\leq\sum_\mu\abs{f(\mathbf{a}_{\mu}^*)}\abs{b_z[\pr(M_\mu|z)]},
\end{align}
with probability $1$. For a derivation of this bound and the formal definition of consistency, see Appendix~\ref{sec:bias-appendix}. Therefore, we can eliminate the asymptotic bias from model averaged results by using an asymptotically unbiased estimator of the model weights, i.e., $b_z[\pr(M_\mu|z)]=0$.

It is worth noting in passing that the effect of bias on model-averaged results can be somewhat subtle.  As discussed briefly in \cite{Jay:2020jkz}, if several models give near-identical estimates $\ev{X}_{M}$ for some expectation value $\ev{X}$, then even the use of a biased model weight estimator will not lead to any significant bias in the estimate for $\ev{X}$ itself.  Nevertheless, we will insist that all of our model weight estimators be asymptotically unbiased.

\clearpage

\section{Kullback-Leibler divergence and information criteria}
\label{sec:kl-div}

The problem of estimating model probabilities can be reformulated in terms of the Kullback-Leibler (K-L) divergence, which measures the deviation of a candidate distribution from an underlying true distribution. The K-L divergence can be seen as a starting point for the standard methods of model fitting and model weight estimation. Framing the problem of model averaging by beginning with the K-L divergence will lead us naturally to the construction of alternative model weight estimators.

In \cite{Jay:2020jkz}, a specific formula for model weight was derived using basic manipulations of probability formulas. (For a detailed discussion of that paper's results and how they can be connected to the present analysis, see \cref{sec:app_last_paper}.)  However, if we view the central problem as estimation of probability distributions over the data $\yd$, then the model weight formula of \cite{Jay:2020jkz} is not unique; alternative methods of dealing with the model parameters can be used to give alternative estimators for the model weights. To understand this concept, we step back to understand how the problems of model selection and model fitting can be fundamentally viewed in terms of the K-L divergence. This approach, and many of the specific information criteria that we will consider as a result, follows closely the work of Zhou \cite{ZhouThesis,zhou2020posterior}.

Suppose that $\yd$ is a random sample of size $N$ drawn independently from an unknown true distribution with probability density function $\pr_{M_{\T}}(z)$. The basic goal of data modeling is to approximate $\pr_{M_{\T}}$ as closely as possible with a model distribution $\pr_{M_{\mu}}$. We may evaluate the ``closeness" of a given model distribution to the true probability density with the K-L divergence \cite{Kullback1951},
\begin{align}
\KL(M_{\mu}) &\equiv \int dz\ \left[\pr_{M_{\T}}(z) \log \pr_{M_{\T}}(z) - \pr_{M_{\T}}(z) \log \pr_{M_{\mu}}(z)\right] \\
\label{eqn:true-prob-meas} &= \int dF_{M_{\T}}(z) \left[ \log \pr_{M_{\T}}(z) - \log \pr_{M_{\mu}}(z)\right] \\
&= E_z \left[ \log \pr_{M_{\T}}(z) \right] - E_z \left[ \log \pr_{M_{\mu}}(z) \right] \label{eqn:kl-div},
\end{align}
where $F_{M_{\T}}(z)$ is the cumulative distribution function for future observation $z$ drawn from $\pr_{M_{\T}}(z)$, and $E_z[...]$ denotes an expectation with respect to the true distribution.  The K-L divergence, which is also known as the relative entropy, measures the information loss in the estimation of $\pr_{M_{\T}}(z)$ with the model distribution $\pr_{M_{\mu}}(z)$. The K-L divergence is positive semi-definite and vanishes if and only if $\pr_{M_{\T}}$ is equivalent in the sense of distributions to $\pr_{M_{\mu}}$. Because the first term in the divergence depends only on the unknown true distribution and not on the candidate model, minimizing the K-L divergence with respect to the model is equivalent to maximizing the quantity $E_z [ \log \pr_{M_{\mu}}(z) ]$.

The K-L divergence can be used as the starting point for a number of standard methods related to modeling data. For example, consider the usual case in which a parameter-dependent version of the model probability density is $\pr_{M_\mu}(z) = \pr(z|\vec{a},M_{\mu})$, i.e., our model probability distributions depend on additional parameters $\vec{a}$. Determination of the best-fit parameters $\vec{a}^*$ for a given model $M_\mu$ can be viewed as an optimization problem over $\pr(z|\vec{a}, M_\mu)$ such that $E_z [\log \pr(z|\vec{a}, M_\mu)]$ is maximized, so that $\KL(M_\mu)$ is minimized. In practice, the true distribution is inaccessible and estimators using a finite sample $\yd$ must be used instead.  A common practice is to estimate  $E_z[\log\pr(z|\vec{a},M_{\mu})]$ by the standardized out-of-sample log likelihood function:
\begin{align}\label{eqn:freq_likelihood}
E_z[\log\pr(z|\vec{a},M_{\mu})]\simeq\frac{1}{N}\sum_i \log\pr(y_i|\vec{a},M_{\mu})=\frac{1}{N}\log\pr(\yd|\vec{a},M_{\mu}).
\end{align}
where as introduced in \cref{subsec:summary}, the symbol $\simeq$ indicates that the right-hand side is an unbiased sample estimator of the quantity on the left.  Directly maximizing this likelihood function gives the quantity $\mathbf{a}^*_{\MLE}$, known as the maximum likelihood estimator (MLE).  The MLE is commonly used in the frequentist literature.  On the other hand, in Bayesian modeling the distribution of the parameters is inferred directly by applying Bayes' theorem to obtain the posterior (i.e., the likelihood weighted by the prior):
\beq\label{eqn:bayes_likelihood}
\log \pr (\vec{a} | \yd, M_\mu) \propto \frac{1}{N} \sum_i \log [\pr(y_i | \vec{a}, M_{\mu}) \pr(\vec{a} | M_\mu)^{1/N}],
\eeq
where the $1/N$ exponent on the prior distribution ensures that this summed version is equivalent to the conventional posterior estimate defined with respect to the full dataset, $(1/N) \log [\pr(\yd | \vec{a}, M_\mu) \pr(\vec{a} | M_\mu)]$.  Maximizing the posterior probability over $\vec{a}$ gives the posterior mode (PM), $\vec{a}^\star_{\PM}$.  

For the case of model averaging or model selection, rather than a single model distribution, we would like to compare a set of models $\{M_{\mu}\}$, identifying $\pr_{M_\mu}(z) = \pr(z | M_\mu)$ for each model $M_\mu$ in the set. The model weights can then be related directly to the probability density in the K-L divergence using Bayes' theorem,
\beq
\pr(M_\mu | z) \propto \pr(z | M_\mu).
\eeq
Note that there is no explicit reference to the model parameters $\vec{a}$ here.  This observation is crucial to a more general treatment of model weights and model averaging. To restate this important idea in words: in the context of the K-L divergence, the model weights are determined by each model's predicted distribution over the data $\pr(z | M_\mu)$. Since $\pr_{M_{\T}}$ is clearly independent of the parameters $\vec{a}$, whatever expression represents the candidate model in the K-L divergence must also be independent of $\vec{a}$.  From this perspective, we are completely free to specify a prescription for dealing with the model parameters $\{ \vec{a} \}$. We may view each possible prescription as a variation on the standard definition of the K-L divergence.  These variations in turn may be used to directly define new information criteria.

In the discussion of bias for information criteria to follow, it will be necessary to consider two matrices defined from the log-likelihood: the Fisher information matrix $I_z$ and the negative Hessian matrix $J_z$, which are defined for a given model as
\begin{align}
(I_z)_{ab}(\vec{a})&\equiv E_z\left[\frac{\partial\log L(z;\vec{a})}{\partial\vec{a}_a}\frac{\partial\log L(z;\vec{a})}{\partial\vec{a}_b}\right], \label{eqn:fisher_asymptotic}\\
(J_z)_{ab}(\vec{a})&\equiv-E_z\left[\frac{\partial^2\log L(z;\vec{a})}{\partial\vec{a}_a\partial\vec{a}_b}\right] \label{eqn:hess_asymptotic},
\end{align}
where $L(z;\vec{a})$ denotes the asymptotic likelihood function---the left-hand side of either \cref{eqn:freq_likelihood} or \cref{eqn:bayes_likelihood}, depending on whether it is being evaluated in a frequentist or Bayesian context, respectively (in the latter case, this is the posterior probability function.)  Given a finite sample of size $N$, unbiased estimators for these two matrices are given by
\begin{align}
(I_N)_{ab}(\vec{a})&\equiv\frac{1}{N-1}\sum_{i=1}^{N}\left[\frac{\partial\log L(x_i;\vec{a})}{\partial\vec{a}_a}\frac{\partial\log L(x_i;\vec{a})}{\partial\vec{a}_b}\right], \label{eqn:fisher_sample}\\
(J_N)_{ab}(\vec{a})&\equiv-\frac{1}{N}\sum_{i=1}^{N}\frac{\partial^2\log L(x_i;\vec{a})}{\partial\vec{a}_a\partial\vec{a}_b}, \label{eqn:hess_sample}
\end{align}
where now $L(x_i;\vec{a})$ corresponds to the right-hand side of either \cref{eqn:freq_likelihood} (frequentist) or \cref{eqn:bayes_likelihood} (Bayesian). We note in passing that more general definitions exist for the negative Hessian and the Fisher information matrices \cite{Konishi1996}.

Finally, we will frequently assume that the parameter prior information does not grow too rapidly with increasing $N$, i.e.,
\begin{align}\label{eqn:prior_assumption}
\lim_{N\rightarrow\infty}N^{-1}\log\pr(\vec{a}|M_{\mu})=0.
\end{align}
So long as this condition holds (e.g., the prior information does not depend on $N$), the PM and MLE become identical as $N \rightarrow \infty$ and the influence of the prior term vanishes relative to the likelihood.
This assumption is common in the statistical analysis literature (see \cite{ZhouThesis,zhou2020posterior,Spiegelhalter2002,Ando2007} among others) and holds in the typical case where priors are independent of the data. In making this assumption, the distinction between the Bayesian and frequentist cases vanishes in the large-$N$ limit. For more a detailed discussion on this assumption, see Appendix 2 of \cite{Ando2007}.

\subsection{Plug-in K-L divergence}
\label{subsec:kl-div-plug}

We turn now to the problem of how to deal with model parameters in the estimation of $E_z[\log \pr_M(z)]$ in \cref{eqn:kl-div}.  A simple approach to dealing with model parameter dependence is to determine a ``best-fit'' value $\vec{a}^*$ and plug in this estimate to construct a predictive density function $\pr(z|\vec{a}^*,M_{\mu})$, which no longer depends on $\vec{a}$. This leads to the {plug-in K-L divergence}:
\begin{align}\label{eqn:kl-div-plug-in}
\KL_{\plug}(M_{\mu}) \equiv E_z[\log\pr_{M_{\T}}(z)]-E_z[\log\pr(z|\vec{a}^*,M_{\mu})].
\end{align}
The exact definition of this estimator depends on the choice of ``best-fit'' estimator $\vec{a}^*$.

In the frequentist literature, it is common to use the maximum likelihood estimator  $\vec{a}_{\MLE}^*$ as the plug-in estimator. In this case, $N^{-1}\log\pr(\yd|\vec{a}_{\MLE}^*,M_{\mu})$ is an asymptotically biased estimator of $E_z[\log\pr(z|\vec{a}_{\MLE}^*,M_{\mu})]$, as discussed in \cite{Takeuchi1976,Stone1977,Shibata1989,Konishi1996}.  Given a finite sample, we may construct an estimator $b_N$ for the asymptotic bias $b_z$, which was done in \cite{Takeuchi1976}:
\begin{align}\label{eqn:bias-tic}
b_{N}^{\plug} = \frac{1}{N}\tr\left[J_N^{-1}(\vec{a}_{\MLE}^*)I_N(\vec{a}_{\MLE}^*)\right],
\end{align}
where $I_N$ and $J_N$ are the sample estimates of the (frequentist) log-likelihood Fisher information and negative Hessian matrices, as defined in \cref{eqn:fisher_sample,eqn:hess_sample} above.  Subtracting $b_N^{\plug}$ gives us an asymptotically unbiased estimator:
\beq
E_z[\log\pr(z|\vec{a}^*_{\MLE},M_{\mu})] \simeq \frac{1}{N}\sum_i \log\pr(y_i|\vec{a}^*_{\MLE},M_{\mu})  -\frac{1}{N}\tr\left[J_N^{-1}(\vec{a}_{\MLE}^*)I_N(\vec{a}_{\MLE}^*)\right].
\eeq
Multiplying by a conventional factor of $-2N$ (see \cref{eqn:IC} and \cref{eqn:freq_likelihood}) then gives the Takeuchi information criterion (TIC):
\begin{align}
\TIC_{\mu}=-2\log\pr(\yd|\vec{a}_{\MLE}^*,M_{\mu})+2\ \tr\left[J_N^{-1}(\vec{a}_{\MLE}^*)I_N(\vec{a}^*_{\MLE})\right].
\end{align}
We emphasize that this, and other information criteria to be introduced, may be viewed as formulas for the model weight $\pr(M_\mu | \yd)$ by way of \cref{eqn:IC_MA}.  To be explicit, the model-averaging version of the TIC is
\beq
\TIC_{\MA,\mu} = -2\log \pr(M_\mu) -2\log\pr(\yd|\vec{a}_{\MLE}^*,M_{\mu})+2\ \tr\left[J_N^{-1}(\vec{a}_{\MLE}^*)I_N(\vec{a}^*_{\MLE})\right],
\eeq
with an implied (unnormalized) model weight of $\pr(M_\mu | \yd) = \exp (-\TIC_{\MA,\mu}/2)$.  Models that minimize the TIC will be favored as they minimize the K-L divergence; this will be true for all of the information criteria discussed.

If we assume further that the true distribution belongs to the family of candidate distributions, then we may make the replacement $\tr\left[J_N^{-1}(\vec{a}_{\MLE}^*)I_N(\vec{a}_{\MLE}^*)\right]\rightarrow k$, where $k$ is the number of parameters (i.e., the dimension of the parameter vector $\vec{a}$). This replacement follows from the equivalence of the asymptotic Fisher matrix $I(\vec{a})$ and the Hessian matrix $J(\vec{a})$, as proven in \cite{Jay:2020jkz,Dixon2018,White1982} among others, so that the trace is over the $k\times k$ identity matrix. With this replacement, the TIC reduces to the Akaike information criterion (AIC) \cite{akaike1973information, akaike1974new,Akaike1998}:
\begin{align}\label{eqn:aic-freq}
\AIC_{\mu}=-2\log\pr(\yd|\vec{a}_{\MLE}^*,M_{\mu})+2k.
\end{align}

We emphasize that the AIC and TIC are frequentist information criteria and thus make no reference to the prior distribution. If we are interested in Bayesian applications, we must modify the derivation above to reflect this. It is shown in \cite{ZhouThesis} that plug-in usage of the posterior mode and removal of asymptotic bias leads to the Bayesian TIC (BTIC):
\begin{align} \label{eqn:btic-gen}
\BTIC_{\mu}=-2\log\pr(\yd|\vec{a}_{\PM}^*,M_{\mu})+2\tr\left[J_N^{-1}(\vec{a}_{\PM}^*)I_N(\vec{a}_{\PM}^*)\right],
\end{align}
where $\vec{a}_{\PM}^*$ is now the posterior mode.  In \cref{eqn:btic-gen}, the log-likelihood Fisher information and negative Hessian matrices are defined using the Bayesian form of \cref{eqn:fisher_sample,eqn:hess_sample}; henceforth, we will always use the Bayesian form of $I_N$ and $J_N$ unless otherwise stated. 

With the further assumption that the candidate models contain the true distribution, we may again make the replacement
 $\tr\left[J_N^{-1}(\vec{a}_{\PM}^*)I_N(\vec{a}_{\PM}^*)\right]\rightarrow k$, recovering a direct Bayesian analogue of the AIC, which we dub the ``Bayesian AIC":
 \begin{align} \label{eqn:atic-gen}
\BAIC_{\mu}=-2\log\pr(\yd|\vec{a}_{\PM}^*,M_{\mu})+2k.
\end{align}
As far as we know, the abbreviation ``BAIC" is so far unused in the statistics literature. The BAIC is not to be confused with ``a Bayesian information criterion'' (ABIC) (often referred to as ``Akaike's Bayesian information criterion" \cite{Konishi2008}), which can be derived from the K-L divergence by marginalizing over the parameter space \cite{Akaike1980,Kitagawa1997}, or with Schwarz's ``Bayesian information criterion" (BIC) \cite{Schwarz1978}, which also has connections to the marginalized K-L divergence. See \cref{subsubsec:kl-div-marg} and Appendix~\ref{sec:app_last_paper} for further discussion of the marginalized K-L divergence and associated information criteria.

Although the BTIC and BAIC are appropriate for use in Bayesian inference, we note that the use of a plug-in estimator implies the existence of a fixed underlying set of model parameters, which is more inline with the frequentist approach to inference.  A more natural Bayesian approach would consider model probability distributions rather than fixed values; this will be the case for the subsequent information criteria.

Unless otherwise stated, we denote the posterior mode $\vec{a}_{\PM}^*$ as $\vec{a}^*$ omitting the subscript from here forward.

\subsubsection{Digression: Marginalized K-L divergence}
\label{subsubsec:kl-div-marg}

Another AIC-like information criterion for Bayesian model averaging is proposed in \cite{Jay:2020jkz}. This information criterion is derived from the marginalized K-L divergence:
\begin{align}\label{eqn:kl-div-marg}
\KL_{\marg}(M_{\mu}) &\equiv E_z[\log\pr_{M_{\T}}(z)]-E_z\left[\log\int d\vec{a}\ \pr(z|\vec{a},M_{\mu}) \pr(\vec{a} | M_{\mu}) \right] \\
&= E_z[\log \pr_{M_{\T}}(z)] - E_z [ \log E_{\vec{a}} [ \pr(z | \vec{a}, M_\mu) ] ].
\end{align}
The expectation value over the parameters with respect to the prior probability distribution is
\beq\label{eqn:prior-expectation}
E_{\vec{a}}[...] \equiv \int d\vec{a}\ \pr(\vec{a} | M_\mu) (...),
\eeq
where we have assumed that the prior distribution $\pr(\vec{a} | M_\mu)$ is normalized.\footnote{In the case of improper priors, the integral in \cref{eqn:kl-div-marg} would be the same but would not be interpreted as an ``expectation value".} Written in this form, it is apparent that the marginalized K-L divergence has a strong dependence on the prior distribution. This is manifestly evident by comparison to \cref{eqn:kl-div}, where the use of $\KL_{\marg}$ is equivalent to the identification
\begin{equation}
\pr_{M_{\mu}}(z) \sim E_{\vec{a}} [\pr(z | \vec{a}, M_\mu)],
\end{equation}
which makes no reference at all to the data sample $\yd$, only to the prior parameter distribution.

Comparing two models using $\KL_{\marg}$ is equivalent to evaluating which model (together with its \emph{prior} parameter distribution) is more effective in describing the observed data.  This may be desirable in specific contexts, but attempting to use the marginalized K-L divergence in more typical cases where the posterior parameter values are of interest can lead to counterintuitive effects such as the Jeffreys-Lindley paradox in which the results under certain choices of prior become fully independent of the data (see Appendix~\ref{sec:app_last_paper} for further discussion).    

By approximating the integral in \cref{eqn:kl-div-marg} to leading order in large $N$ and appealing to the use of a cross-validation method to set the priors \cite{Jay:2020jkz}, one can obtain from $\KL_{\marg}$ what we will refer to as ``ABIC$_{\CV}$", a variation of Akaike's Bayesian information criterion:
\begin{align}\label{eqn:abic_cv}
\ABIC_{\CV,\mu}=-2\log\left[\pr(\yd|\vec{a}^*,M_{\mu})\pr(\vec{a}^*|M_{\mu})\right]+2k.
\end{align}
The ABIC$_{\CV}$, which is just called ``AIC" in \cite{Jay:2020jkz}, is identical to the BAIC except for the use of the posterior rather than the likelihood (for emphasis, both use the posterior mode $\vec{a}_{\PM}^*$ for the plug-in estimator $\vec{a}^*$).  While the ABIC$_{\CV}$ formula is asymptotically equivalent to the BAIC when \cref{eqn:prior_assumption} holds (e.g., the prior is $N$-independent), the use of cross-validation requires the priors to be adjusted as more data is accumulated, giving a prior that depends too strongly on $N$.  The full ABIC (without the use of cross-validation) has not been shown to be asymptotically unbiased, and in fact appears to differ by $O(\log N)$ terms from the (asymptotically unbiased) BAIC at large $N$.  Due to concerns regarding its asymptotic bias, we do not study the marginalized K-L divergence further here.  Appendix~\ref{sec:app_last_paper} contains some further discussion of the marginalized K-L divergence and the connection to the ABIC$_{\CV}$ from \cite{Jay:2020jkz}.

\subsection{Posterior averaged K-L divergence}
\label{subsec:kl-div-post}

Though adaptations like the BTIC and BAIC exist, the plug-in prescription is an inherently frequentist approach as it considers the underlying model parameters fixed. In Bayesian inference, parameter estimates are given as probability distributions. In light of this distinction, it is natural to consider averaging over the posterior distribution to measure deviations from model truth. This prescription gives the posterior averaged K-L divergence:

\beq \label{eqn:kl-div-post}
\KL_{\post}(M_\mu) \equiv E_z[ \log \pr_{M_{\T}}(z)] - E_z[ E_{\vec{a} | \yd} [\log \pr(z | \vec{a}, M_\mu)] ],
\eeq
where the expectation value over parameters with respect to the posterior distribution is 
\beq \label{eqn:post-avg-def}
E_{\vec{a} | \yd} [ ... ] \equiv \frac{\int d\vec{a}\ \pr(\vec{a} | \yd, M_\mu) (...)}{\int d\vec{a}\ \pr(\vec{a} | \yd, M_\mu)} = \frac{\int d\vec{a}\ \pr(\yd | \vec{a}, M_\mu) \pr(\vec{a} | M_\mu) (...)}{\int d\vec{a}\ \pr(\yd | \vec{a}, M_\mu) \pr(\vec{a} | M_\mu)}.
\eeq
With a trivial rewriting of \cref{eqn:kl-div-post} as
\beq \label{eqn:kl-div-post-trivial-rewrite}
\KL_{\post}(M_\mu) \equiv E_z[ \log \pr_{M_{\T}}(z)] - E_z[\log\exp(E_{\vec{a} | \yd} [\log \pr(z | \vec{a}, M_\mu)])],
\eeq
we identify $\exp(E_{\vec{a} | \yd} [\log \pr(z | \vec{a}, M_\mu)])$ as the relevant predictive distribution estimating $\pr_{M_{\mu}}(z)$; there is no common name associated with this distribution.  This rearrangement shows that in the sense of predictive distributions, the posterior averaged K-L divergence is somewhat less natural compared to the posterior predictive K-L divergence defined in \cref{subsec:kl-div-post} below.  We note in passing that unlike the predictive distributions associated with the plug-in or posterior predictive K-L divergences, this predictive distribution is not obviously properly normalized.  This does not have any obvious impact on the derivations to follow, but it may be interesting to explore the normalization of the predictive distribution in future work.

As above, to convert this to a useful information criterion we must approximate the second term in $\KL_{\post}(M_\mu)$ at finite sample size.  One way to do so is to replace the expectation over $z$ by using a sum over the sample data, which in turn will require a bias correction term similar to the BTIC.  This approach, which was proposed by Zhou \cite{ZhouThesis,zhou2020posterior}, gives the posterior averaging information criterion (PAIC):
\begin{align}\label{eqn:paic_unsimp}
\PAIC_{\mu}=-2 E_{\vec{a}|\yd}[\log\pr(\yd|\vec{a},M_{\mu})]+2\tr[J_N^{-1}(\vec{a}^*)I_N(\vec{a}^*)].
\end{align}

Evaluation of the PAIC requires carrying out a full integration of the posterior-weighted likelihood over the parameter space to evaluate $E_{\vec{a} | \yd}$, which may be difficult or impractical.  Historically, alternative ways of estimating $E_z[E_{\vec{a} | \yd}[\log \pr(z | \vec{a}, M_\mu)]]$ appeared in the literature well before the PAIC.  This was first attempted in \cite{Spiegelhalter2002} where the deviance information criterion (DIC) was proposed:
\begin{align}
\DIC&=-2\log \pr(\yd | E_{\vec{a} | \yd}[\vec{a}], M_\mu)+2p_D,
\end{align}
where
\begin{align}
p_D&=-2E_{\vec{a} | \yd}[\log \pr(\yd | \vec{a}, M_\mu)]+2\log \pr(\yd | E_{\vec{a} | \yd}[\vec{a}], M_\mu).
\end{align}
This DIC is defined by analogy to the BAIC where the posterior mean $E_{\vec{a} | \yd}[\vec{a}]$ is an alternative parameter plug-in to the posterior mode $\vec{a}^*$, and $p_D$ is interpreted as an effective number of parameters. $E_z[E_{\vec{a} | \yd}[\log \pr(z | \vec{a}, M_\mu)]]$ arises implicitly in the DIC through $p_D$. 

Note that like the BAIC, the DIC is defined to estimate $\KL_{\plug}$ rather than $\KL_{\post}$. It is only in correcting for asymptotic bias that we see estimates of $E_z[ E_{\vec{a} | \yd} [\log \pr(z | \vec{a}, M_\mu)] ]$ appear. This is what inspired studies of $\KL_{\post}$ and supports the idea that all of the variants of K-L divergence discussed here are equivalent in some sense. 

The DIC has since been criticized for its heuristic derivation and tendency to overfit observed data as it under-penalizes overly complex models \cite{Robert2002}, and thus will not be discussed further here (although a more detailed exploration of the DIC compared to the other ICs defined here could be an interesting direction for future study). A more rigorous alternative to the DIC was studied in \cite{Ando2007} where the Bayesian predictive information criterion (BPIC) was introduced:
\begin{align}\label{eqn:bpic_unsimp}
\nonumber \BPIC_{\mu}=& -2\log\left(\pr(\yd|\vec{a}^*,M_{\mu})\pr(\vec{a}^*|M_{\mu})\right)+2E_{\vec{a}|\yd}[\log\pr(\vec{a}|M_{\mu})]\\
&+2\tr[J_N^{-1}(\vec{a}^*)I_N(\vec{a}^*)]+k.
\end{align}
The BPIC has also been studied in the context of Bayesian model averaging \cite{Ando2008}.

Explicitly, the BPIC trades the integration over the full posterior distribution for an integration over the prior distribution $\pr(\vec{a} | M_\mu)$.  This is accomplished by including a plug-in estimator in the asymptotic bias correction, which cancels the integral of $\log \pr(\yd | \vec{a},M_{\mu})$ from $\KL_{\post}$.
In other words, the BPIC is defined as
\begin{align}
\BPIC_{\mu}=-2E_{\vec{a}|\yd}[\log\pr(\yd|\vec{a},M_{\mu})]+2Nb_N^{\BPIC},
\end{align}
where the asymptotic bias $b_z^{\BPIC}$ is estimated by
\begin{align}
\nonumber Nb_z^{\BPIC}\simeq Nb_N^{\BPIC}=&-\log(\pr(\yd|\vec{a}^*,M_{\mu})\pr(\vec{a}^*|M_{\mu}))+E_{\vec{a}|\yd}[\log(\pr(\yd|\vec{a},M_{\mu})\pr(\vec{a}|M_{\mu}))]\\
&+\tr[J_N^{-1}(\vec{a}^*)I_N(\vec{a}^*)]+\frac{1}{2}k,
\end{align}
in contrast to the PAIC, which is defined as
\begin{align}
\PAIC_{\mu}=&-2 E_{\vec{a}|\yd}[\log\pr(\yd|\vec{a},M_{\mu})]+2Nb_N^{\PAIC},\\
Nb_z^{\PAIC}\simeq&Nb_N^{\PAIC}=\tr[J_N^{-1}(\vec{a}^*)I_N(\vec{a}^*)].
\end{align}
For emphasis, the PAIC and the BPIC are both estimators of \cref{eqn:kl-div-post} and differ only by subleading terms in their bias corrections, a difference that vanishes in the $N\rightarrow\infty$ limit.

The BPIC is easier to evaluate in many situations; we will find below that for the least-squares case, when using approximate expressions for the integrals, the BPIC is much more accurate than the PAIC for smaller sample sizes.  On the other hand, the PAIC does have certain advantages.  Specifically, by using a plug-in estimator in its asymptotic bias correction, the BPIC loses estimation efficiency compared to the PAIC when the posterior is asymmetric or when there is nonzero correlation between parameters; furthermore, the BPIC is not well-defined when the prior distribution is degenerate. For more detail on these cases, see \cite{ZhouThesis,zhou2020posterior}.

As above, under the usual assumption of correct model specification, we may replace the trace in the BPIC and PAIC bias correction terms with the number of parameters $k$; we do not give these variations separate names. As a brief aside (assuming correct model specification for simplicity), we see from \cref{eqn:bpic_unsimp} that the BPIC includes a $3k$ term in contrast to the BAIC's $2k$ term; as shown in Appendix~\ref{sec:paic_formulas}, evaluting the posterior average in \cref{eqn:paic_unsimp} gives rise to an additional $k$ totaling $3k$ for the PAIC as well.  The BPIC and PAIC are still asymptotically equivalent to the BAIC, where throughout the paper we use the term ``asymptotically equivalent'' when referring to information criteria to mean equivalence in the context of model choice; see Appendix~\ref{sec:asymptotic-appendix} for further discussion.  For the sake of concreteness, we hold off on further discussion of the $3k$ term until \cref{subsec:bpic}.

\subsection{Posterior predictive K-L divergence}
\label{subsec:kl-div-pred}

As a final variation on construction of the K-L divergence, we may observe that the way in which the posterior average was constructed in \cref{eqn:kl-div-post} is not unique. Specifically, the second expectation value can be moved inside the logarithm,\footnote{We note in passing that a similar rearrangement may be done to the marginalized K-L divergence, defining $\KL'_{\marg}(M_\mu) \equiv E_z[\log \pr_{M_{\T}}(z)] - E_z[ E_{\vec{a}}[ \log \pr(z | \vec{a}, M_\mu)]]$. As far as we know the resulting IC from this definition has not been studied in the literature, but it is not obvious that it has any advantage compared to the other ICs discussed so far, and it is likely to suffer from the same difficulties as the marginalized K-L divergence.  Based on the Jensen's inequality argument given in the text below, this IC would also perform worse than $\KL_{\marg}$ by transposing the logarithm in this way.} defining the posterior predictive K-L divergence:
\beq \label{eqn:kl-div-post-pred}
\KL_{\pred}(M_\mu) \equiv E_z[\log \pr_{M_{\T}}(z)] - E_z[ \log E_{\vec{a} | \yd} [ \pr(z | \vec{a}, M_\mu) ]].
\eeq
The name ``posterior predictive'' follows from the observation that we may rewrite
\begin{align}
E_{\vec{a} | \yd} [\pr(z | \vec{a}, M_\mu) ] &\propto \int d\vec{a}\ \pr(z | \vec{a}, M_\mu) \pr(\yd | \vec{a}, M_\mu) \pr (\vec{a} | M_\mu) \\
&\propto \int d\vec{a}\ \pr(z | \vec{a}, M_\mu) \pr(\vec{a} | \yd, M_\mu) \\
&\equiv \pr(z | \yd, M_\mu),
\end{align}
which is the predictive distribution for future observation $z$ obtained by averaging the model parameters over the posterior distribution. Though we will see that the commutation of the expectation and the log will add some computational complexity in practice, the use of the posterior predictive distribution as an estimator of $\pr_{M_{\mu}}(z)$ makes $\KL_{\pred}$ somewhat more natural in Bayesian inference than $\KL_{\post}$ (cf. \cref{eqn:kl-div-post-trivial-rewrite}). A less heuristic motivation for $\KL_{\pred}$ over $\KL_{\post}$ follows from Jensen's inequality \cite{Jensen1906,Hunter2001}. Specifically, we have that for a general expectation operator $E[\dots]$ and random variable $X$
\begin{align}
E[\log X]\leq \log E[X].
\end{align}
Therefore,
\begin{align} \label{eqn:KL-ineq}
\KL_{\pred}\leq\KL_{\post}.
\end{align}

Since the K-L divergence is positive semi-definite, this in turn implies that $\KL_{\pred}$ will be closer to zero.  In other words, minimizing the posterior averaged K-L divergence with respect to the set of models and parameter values can never do better than minimizing the posterior predictive K-L divergence.  It is meaningful to compare these two K-L divergences by framing them both in terms of the non-parametric K-L divergence \cref{eqn:kl-div}.  The inequality above then implies that in terms of closeness to the true distribution $\pr_T(z)$, using the posterior predictive $\log E_{\vec{a}|\yd}[\pr(z|\vec{a},M_\mu)]$ as our choice for the non-parametric $\pr_{M_\mu}(z)$ will never underperform the posterior average $E_{\vec{a}|\yd}[\log \pr(z|\vec{a}, M_\mu)]$ for a given choice of model $M_\mu$.

One bias-corrected information criterion corresponding to $\KL_{\pred}$ is the posterior predictive information criterion (PPIC):
\begin{align}\label{eqn:ppic-unsimp}
\PPIC_{\mu}=-2\sum_{i=1}^{N}\log\pr(y_i|\yd,M_{\mu})+2\tr[J_N^{-1}(\vec{a}^*)I_N(\vec{a}^*)].
\end{align}
The PPIC was proposed in \cite{ZhouThesis} as an ad-hoc information criterion based on certain formulations of Bayes factors.  However, the first term of the PPIC appears earlier in the literature; the first instance we are aware of is \cite{Konishi1996}.  Other information criteria which include the same first term include the predictive information criterion (PIC) \cite{Kitagawa1997,Konishi2008} and the Watanabe-Akaike information criteria (WAIC) \cite{Gelman2014}. These information criteria differ from the PPIC in their bias corrections: the PIC lacks a simple general definition for the bias term, and the WAIC includes additional posterior averages, resulting in a higher complexity.
We will focus on the PPIC here as its bias correction is of the same form as the other information criteria discussed.  We show in Appendix~\ref{sec:asymptotic-appendix} that the PPIC is asymptotically equivalent to the BAIC.

Although the modification of the K-L divergence to obtain the PPIC rather than the PAIC seems relatively minor, we will find in practice that the PPIC is uniquely sensitive to information encoded in the individual fluctuations within the sample $\yd$, and as such can be particularly effective for certain problems.

We note in passing that some information criteria in the literature can be derived using a combination of the various K-L divergence formulations discussed here (e.g., the WAIC can be written as $2\PAIC-\PPIC-2\tr[J_N^{-1}(\vec{a}^*)I_N(\vec{a}^*)]$). It is unclear to us whether doing so has any theoretical or practical motivations, hence we ignore these alternatives and present only the more natural information criteria defined above.

\section{Specialization to least-squares regression}
\label{sec:least_squares}

In this section, we specialize our discussion of Bayesian model averaging, the K-L divergence, and information criteria to least-squares regression, which is of primary interest in the context of lattice simulations. We start with a brief overview of least-squares fitting and the relevant notation. The BAIC is discussed as a reformulation of the AIC-like information criterion proposed in \cite{Jay:2020jkz}. We then discuss an asymptotic integral approximation known as Laplace's method that will be needed in the subsequent sections. Next, we return to some of the aforementioned information criteria (BPIC, PAIC, and PPIC) and give approximations for each in the case of least-squares fitting. Lastly, we discuss improvements to the information criteria approximations.

\subsection{Least-squares fitting}
\label{subsec:least-sqr-fit}

The discussion thus far has been completely general with regards to the probability distributions appearing in the K-L divergence, information criteria, and model averaging formulas. We now specialize our discussion to the case of least-squares regression of a model $M_\mu$ with parameters $\vec{a}$ to a set of data $\yd$. The likelihood function is
\begin{align}
\pr(\yd|\vec{a},M_{\mu})=\prod_{i=1}^{N}\frac{1}{(2\pi)^{d/2}(\det\Sigma)^{1/2}}\exp\left[-\frac{1}{2}\chi_i^2\right],
\end{align}
where
\begin{align}\label{eqn:chi2_i}
\chi_i^2\equiv\left(y_i-f_{\mu}(\vec{a})\right)^T\Sigma^{-1}\left(y_i-f_{\mu}(\vec{a})\right),
\end{align}
is the standard chi-squared goodness of fit statistic, which involves the data sample $y_i$, the model function $f_{\mu}(\vec{a})$ corresponding to the model $M_{\mu}$, and the covariance matrix between the individual samples $\Sigma=\frac{1}{N-1}\sum_{i=1}^{N}\left(y_i-\bar{y}\right)\left(y_i-\bar{y}\right)^T$; we assume the samples are drawn independently from some underlying distribution. The dimension of a single observation vector $y_i$ is denoted by $d$, and the number of independent observations drawn from the true distribution is $N$.

As for the prior distribution, a common choice is to use a multivariate Gaussian \cite{Lepage2002,Schindler2009},
\begin{align}\label{eqn:param-prior}
\pr(\vec{a}|M_{\mu})=\frac{1}{(2\pi)^{k/2}(\det\tilde{\Sigma})^{1/2}}\exp\left[-\frac{1}{2}(\vec{a}-\tilde{\vec{a}})^T\tilde{\Sigma}^{-1}(\vec{a}-\tilde{\vec{a}})\right],
\end{align}
where $k$ is the number of fit parameters in model $M_{\mu}$, $\Tilde{\Sigma}$ is the prior covariance matrix, and $\tilde{\vec{a}}$ is the prior central value. We define the ``prior chi-squared statistic''
\begin{equation}
\tilde{\chi}^2 \equiv -2 \log \pr(\vec{a} | M_\mu)
\end{equation}
for later use.  In the multivariate Gaussian case, $\tilde{\chi}^2\equiv(\vec{a}-\tilde{\vec{a}})^T\tilde{\Sigma}^{-1}(\vec{a}-\tilde{\vec{a}}) + ({\rm const.})$, but the approximate formulas derived below apply in general.  Unless otherwise stated, we will assume that \cref{eqn:prior_assumption} holds, i.e., that the prior information grows sufficiently slowly with the sample size.

Since we are only considering the case of a fixed data set, the overall normalization of the likelihood function $\pr(\yd|\vec{a},M_{\mu})$ will be the same for all models and can be ignored.\footnote{The problem of data subset selection is treated as a model variation problem, so that this normalization factor remains irrelevant; see \cref{sec:data_subset_select}.  See also further discussion of this issue in \cite{Neil:2023pgt}. }  On the other hand, in the presence of models with varying numbers of parameters the normalization of prior distribution $\pr(\vec{a}|M_{\mu})$ may not be omitted.

The best-fit point $\vec{a}^*$ is the posterior mode, which maximizes the posterior or, equivalently, minimizes the negative log posterior:
\begin{align}\label{eqn:chi2_aug}
-2\log\left[\pr(\yd|\vec{a},M_{\mu})\pr(\vec{a}|M_{\mu})\right]-(N-1)d=\hat{\chi}^2(\vec{a})+\tilde{\chi}^2(\vec{a})\equiv\chi_{\aug}^2(\vec{a}),
\end{align}
where
\begin{align}\label{eqn:chi2_hat}
\hat{\chi}^2(\vec{a})\equiv\sum_{i=1}^{N}\chi_i^2(\vec{a})-(N-1)d=\left(\bar{y}-f_{\mu}(\vec{a})\right)^T\hat{\Sigma}^{-1}\left(\bar{y}-f_{\mu}(\vec{a})\right),
\end{align}
and $\hat{\Sigma}\equiv\Sigma/N$ is the standard-error covariance matrix. $\chi_{\aug}^2$ is the so-called augmented chi-squared function \cite{Lepage2002}.  The $(N-1)d$ term appears when converting between the use of the sample-based $\sum_i \chi_i^2$ and the mean-based $\hat{\chi}^2$, which subtracts it by convention.  For data with a constant dimension over all models, the $(N-1)d$ term is constant and thus can be ignored.  However, it will play an important role in \cref{sec:data_subset_select} where we consider model averaging over different data subsets, i.e., variable $d$.

\subsection{BAIC}
\label{subsec:aic}

In the context of least-squares regression, the BAIC takes the form
\begin{align}\label{eqn:baic}
\BAIC_{\mu}\equiv&\hat{\chi}^2(\vec{a}^*) +2k.
\end{align}
We reiterate here that unless noted otherwise, throughout this work the plug-in estimator $\vec{a}^*$ is the posterior mode $\vec{a}_{\PM}^*$. For common applications in lattice simulations with weakly informative priors (so that $\tilde{\chi}^2$ is negligible compared to $\hat{\chi}^2$), the BAIC is nearly identical to the ABIC$_{\CV}$:
\begin{align}\label{eqn:abic_cv-least-sqr}
\ABIC_{\CV,\mu}\equiv& \chi_{\aug}^2(\vec{a}^*)+2k.
\end{align}
This is presented as simply the ``AIC" in \cite{Jay:2020jkz}. While the choice between BAIC and ABIC$_{\CV}$ should be inconsequential for most lattice applications when the priors are relatively uninformative, we omit further analysis of the ABIC$_{\CV}$ due to its lack of solid theoretical foundation (see Appendix~\ref{sec:app_last_paper}).

\subsection{Laplace's method}
\label{subsec:laplace}

For evaluation of the subsequent information criteria, we will need integrals of the form
\begin{align}\label{eqn:gen_int}
\mathcal{I}[\psi]=\int d\vec{a}\exp\left[-\frac{1}{2}\chi_{\aug}^2(\vec{a})\right]\psi(\vec{a}).
\end{align}
In the case of nonlinear least squares, this expression cannot be computed analytically in general.  One option is numerical evaluation of the integrals, but this can be relatively expensive as part of a fitting analysis and provides an additional source of numerical instability to deal with.  Our focus instead will be on the use of a closed-form approximation known in the asymptotics literature as Laplace's method. Specifically, we will write a next-to-leading-order (NLO) perturbation expansion in the inverse sample size $N^{-1}$ for \cref{eqn:gen_int}, which becomes increasingly accurate as $N\rightarrow\infty$. We have implemented integrals of the form in \cref{eqn:gen_int} numerically and verified the accuracy of our approximation.

The details of this approximation are summarized in Appendix~\ref{sec:app_laplace_method}. The main result is
\begin{align}\label{eqn:nlo_lap_approx}
\nonumber \mathcal{I}[\psi]\approx&(2\pi)^{k/2}|\Sigma^*|^{1/2}\exp\left[-\frac{1}{2}\chi_{\aug}^2(\vec{a}^*)\right]\\
\nonumber &\cdot\left(\psi(\vec{a}^*)+\frac{1}{2}H_{ba}(\Sigma^*)_{ab}-\frac{1}{2}g_dT_{cba}(\Sigma_2^*)_{abcd}-\frac{1}{2}\psi(\vec{a}^*)F_{dcba}(\Sigma_2^*)_{abcd}\right.\\
&\hspace{1em}\left.+\frac{1}{8}\psi(\vec{a}^*)T_{fed}T_{cba}(\Sigma_3^*)_{abcdef}\right),
\end{align}
where the inverse parameter covariance matrix is
\begin{equation}
\label{eqn:covariance}
(\Sigma^*{}^{-1})_{ab} = \frac{1}{2} \left. \frac{\partial^2 \chi_{\rm aug}^2}{ \partial a_a \partial a_b} \right|_{\vec{a}=\vec{a}^*},
\end{equation}
the higher-order contractions of the covariance matrix are
\begin{align}
\label{eqn:covariance_contractions}
(\Sigma_2^*)_{abcd}\equiv3(\Sigma^*)_{ab}(\Sigma^*)_{cd}\qquad (\Sigma_3^*)_{abcdef}\equiv9(\Sigma^*)_{ab}(\Sigma^*)_{cd}(\Sigma^*)_{ef}+6(\Sigma^*)_{ad}(\Sigma^*)_{be}(\Sigma^*)_{cf},
\end{align}
and the remaining tensors are given by
\begin{align}
\label{eqn:TF}
T_{abc}\equiv&\left.\frac{1}{6}\frac{\partial^3\chi_{\aug}^2}{\partial a_a\partial a_b\partial a_c}\right|_{\vec{a}=\vec{a}^*}, &
F_{abcd}\equiv&\left.\frac{1}{24}\frac{\partial^4\chi_{\aug}^2}{\partial a_a\partial a_b\partial a_c\partial a_d}\right|_{\vec{a}=\vec{a}^*},\\
\label{eqn:gH}
g_a\equiv&\left.\frac{\partial\psi}{\partial a_a}\right|_{\vec{a}=\vec{a}^*}, &
H_{ab}\equiv&\left.\frac{\partial^2\psi}{\partial a_a\partial a_b}\right|_{\vec{a}=\vec{a}^*}.
\end{align}
Note the use of Einstein summation notation for the tensor contractions. This result is in agreement with a special case of a more general integral computed in \cite{Kirwin2010}.

In the cases of interest, the integral \cref{eqn:gen_int} will appear with the following normalization:
\begin{align}
\frac{\mathcal{I}[\psi]}{\mathcal{I}[1]}=\frac{\int d\vec{a}\exp\left[-\frac{1}{2}\chi_{\aug}^2(\vec{a})\right]\psi(\vec{a})}{\int d\vec{a}\exp\left[-\frac{1}{2}\chi_{\aug}^2(\vec{a})\right]}.
\end{align}
As shown in Appendix~\ref{sec:app_laplace_method}, normalized integrals of this form can be approximated by first applying Laplace's method \cref{eqn:nlo_lap_approx} to both the numerator and denominator followed by a geometric expansion.  Keeping terms to NLO gives
\begin{align}\label{eqn:nlo_lap_approx_w_geo_exp}
\frac{\mathcal{I}[\psi]}{\mathcal{I}[1]}\approx\psi(\vec{a}^*)+\frac{1}{2}H_{ba}(\Sigma^*)_{ab}-\frac{1}{2}g_dT_{cba}(\Sigma_2^*)_{abcd}.
\end{align}
This geometric expansion maintains the same order of accuracy and is used in the probability literature \cite{Tierney1986}.

For the case of linear least squares, the $\chi_{\aug}^2$ function can be written as a quadratic form in the fit parameter vector $\vec{a}$ in which case the tensors $T$ and $F$ are identically zero. Furthermore, if $\psi$ is quadratic in $\vec{a}$ so that its higher derivatives vanish, then the ``approximations'' in \cref{eqn:nlo_lap_approx,eqn:nlo_lap_approx_w_geo_exp}  are in fact exact.  This will be the case for linear fit models and the BPIC, but not for the PPIC in which $\psi$ has exponential form.  For linear fit models, the PPIC integrals are Gaussian and can be computed exactly, but the expression is unwieldy and since this only works for linear models, we do not pursue it further here.

We emphasize that the rationale for this approximation is based on expansion in the inverse sample size $N^{-1}$.  To verify that the order of approximation is consistent, it is useful to note the $N$-dependence of these tensors: $\Sigma^*=O(N^{-1})$, $\Sigma_2^*=O(N^{-2})$, $\Sigma_3^*=O(N^{-3})$, $T,F=O(N)$, and $g,H=O(\psi(\vec{a}^*))$ as $N\rightarrow\infty$.  Thus, the approximation in \cref{eqn:nlo_lap_approx} is accurate to $O(N^{-1}\psi(\vec{a}^*))$.  We will consider cases where $\psi(\vec{a}^*)=O(1)$ and $\psi(\vec{a}^*)=O(N)$.

\subsection{BPIC}
\label{subsec:bpic}

In \cref{subsec:kl-div-post}, we introduced two other information criteria, the BPIC and PAIC, based on the posterior averaged K-L divergence in \cref{eqn:kl-div-post}. Here we specialize our discussion of the BPIC and PAIC to the case of least-squares regression (see \cref{subsec:least-sqr-fit}) with correct model specification (so that we may replace $\tr\left[J^{-1}(\vec{a}^*)I(\vec{a}^*)\right]\rightarrow k$). Using the NLO Laplace approximation discussed in \cref{subsec:laplace}, we give a computationally efficient approximation of the BPIC. We will not pursue the PAIC further in the body of the text due to the lower order of accuracy of the NLO Laplace approximation in this case (see discussion below); the relevant formulas for the PAIC are summarized in Appendix~\ref{sec:paic_formulas}.

First, we consider the BPIC. In the cases of interest, \cref{eqn:bpic_unsimp} reduces to (up to constant terms)
\begin{align}\label{eqn:bpic_pre_int}
\BPIC_{\mu}=\chi_{\aug}^2(\vec{a}^*)-E_{\vec{a}|\yd}\left[\tilde{\chi}^2(\vec{a})\right]+3k,
\end{align}
where
\begin{align}\label{eqn:post-avg}
E_{\vec{a}|\yd}\left[\dots\right]=\frac{\int d\vec{a}\exp\left[-\frac{1}{2}\chi_{\aug}^2(\vec{a})\right](\dots)}{\int d\vec{a}\exp\left[-\frac{1}{2}\chi_{\aug}^2(\vec{a})\right]}.
\end{align}
Using the NLO Laplace approximation with the geometric expansion simplification given in \cref{eqn:nlo_lap_approx_w_geo_exp}, we obtain
\begin{align}\label{eqn:bpic_int_simp}
E_{\vec{a}|\yd}\left[\tilde{\chi}^2(\vec{a})\right]\approx\tilde{\chi}^2(\vec{a}^*)+\frac{1}{2}\tilde{H}_{ba}(\Sigma^*)_{ab}-\frac{1}{2}\tilde{g}_dT_{cba}(\Sigma_2^*)_{abcd}.
\end{align}
where
\begin{align}
\tilde{g}_a\equiv&\left.\frac{\partial\tilde{\chi}^2}{\partial a_a}\right|_{\vec{a}=\vec{a}^*}, &
\tilde{H}_{ab}\equiv&\left.\frac{\partial^2\tilde{\chi}^2}{\partial a_a\partial a_b}\right|_{\vec{a}=\vec{a}^*}.
\end{align}
Substituting \cref{eqn:bpic_int_simp} into \cref{eqn:bpic_pre_int} gives
\begin{align}\label{eqn:bpic}
\BPIC_{\mu}\approx\hat{\chi}^2(\vec{a}^*)-\frac{1}{2}\tilde{H}_{ba}(\Sigma^*)_{ab}+\frac{1}{2}\tilde{g}_dT_{cba}(\Sigma_2^*)_{abcd}+3k.
\end{align}

An interesting feature of the BPIC is the last term in \cref{eqn:bpic_unsimp}, which gives the $3k$ term in \cref{eqn:bpic} as opposed to the $2k$ term in the BAIC.   The additional $k$ in \cref{eqn:bpic_unsimp} comes from the posterior averaging prescription (this is seen explicitly for the PAIC as shown in Appendix~\ref{sec:paic_formulas}).  As a result, the BPIC tends to favor more parsimonious models than the BAIC.  While this may seem like an advantage of the BPIC, we emphasize that additional parsimony comes at the cost of larger K-L divergence as discussed in \cref{subsec:kl-div-pred}.  Despite this difference, the BPIC remains asymptotically equivalent to the BAIC in the limit of infinite sample size, as shown in Appendix~\ref{sec:asymptotic-appendix}.

It is worth discussing a limit in which the other $O(1)$ terms in the BPIC cancel the additional $k$ (in this case, equality holds for \cref{eqn:KL-ineq}). Specifically, consider the case of infinitesimal prior widths, i.e., the prior information is infinitely constraining.  In this case, $\Sigma^*$ goes to $\frac{1}{2}\tilde{H}$, and the trace term in \cref{eqn:bpic} cancels the additional $k$ exactly (of course, there will be additional asymptotic bias from the $\hat{\chi}^2$ unless the prior center value is the true model parameter value).  The third term containing $\tilde{g}$ goes to zero, since in this limit $\vec{a}^* \rightarrow \tilde{\vec{a}}$.  In the case of finite prior widths, results will depend on the observed data and the additional $k$ persists. This behavior demonstrates that the additional dependence on the observed information over the prior information manifests itself by favoring parsimonious models  more than if there were no posterior averaging, as is the case for the BAIC.

As an aside, consider the PAIC. From \cref{eqn:paic_unsimp}, once again assuming correct model specification the PAIC is given by
\begin{align} \label{eqn:paic_expectation}
\PAIC_{\mu}=E_{\vec{a}|\yd}[\hat{\chi}^2(\vec{a})]+2k,
\end{align}
where $E_{\vec{a}|\yd}[\dots]$ is defined in \cref{eqn:post-avg}. We could proceed by attempting to approximate the expectation as we did in above for the BPIC.  However, a complication arises from the fact that $\hat{\chi}^2$ is itself $O(N)$.  This means that working to the same order in inverse sample size, $O(N^{-1})$, would require a next-to-next-to-leading order (NNLO) evaluation of the integral. This endeavor would require a notable increase in mathematical complexity. As discussed in \cref{subsec:kl-div-post}, the PAIC only differs from BPIC in subleading terms in the bias correction. Since the PAIC will be a lower order or more complicated version of the BPIC in practice with similar theoretical motivations, it suffices for our present purposes to omit further discussion of the PAIC. The corresponding results for the PAIC are given in Appendix~\ref{sec:paic_formulas}; these may be used in cases were the BPIC is not well defined, as discussed in \cref{subsec:kl-div-post}; see also the discussion in \cite{ZhouThesis,zhou2020posterior}.

\subsection{PPIC}
\label{subsec:ppic}

Finally, we turn to the PPIC, defined in \cref{eqn:ppic-unsimp}.  The PPIC involves the posterior predictive distribution, which can be rearranged to the form
\begin{align}
\pr(y_i | \yd, M_\mu) &= \int d\vec{a}\ \pr(y_i | \vec{a}, M_\mu) \pr(\vec{a} | \yd, M_\mu) \\
&= \frac{ \int d\vec{a}\ \pr(y_i | \vec{a}, M_\mu) \pr(\yd | \vec{a}, M_\mu) \pr(\vec{a} | M_\mu) }{\int d\vec{a}\ \pr(\yd | \vec{a}, M_\mu) \pr(\vec{a} | M_\mu) }.
\end{align}
Note that this contains both the combined likelihood for the entire dataset $\pr(\yd | \vec{a}, M_\mu)$ as well as the likelihood function for the $i$-th single data sample $\pr(y_i | \vec{a}, M_\mu)$.  In the case of least-squares regression with correct model specification, the PPIC then becomes
\begin{align}\label{eqn:ppic_least_squares_int}
\PPIC_{\mu}=-2\sum_{i=1}^{N}\log\frac{\int d\vec{a}\exp\left[-\frac{1}{2} \chi_{\aug}^2(\vec{a}) \right] \exp \left[-\frac{1}{2} \chi_i^2(\vec{a})\right]}{\int d\vec{a}\exp\left[-\frac{1}{2}\chi_{\aug}^2(\vec{a})\right]}+2k.
\end{align}
Here, we must compute $N$ integrals, one for each observation. Using \cref{eqn:nlo_lap_approx_w_geo_exp}, we obtain
\begin{align}\label{eqn:ppic}
\nonumber \PPIC_{\mu}\approx&\hat{\chi}^2(\vec{a}^*)+2k\\
&-2\sum_{i=1}^{N}\log\left[1+\frac{1}{2}\left(\frac{1}{4}(g_i)_b(g_i)_a-\frac{1}{2}(H_i)_{ba}\right)(\Sigma^*)_{ab}+\frac{1}{4}(g_i)_dT_{cba}(\Sigma_2^*)_{abcd}\right],
\end{align}
where
\begin{align}\label{eqn:indv-derivs}
(g_i)_a\equiv&\left.\frac{\partial\chi_i^2}{\partial a_a}\right|_{\vec{a}=\vec{a}^*}, &
(H_i)_{ab}\equiv&\left.\frac{\partial^2\chi_i^2}{\partial a_a\partial a_b}\right|_{\vec{a}=\vec{a}^*}.
\end{align}

We see from \cref{eqn:ppic} that through the final log term, the PPIC relies on information from each individual observation rather than solely on averaged and prior statistics like the other information criteria discussed. As we will see in \cref{subsec:ex2}, this sensitivity to sample fluctuations gives the PPIC the ability to parse out models with poor parameter estimates, a very attractive quality in model averaging.

\subsection{Superasymptotics and optimal truncation}
\label{subsec:super}

In the preceding subsections, approximate expressions for the BPIC and PPIC were obtained using the NLO Laplace approximation.  In the limit of large sample size $N$, this approximation should become increasingly accurate and the size of the subleading terms in $N^{-1}$ should become negligible.  However, in practice these information criteria will be computed and used at fixed, finite sample size.  For a given value of $N$, it is possible for the coefficients of our expansion to be such that the subleading terms are larger than the leading terms.  For example, there is no reason to suspect the gradients appearing in the NLO Laplace approximations should remain small in cases where the candidate model is unable to accurately represent the data.  While such poor models will likely be rejected based on their $\chi_{\aug}^2$ values alone, numerical issues can arise in model averaging when subleading terms are dominant.  In particular, this effect can cause logarithms with negative real arguments to arise in the PPIC.

As discussed previously, the Laplace approximation discussed in \cref{subsec:laplace} and derived in Appendix~\ref{sec:app_laplace_method} is an NLO perturbation expansion. This type of fixed-order expansion is known in the asymptotics literature as a Poincar\'e expansion \cite{Poincare1886,Berry1991a,Berry1991b,Boyd1999,Paris2011}. Another type of expansion is the ``superasymptotic" expansion.  First proposed by Sir George Gabriel Stokes for a similar integral approximation problem \cite{Stokes1847}, superasymptotics rely on the fact that an asymptotic series need not converge to give an accurate approximation with finitely many terms. Ignoring the case of singular perturbation expansions for simplicity, this means as additional terms are added to the asymptotic series, a formally divergent regular perturbation expansion has a ``convergent" part where terms decrease in magnitude algebraically in the perturbation parameter and a ``divergent" part which typically grows with additional terms, causing the series to diverge. A superasymptotic expansion is one that is ``optimally truncated" after the term of minimum modulus \cite{Berry1991a,Berry1991b,Boyd1999,Paris2011}.

While superasymptotics have been applied to an array of problems (see \cite{Costin1999} and references therein), it is often used for the method of steepest descent \cite{Berry1991a}. Since Laplace's method is a special case of the method of steepest descent, the use of superasymptotics is well suited for our purposes. In principle, this approach can achieve $O(\exp(-N))$ accurate integral approximations. (Since this error is exponentially small rather than algebraically small, superasymptotics is sometimes referred to as ``asymptotics beyond all orders" \cite{Kruskal1991}.) While we cannot guarantee this level of accuracy due to other sources of error in the derivation of each criteria (e.g., subleading terms in the bias corrections, see \cite{Ando2007,ZhouThesis,zhou2020posterior} for details), it does suggest the power of optimal truncation.  Furthermore, superasymptotic expansions are known to work well even when the perturbation parameter ($N^{-1}$ here) approaches $O(1)$ \cite{Boyd1999}, which will be the case for small sample sizes. Possible issues could arise with optimal truncation due to the high-dimensionality of the integrals considered, but this is unlikely assuming that the extremum in $\chi_{\aug}^2$ is a simple global minimum \cite{Paris2011}.

To benefit from these ideas, we propose using the Poincar\'e expansions previously developed, unless the second term in the NLO Laplace approximation is larger than the first.  If the second term is larger, then superasymptotics suggest that optimal truncation should leave only the leading order term.  Note that this prescription does not apply for the case of linear least squares for the BPIC, in which case the NLO ``approximations'' are exact.

Under the use of optimal truncation, a combined approximate formula for the BPIC becomes (suppressing the tensor indices for compactness)
\begin{align}\label{eqn:bpic_opt_trunc}
\BPIC_{\mu} \approx &\begin{cases}
\hat{\chi}^2(\vec{a}^*)-\frac{1}{2}\tilde{H}(\Sigma^*)+\frac{1}{2}\tilde{g} T (\Sigma_2^*)+3k, & |\frac{1}{2}\tilde{H} \Sigma^* - \frac{1}{2}\tilde{g} T \Sigma_2^*| < \hat{\chi}^2(\vec{a}^*), \\ 
\hat{\chi}^2(\vec{a}^*)+3k, & {\rm otherwise}.
\end{cases}
\end{align}

In the case of the PPIC, there are $N$ separate integrals for each of the data samples.  We can consider optimal truncation case-by-case for each individual integration:
\begin{align}
\frac{\int d\vec{a}\exp\left[-\frac{1}{2} \chi_{\aug}^2(\vec{a}^*) \right] \exp \left[-\frac{1}{2} \chi_i^2(\vec{a})\right]}{\int d\vec{a}\exp\left[-\frac{1}{2}\chi_{\aug}^2(\vec{a}^*)\right]} = \begin{cases}
\exp[-\frac{1}{2} \chi_i^2(\vec{a}^*)] (1 + \SL_i), & |\SL_i| < 1, \\
\exp[-\frac{1}{2} \chi_i^2(\vec{a}^*)], & {\rm otherwise,}
\end{cases}
\end{align}
where $\SL_i$ denotes the subleading terms associated with the $i$-th data sample,
\begin{align}
\SL_i=\frac{1}{2}\left(\frac{1}{4}(g_i)_b(g_i)_a-\frac{1}{2}(H_i)_{ba}\right)(\Sigma^*)_{ab}+\frac{1}{4}(g_i)_dT_{cba}(\Sigma_2^*)_{abcd}.
\end{align}
The form of the PPIC under optimal truncation is therefore
\begin{align}
\label{eqn:ppic_opt_trunc} \PPIC_{\mu}\approx&\hat{\chi}^2(\vec{a}^*)-2\sum_{\{i:|\SL_i|<1\}}\log\left(1+\SL_i\right)+2k.
\end{align}
Adopting this prescription thus eliminates the possibility of negative arguments within the logarithms.

There is also a wealth of literature on hyperasymptotics where the divergent part of the asymptotic series is used to obtain orders of accuracy superior to even those of superasymptotics \cite{Dingle1973,Berry1991a,Berry1991b,Howls1997,Boyd1999,Paris2011}. Since other sources of error would diminish the efficacy of such a procedure here, we advocate for the use of the superasymptotic schema described above rather than develop a hyperasymptotic one. This also maintains relative simplicity in the implementation of the information criteria.

We emphasize that the integral expansions we have carried out above are done primarily for ease of calculation and to reveal useful details about the structure of the various information criteria.  The only obstacles, in principle, to direct evaluation of the integral versions of each IC are the lack of a general analytic solution and the computational cost associated with accurate numerical evaluation.  However, we find in practice in our numerical tests that the NLO Laplace approximations with optimal truncation yield essentially identical results to direct integration with much lower computational cost.  We will discuss this comparison further in \cref{subsec:ex2}.

\section{Data subset selection}
\label{sec:data_subset_select}

As part of a lattice field theory analysis (or in Bayesian model averaging applications more broadly), it is often desirable to additionally select a subset of the data beyond which the model is not applied, i.e., selecting $d_{\C}$ dimensions of the samples $y_i$ to be ignored and fitting models only to the other $d_{\K}\equiv d-d_{\C}$ dimensions.  The subscript ``$\C$" refers to the ``cut" portion of the data and the subscript ``$\K$" refers to the ``kept" portion of the data.  A simple and common example of such a procedure in the context of lattice field theory is fitting a two-point correlation function $C(t)$ for the ground-state energy. The full model describing $C(t)$ involves an exponential decay series
\begin{align}
C(t)=\sum_{m=0}^{\infty}A_me^{-E_mt},
\end{align}
where $\{E_m\}$ increases monotonically with $m$. If only the first few states are of interest (as is often the case), it is sufficient to apply the model to times with $t\geq t_{\min}$ for some $t_{\min}$ after which the more rapidly decaying modes become negligible. Choosing $t_{\min}$ has (historically) often been done manually, although outside of model averaging there have been a variety of methods used for determination of $t_{\min}$ and/or estimation of associated systematic errors, see e.g.~\cite{BMW:2014pzb, LatticeStrongDynamics:2018hun,NPLQCD:2020ozd}.

Though the problem described is one of data subset selection, it can be reformulated as one of model selection. The key to doing so is to define a joint model that describes the full data set. First, choose a subset of the data to which the model $M_{\mu}$ is fit. Next, fit the remaining data to a ``perfect" model $M_{\perf}$ with zero degrees of freedom (with the use of the approximate formulas we give below, $M_{\perf}$ need not be constructed in practice.)  An example of such a perfect model is a polynomial of degree $d_{\C}-1$; in this example, fitting the data for model parameters is equivalent to finding the polynomial interpolant of the data where the differences between the model and sample means vanish identically.

To give a more explicit construction, we first define $P$ as the partition of each observation vector into $y_i=\begin{pmatrix} y_{\C,i} & y_{\K,i}\end{pmatrix}^T$, where $y_{\K,i}\in\mathbb{R}^{d_{\K}}$ is to be modeled by $M_{\mu}$ and $y_{\C,i}\in\mathbb{R}^{d_{\C}}$ is to be modeled by $M_{\perf,P}$.  We can similarly divide up the inverse sample standard-error covariance matrix as
\begin{align}\label{eqn:block-sigma-hat}
\hat{\Sigma}^{-1}=\begin{pmatrix} (\hat{\Sigma}^{-1})_{\C} & (\hat{\Sigma}^{-1})_{\off} \\ (\hat{\Sigma}^{-1})_{\off}^T & (\hat{\Sigma}^{-1})_{\K}\end{pmatrix},
\end{align}
where the subscript ``$\off$" stands for ``off-block-diagonal."

We then define the corresponding partitioned model $\phi_{M,P}(\vec{a})$ as
\begin{align}
(y_i-\phi_{M,P}(\vec{a}))_x=\begin{cases}
(y_i-\vec{a}_{\C})_x, & (y_i)_x\in y_{\C,i},\\
(y_i-f_M(\vec{a}_{\K}))_x, & (y_i)_x\in y_{\K,i},
\end{cases}
\end{align}
We note for later use that the cut part of the best-fit parameter $\vec{a}_{\C}^*$ is simply the mean of the cut data, i.e., $\vec{a}_{\C}^*=\bar{y}_{\C}$. Even though $\vec{a}_{\C}^*$ are known \textit{a priori}, we cannot take the cut parameter priors to be too constraining as this would violate \cref{eqn:prior_assumption} and thus not guarantee asymptotic unbiasedness of the information criteria.  Therefore, we will take the cut parameter priors to be infinitely diffuse, i.e., $(\tilde{\Sigma}_{\C})^{-1}\rightarrow0$, which is the limit where predictions rely solely on the data.

Based on these definitions, we can define a partition of the chi-squared function
\begin{align}
\hat{\chi}^2(\vec{a})=&\left(\bar{y}-\phi_{M,P}(\vec{a})\right)^T\hat{\Sigma}^{-1}\left(\bar{y}-\phi_{M,P}(\vec{a})\right)\\
=&\begin{pmatrix}\bar{y}_{\C}-\vec{a}_{\C}\\\bar{y}_{\K}-f_M(\vec{a}_{\K})\end{pmatrix}^T\begin{pmatrix} (\hat{\Sigma}^{-1})_{\C} & (\hat{\Sigma}^{-1})_{\off} \\ (\hat{\Sigma}^{-1})_{\off}^T & (\hat{\Sigma}^{-1})_{\K}\end{pmatrix}\begin{pmatrix}\bar{y}_{\C}-\vec{a}_{\C}\\\bar{y}_{\K}-f_M(\vec{a}_{\K})\end{pmatrix}\\
\equiv&\hat{\chi}_{\C}^2(\vec{a}_{\C})+\hat{\chi}_{\K}^2(\vec{a}_{\K})+2\hat{\chi}_{\off}^2(\vec{a}_{\C},\vec{a}_{\K}),
\end{align}
with analogous definitions for partitions of $\tilde{\chi}^2$ and $\chi_{\aug}^2$.

The ABIC$_{\CV}$ (see \cref{sec:app_last_paper}) is derived under this construction for the case of least-squares regression in \cite{Jay:2020jkz}:
\begin{align}\label{eqn:abic_cv-cut}
\ABIC_{\CV,\mu,P}=\chi_{\aug,\K}^2(\vec{a}^*)+2k+2d_{\C},
\end{align}
where $\chi_{\aug}^2(\vec{a}^*)$ is evaluated only for $M_{\mu}$, as the contribution from $M_{\perf,P}$ vanishes. Note that this result holds even without taking the infinitely diffuse cut prior limit, and without any assumptions on the structure of the correlations between the $y_{\C,i}$ and $y_{\K,i}$ partitions. The derivation for the BAIC is similar, giving
\begin{align}\label{eqn:baic_cut}
\BAIC_{\mu,P}=\hat{\chi}_{\K}^2(\vec{a}^*)+2k+2d_{\C},
\end{align}
which should also hold for any cut prior width satisfying \cref{eqn:prior_assumption}.  We will rederive this result below. 

A subtle point which appears here is the distinction between the sub-blocks of the inverse covariance matrix, e.g., $(\hat{\Sigma}^{-1})_{\K}$, and the inverse of a sub-block, e.g., $(\hat{\Sigma}_{\K})^{-1}$.  The former quantity contains indirect contributions from the cut portion of the data.  If we use the kept data exclusively to compute $\BAIC_{\mu,P}$ above, then this is equivalent to making the approximation
\begin{align}\label{eqn:sub-block-approx}
(\hat{\Sigma}^{-1})_{\K}\approx(\hat{\Sigma}_{\K})^{-1},
\end{align}
which is commonly used in the lattice community \cite{Jay:2020jkz}.  This approximation can avoid numerical instabilities that may occur when inverting the full $\hat{\Sigma}^{-1}$, particularly when $d_{\K} \ll d$.  In fact, this approximation is better than it may seem.  Even if $\hat{\Sigma}$ is estimated unbiasedly, simply inverting to find $\hat{\Sigma}^{-1}$ will introduce some finite-$N$ bias (that vanishes asymptotically).  The corrected estimator is \cite{Hartlap2007,Anderson2003}
\begin{align}
\hat{\Sigma}_{\rm BC}^{-1}=\frac{N-d-2}{N-1}\hat{\Sigma}^{-1}=\frac{N-d_{\C}-d_{\K}-2}{N-1}\hat{\Sigma}^{-1}.
\end{align}
where the subscript ``{\rm BC}" denotes the bias corrected inverse. The analogous expression for $(\hat{\Sigma}_{\K})_{\BC}^{-1}$ is
\begin{align}
(\hat{\Sigma}_{\K})_{\BC}^{-1}=\frac{N-d_{\K}-2}{N-1}(\hat{\Sigma}_{\K})^{-1}.
\end{align}
So, when $d_{\C}$ is large (i.e., when \cref{eqn:sub-block-approx} would be a poor approximation), $(\hat{\Sigma}_{\K})^{-1}$ will in fact give a less biased result than $(\hat{\Sigma}^{-1})_{\K}$ at finite $N$.

The distinction between $(\hat{\Sigma}^{-1})_{\K}$ and $(\hat{\Sigma}_{\K})^{-1}$ will become negligible in the case of weak long-range correlations, that is to say, when the off-block-diagonal elements of the sample covariance $\hat{\Sigma}_{\off}$ are small (in the sense of induced operator norm) relative to the elements of $\hat{\Sigma}_{\C}$ and $\hat{\Sigma}_{\K}$.  For the BPIC and PPIC, unlike the BAIC, there will be additional contributions to the ``perfect model'' IC when the long-range correlations are not negligible.  In order to have tractable approximate formulas for these criteria, we will assume weak long-range correlations so that the correlation matrix is approximately block-diagonal, $\Sigma^{-1}\approx\diag\left(\Sigma_{\K}^{-1},\Sigma_{\C}^{-1}\right)$.  For data where this assumption is badly violated, we advocate the use of the BAIC for subset selection, or one may explicitly construct a piecewise model including the perfect model and perform joint fits to the data as a whole.\footnote{Direct estimation of $\hat{\chi}_{\off}^2$ and its derivatives may be challenging; strongly-correlated covariance matrices can have large condition numbers, especially at small sample sizes.  Careful treatment of the covariance matrix (e.g., regularization via singular value decomposition) before inversion is essential in this case.}


One way to derive the perfect model formulas is to explicitly plug in the partitions for $\hat{\chi}^2, \tilde{\chi}^2,$ and $\chi_i^2$ to the definitions of the PPIC and BPIC and integrate over the perfect-model parameters $\vec{a}_{\C}$.  This derivation is shown in \cref{sec:app-subset-alt}.  Here, we take an alternative and simpler approach, which is to work with the K-L divergences directly using a particularly simple choice of the perfect model.

In the following derivations, we use the assumption of negligible long-range correlations described above to separate each definition of the K-L divergence as $\KL = \KL_{\K} + \KL_{\C}$; we are able to do this decomposition by Theorem 3.1 from \cite{Kullback1951} (see discussion in \cite{Neil:2023pgt}).  Explicit calculation of the second term in $\KL_{\C}$ as defined by equations \cref{eqn:kl-div-plug-in}, \cref{eqn:kl-div-post}, and \cref{eqn:kl-div-post-pred} will give us exact results for the associated ICs for the perfect model.  These can then be combined with the formulas for the ICs derived above on the kept portion of the data.

For the remainder of this section, unless otherwise noted we focus on a single data set of size $d_{\C}$ and ignore the kept data.  We assume a specific perfect model $M_{\perf}$ construction of the form $f(x) = \vec{a}$ and $\vec{a}^* = \bar{y}$, i.e. a model which is defined piecewise for each value in the vector $\bar{y}$.  The number of model parameters is manifestly $k = d_{\C}$.

Given a data sample $\yd$ of size $N$, the predicted least-squares likelihood function for a single future observation $z$ is
\begin{equation}
\pr(z|\mathbf{a}, M_{\perf}) = \frac{1}{(2\pi)^{d_{\C}/2} (\det \Sigma)^{1/2}} \exp \left[ -\frac{1}{2} (z-\vec{a})^T \Sigma^{-1} (z-\vec{a})\right],
\end{equation}
where $\Sigma$ is the sample covariance matrix. Note that technically, this means the likelihood function should be written as is $\pr(z|\mathbf{a}, M_{\perf}, \yd)$ since $\Sigma$ depends on $\yd$, although in the large-$N$ limit $\Sigma \rightarrow \Sigma_{\T}$, the true covariance matrix.

Dropping constant factors from the normalization (they will not be constant as $d_{\C}$ is varied, but they will combine with similar normalization factors from the kept data to become overall constants), then, we have
\begin{equation}
E_z [\log \pr(z|\vec{a}^*, M_{\perf})] = -\frac{1}{2} E_z \left[(z-\bar{y})^T \Sigma^{-1} (z-\bar{y})\right]
\end{equation}
In order to simplify further, suppose that the true model is represented by a vector $\mu_{\T}$, and data $y$ are generated from a multivariate Gaussian random process with true covariance $\Sigma_{\T}$.  (In general, we could work with a sample estimator of this probability instead so that the central limit theorem applies, leading to the same form.)  Then the ``true model'' probability distribution is
\begin{equation}
\pr_{\T}(z) = \frac{1}{(2\pi)^{d_{\C}/2} (\det \Sigma_{\T})^{1/2}} \exp \left[-\frac{1}{2} (z-\mu_{\T})^T \Sigma_{\T}^{-1} (z-\mu_{\T}) \right].
\end{equation}
Putting these together and using the formula \cref{eqn:gaussian_integral_I2} derived in \cref{sec:app_laplace_method}, we have
\begin{align}
E_z [\log \pr(z|\vec{a}^*, M_{\perf})] &= \int dz\ \pr_{\T}(z) \left(-\frac{1}{2} (z-\bar{y})^T \Sigma^{-1} (z-\bar{y}) \right) \\
&= -\frac{1}{2} \tr [\Sigma^{-1} \Sigma_{\T}] -\frac{1}{2} (\mu_{\T} - \bar{y})^T \Sigma^{-1} (\mu_{\T} - \bar{y}).
\end{align}
Considering the second term first, based on a result by White \cite{Jay:2020jkz,White1982} (or for this particular model, simply invoking the central limit theorem), the difference $\sqrt{N}(\mu_{\T} - \bar{y})$ is normally distributed as $N \rightarrow \infty$, with mean zero and covariance $C = J^{-1} I J^{-1}$.  Due to the simple structure of this perfect model, we have $J = I = \Sigma_{\T}$, giving the result
\begin{equation}
E_z [\log \pr(z|\vec{a}^*, M_{\perf})] = -\frac{1}{2} \tr[\Sigma^{-1} \Sigma_{\T}] - \frac{1}{2N} \tr[\Sigma^{-1} \Sigma_{\T}] \rightarrow -\frac{d_{\C}}{2} - \frac{d_{\C}}{2N},
\end{equation}
where we have simplified using the fact that $\Sigma \rightarrow \Sigma_{\T}$, i.e., it is a consistent estimator of the true covariance.  In terms of information criteria, this translates to
\begin{equation}
-2N E_z [\log \pr(z|\vec{a}^*, M_{\perf})] \simeq \BAIC_{\perf} = \hat{\chi}^2 + 2k + (N-1)d_{\C} = (N+1)d_{\C},
\end{equation}
where $\hat{\chi}^2 = 0$ identically, the number of parameters $k = d_{\C}$, and we are being careful to keep the overall factor of $(N-1)d_{\C}$ that appears in the definition of $\hat{\chi}^2$ in terms of sample means, see the discussion around \cref{eqn:chi2_aug}.

Moving on to the second definition of the KL divergence, we have to simplify the posterior average.  Using \cref{eqn:post-avg-def}, we have
\begin{align}
E_{\vec{a}|\yd} [ \log \pr(z|\vec{a}, M_{\perf})] &= -\frac{1}{2Z} \int d\vec{a} \exp \left[-\frac{1}{2} \chi_{\rm aug}^2(\vec{a}) \right] (z-\mathbf{a})^T \Sigma^{-1} (z-\mathbf{a}) \\
&= -\frac{1}{2Z} \int d\vec{a} \exp \left[ (\bar{y} - \vec{a})^T (\Sigma^*)^{-1} (\bar{y} - \vec{a}) \right] (z-\mathbf{a})^T \Sigma^{-1} (z-\mathbf{a}),
\end{align}
where due to the infinitely diffuse priors $\Sigma^* = \hat{\Sigma} = \Sigma/N$.  The normalizing factor $Z$ is
\begin{equation}
Z \equiv \int d\vec{a}  \exp \left[-\frac{1}{2} \chi_{\rm aug}^2(\vec{a}) \right].
\end{equation}
Using \cref{eqn:gaussian_integral_I2} once again, we find the result

\begin{equation}
E_{\vec{a}|\yd} [ \log \pr(z|\vec{a}, M_{\perf})] = -\frac{1}{2} \left( (z-\bar{y})^T \Sigma^{-1} (z-\bar{y}) + \tr [\Sigma^{-1} \Sigma^*] \right).
\end{equation}
The first term is precisely the plug-in log likelihood, while the second term reduces to a constant, $\tr[\Sigma^{-1} \Sigma^*] = \frac{1}{N} \tr[\Sigma^{-1} \Sigma] = d_{\C}/N$.  Thus, we find
\begin{equation}
E_z[E_{\vec{a}|\yd} [ \log \pr(z|\vec{a}, M_{\perf})]] = E_z [\log \pr(z|\vec{a}^*, M_{\perf})] - \frac{d_{\C}}{2N},
\end{equation}
which in terms of information criteria translates to 
\begin{equation}
\BPIC_{\perf} = \BAIC_{\perf} + d_{\C}.
\end{equation}
Although we do not focus on the PAIC (see \cref{sec:paic_formulas}), since the PAIC and BPIC both estimate the same posterior averaged K-L divergence, this also implies that $\PAIC_{\perf} = \BAIC_{\perf} + d_{\C}$.

Finally, we consider the posterior predictive KL divergence, for which the posterior average and the log are transposed.  To evaluate this, we first need the posterior average without the log,
\begin{equation}
E_{\vec{a}|\yd} [\pr(z|\vec{a}, M_{\perf})] = \frac{1}{Z} \int d\vec{a} \exp \left[-\frac{1}{2} (\bar{y} - \vec{a})^T (\Sigma^\star)^{-1} (\bar{y} - \vec{a}) - \frac{1}{2} (z-\vec{a})^T \Sigma^{-1} (z-\vec{a}) \right].
\end{equation}
Applying \cref{eqn:gaussian_integral_I3}, the result of the integration is
\begin{align}
E_{\vec{a}|\yd} [\pr(z|\vec{a}, M_{\perf})] &= \left(\frac{N}{N+1}\right)^{d_{\C}/2}  \exp \left[ -\frac{1}{2} \frac{1}{N+1} (z-\bar{y})^T (\Sigma^*)^{-1} (z-\bar{y}) \right].
\end{align}
Taking the log and then the $z$-expectation, we find the result:
\begin{align}
E_z[\log E_{\vec{a} | \yd}[\pr(z | \vec{a}, M_{\perf})]] &= \frac{d_{\C}}{2} \log{\frac{N}{N+1}} -\frac{1}{2} \frac{N}{N+1} E_z[(z-\bar{y})^T \Sigma^{-1} (z-\bar{y})] \\
&= \frac{N}{N+1} E_z[\log \pr(z|\vec{a}^*, M_{\perf})] - \frac{d_{\C}}{2} \log\left(1 + \frac{1}{N}\right),
\end{align}
or in terms of information criteria once more, and dropping the $N$-dependent constant term since it will cancel out in any model averages,
\begin{align}
\PPIC_{\perf} &= \frac{N}{N+1} \BAIC_{\perf} + Nd_{\C} \log \left(1  + \frac{1}{N} \right) \\
&= \frac{N}{N+1} (N+1) d_{\C} + Nd_{\C} \log \left(1 + \frac{1}{N} \right) \\
&\approx (N+1) d_{\C} - \frac{d_{\C}}{2N} = \BAIC_{\perf} - \frac{d_{\C}}{2N}.
\end{align}
When used in data subset selection, there is an additional factor of $(N-1)d_{\K}$ that arises from the definition of chi-squared over the kept part of the data.  This combines with $(N-1)d_{\C}$ to give an overall shift of $(N-1)d$, which is constant and may be dropped.  Doing so, we find our final results for the contribution of the cut portion of the data to each information criterion:
\begin{align}
\Delta_{P} \BAIC &= 2d_{\C}, \\
\Delta_{P} \BPIC &= 3d_{\C}, \\
\Delta_{P} \PPIC &= d_{\C} + Nd_{\C} \log \left(1 + \frac{1}{N} \right) \approx 2d_{\C} - \frac{d_{\C}}{2N}.
\end{align}
where $\Delta_{P}$ denotes the change in the overall model-averaging formulas due to the cut data in data subset selection.

Putting this together with the formulas for the kept data, we have our final results for the three ICs in the presence of data subset selection:
\begin{align}
\label{eqn:baic_subset} \BAIC_{\mu,P}=&\BAIC_{\mu}+\Delta_P\BAIC=\hat{\chi}^2(\vec{a}^*)+2k+2d_{\C},\\
\label{eqn:bpic_subset} \BPIC_{\mu,P}=&\BPIC_{\mu}+\Delta_P\BPIC\approx\hat{\chi}^2(\vec{a}^*)-\frac{1}{2}\tilde{H}_{ba}(\Sigma^*)_{ab}+\frac{1}{2}\tilde{g}_dT_{cba}(\Sigma_2^*)_{abcd}+3k+3d_{\C},\\
\PPIC_{\mu,P}=&\PPIC_{\mu}+\Delta_P\PPIC\\
 \label{eqn:ppic_subset} \nonumber \approx&\hat{\chi}^2(\vec{a}^*)+2k+d_{\C} + Nd_{\C} \log\left(1 + \frac{1}{N} \right)  \nonumber \\
&-2\sum_{i=1}^{N}\log\left[1+\frac{1}{2}\left(\frac{1}{4}(g_i)_b(g_i)_a-\frac{1}{2}(H_i)_{ba}\right)(\Sigma^*)_{ab}+\frac{1}{4}(g_i)_dT_{cba}(\Sigma_2^*)_{abcd}\right]. 
\end{align}
where $\hat{\chi}^2$ and all other quantities are evaluated only for the kept data.  In cases of optimal truncation, \cref{eqn:bpic_opt_trunc,eqn:ppic_opt_trunc} should be used for the BPIC and PPIC, respectively, with the addition of $\Delta_P \BPIC$ and $\Delta_P \PPIC$, respectively.  We remind the reader that for use with model averaging, the factor $-2 \log \pr(M_\mu)$ should be added to all ICs as in \cref{eqn:IC_MA}, although this factor may be ignored completely if $\pr(M_\mu)$ is flat (independent of $\mu$.)

An alternative derivation for these formulas starting from the level of information criteria rather than K-L divergence is provided in \cref{sec:app-subset-alt}.  In addition to the relative simplicity of the K-L divergence approach taken here, we are able to obtain exact results for $\KL_{\C}$, whereas starting from the ICs neglects higher-order bias corrections (see discussion in \cref{sec:app-subset-alt}).

As a final remark on the data subset selection procedure outlined above, it is worth discussing the full bias correction, i.e. the case in which $\tr\bracket{J_N^{-1}(\vec{a}^*)I_N(\vec{a}^*)}$ is used rather than replacing the trace by the number of parameters.  While the $d_{\C}$ contributions computed above are exact for the perfect model, in general the replacement of $\tr\bracket{(J_N^{-1}(\vec{a}^*))_{\K} (I_N(\vec{a}^*))_{\K}}\rightarrow k$ on the kept data may not hold (as discussed in \cref{subsec:kl-div-plug}.)  In particular, long-range correlations will correct all information criteria through contributions to this trace, as $\hat{\Sigma}_{\off}^{-1}$ appears in the analytical expressions for $(J_N^{-1})_{\K}$ and $(I_N)_{\K}$. This can lead to numerical instabilities that will be more significant than the bias correction introduced by the full trace \cite{Vrieze2012,Shibata1983,Shibata1989}. In such cases, the simplified bias corrections should be used even when the true model is not in the family of candidate models.  In general, bias can be reduced by expanding the space of candidate models, ideally to include the true model.  In future work, it would be interesting to explore the use of more robust methods for estimation of these matrices, such as shrinkage \cite{ledoit2004well,ledoit2012nonlinear,ledoit2017direct,Rinaldi:2019thf,FermilabLattice:2022gku}.

\section{Numerical tests}
\label{sec:num-tests}

In this section, we give several numerical examples of model averaging with the various information criteria derived above and comparing their performance to fixed-model parameter estimation procedures.\footnote{The code used to generate the examples in \cref{subsec:ex1,subsec:ex2} is publicly available at \url{https://github.com/jwsitison/improved_model_avg_paper}.}  While tests with the $\ABIC_{\CV}$ were conducted, we omit any numerical results due to its similar performance to the $\BAIC$ in the following examples. All Bayesian least squares fits were performed using the \texttt{lsqfit} package in Python \cite{Lepage2002,lsqfitGitHub}, which uses the Gaussian random variable data type from \texttt{gvar} \cite{gvarGitHub}.

\subsection{Example 1: Polynomial models}
\label{subsec:ex1}

Consider a simple toy problem where the ``true model" is a quadratic polynomial:
\begin{align}
f_{\T}(x)=1.80-0.53\left(\frac{x}{16}\right)+0.31\left(\frac{x}{16}\right)^2.
\end{align}
A set of $N$ samples are generated on $x\in\{1,2,\dots,15\}$ using $f_{\T}$ at each point multiplied by uncorrelated noise $1+\eta(x)$, where $\eta(x)$ is drawn from a Gaussian with mean $\bar{\eta}=0.0$ and variance $\sigma_{\eta}^2=1.0$. To be explicit, the mock data are drawn from $y(x)=(1+\eta(x))f_{\T}(x)$.

Our space of candidate models are polynomials labeled by their degree $\mu\in\{0,1,\dots,5\}$:
\begin{align}
f_{\mu}(x)=\sum_{m=0}^{\mu}a_m\left(\frac{x}{16}\right)^m.
\end{align}
We take uniform model priors $\pr(M_{\mu})=1/6$ corresponding to minimal prior information on the functional form of the true model (except that it can be approximated by a polynomial). We consider the case of moderately unconstrained parameter priors of the Gaussian form given in \cref{eqn:param-prior} with mean zero and width 10.

We use the previously developed model averaging procedures to determine the parameter estimate and error for $a_0$. Since the model functions are linear in the parameters, we use the following forms of the information criteria to determine the model weights: 
\begin{align}
\label{eqn:baic_linear} \BAIC_{\mu}=&-2\log\pr(M_{\mu})+\hat{\chi}^2(\vec{a}^*)+2k,\\
\label{eqn:bpic_linear} \BPIC_{\mu}=&-2\log\pr(M_{\mu})+\hat{\chi}^2(\vec{a}^*)-\frac{1}{2}\tilde{H}_{ba}(\Sigma^*)_{ab}+3k,\\
\label{eqn:ppic_linear} \nonumber \PPIC_{\mu}\approx&-2\log\pr(M_{\mu})+\hat{\chi}^2(\vec{a}^*)+2k\\
&-2\sum_{i=1}^{N}\log\left[1+\frac{1}{2}\left(\frac{1}{4}(g_i)_b(g_i)_a-\frac{1}{2}(H_i)_{ba}\right)(\Sigma^*)_{ab}\right].
\end{align}
Due to the linearity of the model function in this example, the NLO Poincar\'e expansion formula is exact for the BPIC; the superasymptotic schema discussed in \cref{subsec:super} are applied only for the PPIC, although in practice truncation does not occur in this test.   Furthermore, the BPIC and PPIC are simplified for a linear model function since the tensor $T$ is zero, see \cref{subsec:laplace}.  In this example the data set is held fixed, so we drop all terms depending on $d_{\C}.$

The model-averaged results are summarized in \cref{tab:poly} and shown in \cref{fig:poly-fixed-n}; we also report the $Q$-value of the fit (a Bayesian analogue of the $p$-value, see Appendix B of \cite{FermilabLattice:2016ipl}), which gives a measure of the fit quality. The model-averaged results are consistent with model truth but with a larger uncertainty than the individual fit to the correct model with $\mu=2$.  The larger error with model averaging is an inherent feature, reflective of a bias-variance tradeoff; in the face of model uncertainty, model averaging hedges against the possibility of biased results due to selection of the wrong model, at the cost of increased error with a given data sample.  See the further discussion in \cref{sec:conclusion}.  In the top panel of \cref{fig:poly-fixed-n}, the advantage of model averaging over model selection is evident as the model probabilities happen to favor the $\mu = 1$ linear model, which is in fact incorrect.  As the sample size $N$ increases, this model will eventually be ruled out and the model weight will peak at the true model $\mu = 2$ as seen in the bottom panel of \cref{fig:poly-fixed-n}.

\begin{table}[!htbp]
\begin{center}
\subfloat{\begin{tabular}{|>{$}c<{$}|>{$}c<{$}>{$}c<{$}>{$}c<{$}>{$}c<{$}>{$}c<{$}>{$}c<{$}|>{$}c<{$}|}
\cline{1-7}
	& \mu=0	& \mu=1	& \mu=2	& \mu=3	& \mu=4	&\mu=5	\\
\cline{1-7}
a_0	& 1.587(32)	& 1.803(67)	& 1.89(11)	& 2.01(16)	& 1.98(17)	& 1.94(18)	\\
a_1	&   & -0.41(11)	& -0.88(50)	& -2.2(1.3)	& -1.6(1.5)	& -1.0(1.8)	\\
a_2	& 	& 	& 0.44(46)	& 3.6(3.0)	& 0.4(5.0)	& -1.0(5.5)	\\
a_3	& 	& 	& 	& -2.1(2.0)	& 3.4(7.1)	& -3.0(3.7)	\\
a_4	& 	& 	& 	& 	& -3.0(3.7)	& 1.5(7.7)	\\
\cline{8-8}
a_5	& 	& 	& 	& 	& 	& -3.1(4.7)	& \avg{a_0}	\\
\hline
\hat{\chi}^2	& 28.85	& 15.17	& 14.23	& 12.88	& 12.23	& 11.79  & 	\\
Q\text{-value}	& 0.02	& 0.44	& 0.50	& 0.59	& 0.64	& 0.67  & 	\\
\BAIC			& 30.85	& 19.17	& 20.23	& 20.88	& 22.22	& 23.79  & 	\\
\pr(M_{\mu}|\yd)_{\BAIC}	& 0.00	& 0.43	& 0.25	& 0.18	& 0.09	& 0.04  & 1.89(14)	\\
\BPIC	        & 31.85	& 21.17	& 23.23	& 24.73	& 26.30	& 28.13  & 	\\
\pr(M_{\mu}|\yd)_{\BPIC}	& 0.00	& 0.61	& 0.22	& 0.10	& 0.05	& 0.02  & 1.85(12)	\\
\PPIC			& 30.85	& 19.18	& 20.24	& 20.89	& 22.23	& 23.80  & 	\\
\pr(M_{\mu}|\yd)_{\PPIC}	& 0.00	& 0.43	& 0.25	& 0.18	& 0.09	& 0.04  & 1.88(14)	\\
\hline
\end{tabular}}
\end{center}
\caption{Individual best-fit results with information criteria values and corresponding model weights for $N=160$.}
\label{tab:poly}
\end{table}

\begin{figure}[!htbp]
\subfloat{\includegraphics[width=0.8\textwidth]{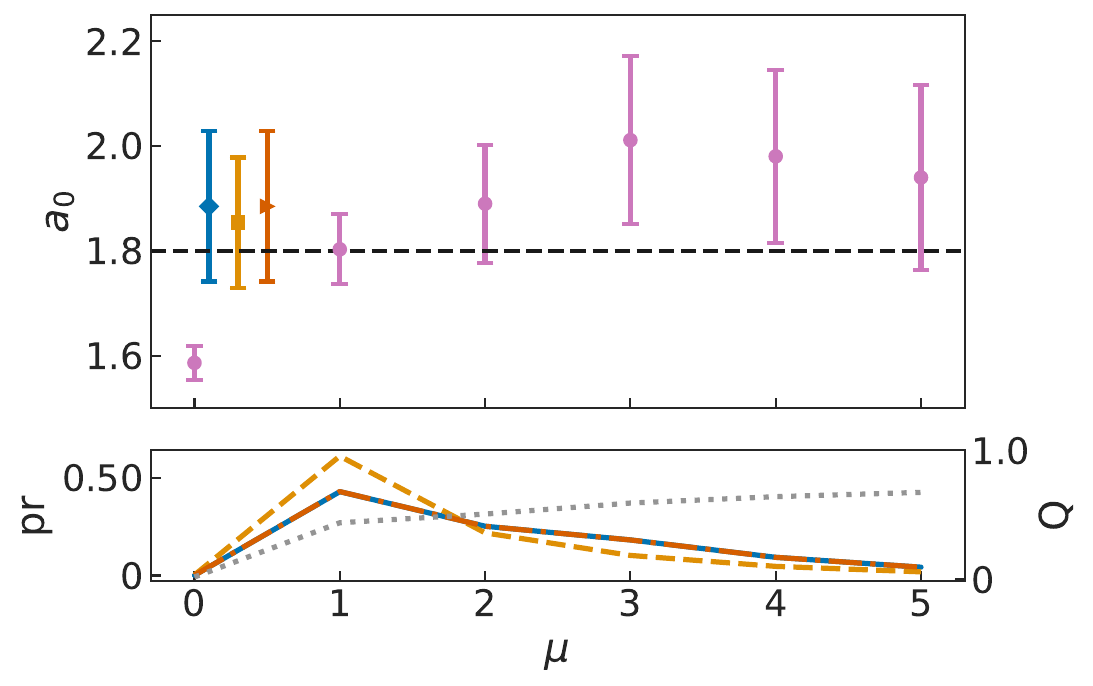}}
\hfill
\subfloat{\includegraphics[width=0.8\textwidth]{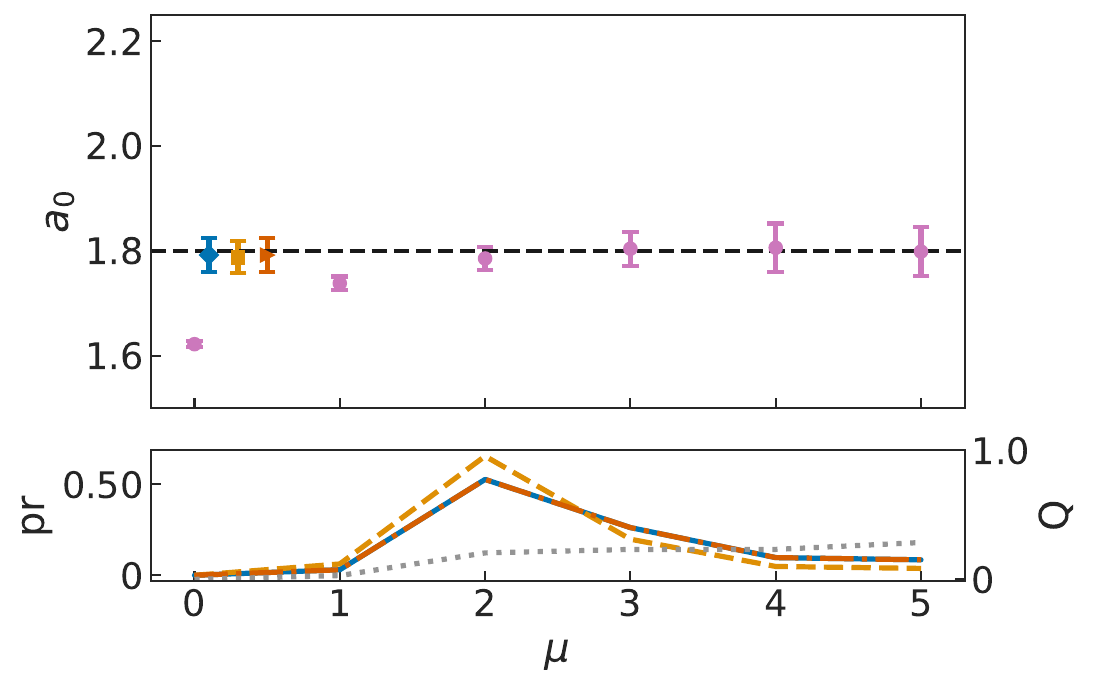}}
\caption{Fit results at $N=160$ (top panel) and $N=5120$ (bottom panel) for $f_{\mu}(x)$ (purple {\color{indv}\LARGE\textbullet}) and model-averaged results with the BAIC (blue {\color{baic}\ding{117}}), BPIC and PAIC (yellow {\color{bpic}$\blacksquare$}) and PPIC (red {\color{ppic}$\blacktriangleright$}) compared to to the known value $a_0=1.80$ (black dashed line). The lower inset shows the standard $Q$-value (grey dotted curve) and the model weights $\pr(M_{\mu}|\yd)$ from the BAIC (blue solid curve), BPIC and PAIC (yellow dashed curve), and PPIC (red dash-dotted curve).}
\label{fig:poly-fixed-n}
\end{figure}

Note that in this case, the models are ``nested'' in the sense that any $\mu > 2$ can capture the true model by setting higher-order $a_m$ to zero.  This means that even as $N \rightarrow \infty$, the model probability $\pr(M_\mu | \yd)$ for $\mu > 2$ will never go to zero, although models with higher complexity will be penalized by the $+2k$ bias-correction term so that peak model probability will be at $\mu=2$.

Observe that the BPIC has less uncertainty than the BAIC or PPIC. This is because the BPIC favors simpler models even more so than the BAIC or PPIC as a result of the additional $k$ that comes from the posterior averaging of the K-L divergence when parameter priors are sufficiently diffuse, as discussed in \cref{subsec:bpic}. The additional parsimony implied by the BPIC comes with a larger K-L divergence for the model distribution. While this might cause concern that the BPIC may actually underestimate the model uncertainty in this example, in practice this does not appear to be the case (as seen in this example by comparing the BPIC-averaged parameter uncertainty with that for the true $\mu=2$ model).

\begin{figure}[!htbp]
\centering
\includegraphics[width=0.8\textwidth]{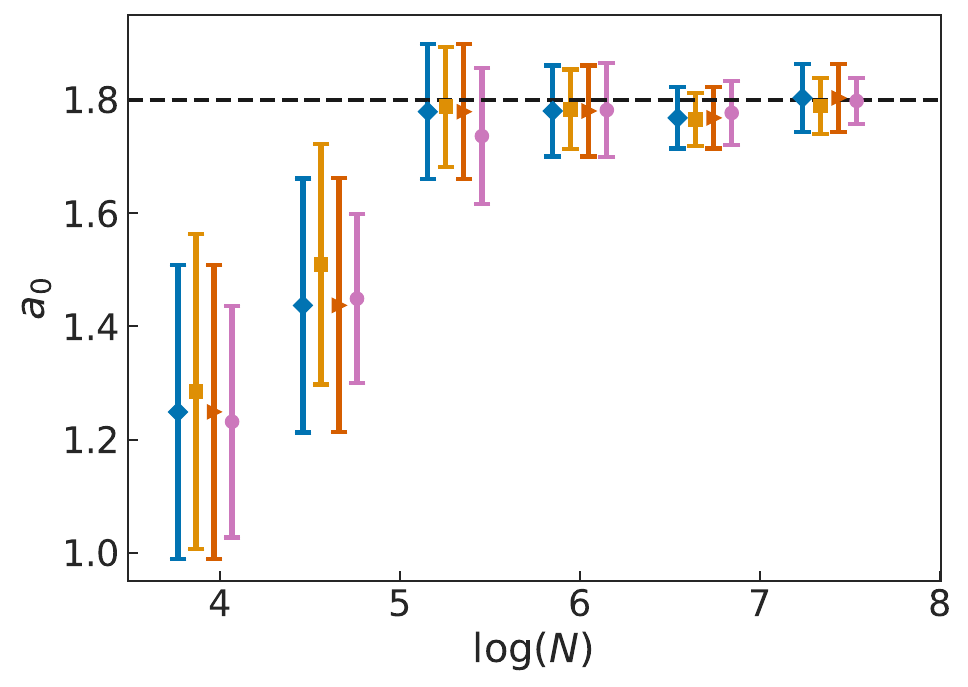}
\caption{$N$-dependent scaling of the various estimates of the intercept $a_0$. The true value (black dashed line) is $a_0=1.80$. The model-averaged results using the BAIC (blue {\color{baic}\ding{117}}), the BPIC and PAIC (yellow {\color{bpic}$\blacksquare$}), and PPIC (red {\color{ppic}$\blacktriangleright$}) are consistent with both model truth and the fit results of the correct quadratic model (purple {\color{indv}\LARGE\textbullet}).}
\label{fig:poly-n-dep}
\end{figure}

We repeat the previous numerical test with several values of $N=40, 80, 160, 320, 640,$ $1280$; the final estimates for $a_0$ are in \cref{fig:poly-n-dep}. The model-averaged results using the BAIC, BPIC, and PPIC are consistent with model truth in all cases. As in the fixed-$N$ study, the uncertainties in the model-averaged estimates are typically larger than that of the fixed quadratic estimate; this is because model averaging has two sources of error---(1) parameter uncertainty in individual fits and (2) variance across the individual fit means---whereas using a fixed model only has the former.

\FloatBarrier
\subsection{Example 2: Exponential model and subset selection}
\label{subsec:ex2}

To test the data subset selection procedure developed in \cref{sec:data_subset_select}, we consider another toy problem meant to resemble a two-point correlation function (see \cref{sec:data_subset_select} for a brief discussion of two-point correlators). For this example, we will set model truth to a two-state exponential:
\begin{align}\label{eqn:toy-model-truth}
f_{\T}(t)=2.0e^{-0.80t}+10.4e^{-1.16t}.
\end{align}
To generate synthetic data, we multiply $f_{\T}$ by correlated noise $1+\eta(t)$, where $\eta(t)$ is drawn from a Gaussian with mean $\bar{\eta}=0.0$ and standard deviation $\sigma_{\eta}\in\{0.3,0.003\}$, as well as an uncorrelated noise floor $\theta(t)$ drawn from a Gaussian with mean $\bar{\theta}=0.0$ and standard deviation $\sigma_{\theta}\in\{0,10^{-5}\}$, i.e., the synthetic data are generated from $y(t)=(1+\eta(t))f_{\T}(t)+\theta(t)$. The correlation matrix used to generate $\eta(t)$ takes the form $\rho_{xy}=\rho^{|t_x-t_y|}$ so that $\rho$ equals one on the diagonal and decreases as a power law as the temporal separation between points increases (similar to a real lattice QCD correlation function). We fix the correlation coefficient to $\rho=0.6$. We generate $N$ mock data samples on $t\in\{1,2,\dots,31\}$; the initial time is omitted from the analysis due to the certain excited state contamination at $t=0$. Other sets of parameters with $\sigma_{\theta}=0.0$ were considered in \cite{Jay:2020jkz} with and without correlation on $\eta(t)$ to no qualitative effect; the same is true for $\sigma_{\theta}>0$ except for very large values of $\rho$ where numerical issues lead to unreliable simulations.

We note that in terms of the observed signal-to-noise ratio in the mock data, the case without the noise floor $\sigma_\theta = 0$ has roughly constant signal-to-noise, analogous to a pion two-point correlation function \cite{Lepage:1989hd}.  On the other hand, with $\sigma_\theta > 0$ a threshold in $t$ is introduced at which the signal-to-noise drops exponentially, with the signal being completely overwhelmed at large $t$ where $\sigma_\theta \gg f_{\T}(t)$.  This more closely resembles the behavior of something like a nucleon two-point function, with extremely poor signal-to-noise at large time separation reflective of a sign problem \cite{Grabowska:2012ik,Wagman:2016bam}.  These resemblances to real data are, at best, qualitative; we do not claim to use a realistic noise model for this toy example.  We will consider the application of our methods to real lattice QCD data in \cref{subsec:ex3}.

Initially, we consider a single candidate model that consists of a single exponential term:
\begin{align}\label{eqn:exp-mod-func}
f_1(t)=A_0e^{-E_0t}.
\end{align}
This model is fit to data $(y_i)_x$ corresponding to $t_x\in[t_{\min},31]$ for $t_{\min}\in\{1,2,\dots,28\}$. Model-averaged results for the ground-state energy $E_0$ are obtained using
\begin{align}
\BAIC_{\mu,d_{\C}}=&-2\log\pr(M_{\mu})+\hat{\chi}^2(\vec{a}^*)+2k+2d_{\C},\\
\BPIC_{\mu,d_{\C}}\approx&-2\log\pr(M_{\mu})+\hat{\chi}^2(\vec{a}^*)-\frac{1}{2}\tilde{H}_{ba}(\Sigma^*)_{ab}+\frac{1}{2}\tilde{g}_dT_{cba}(\Sigma_2^*)_{abcd}+3k+3d_{\C},\\
\nonumber \PPIC_{\mu,d_{\C}}\approx&-2\log\pr(M_{\mu})+\hat{\chi}^2(\vec{a}^*)+2k+d_{\C} + Nd_{\C} \log \left(1 + \frac{1}{N} \right) \\
\label{eqn:ppic-ex2}&-2\sum_{i=1}^{N}\log\left[1+\frac{1}{2}\left(\frac{1}{4}(g_i)_b(g_i)_a-\frac{1}{2}(H_i)_{ba}\right)(\Sigma^*)_{ab}+\frac{1}{4}(g_i)_dT_{cba}(\Sigma_2^*)_{abcd}\right],
\end{align}
as derived in \cref{sec:data_subset_select}. In this example, $d_{\C}=t_{\min}$ entirely determines model complexity as $k=2$ is fixed across the space of models. Unlike the polynomial example, the integrals used to obtain the information criteria are not computed exactly as the model function \cref{eqn:exp-mod-func} is nonlinear in the model parameters. To improve the accuracy of these results, we implement the superasymptotic schema described in \cref{subsec:super} for the BPIC and PPIC.

\begin{figure}[!htbp]
\subfloat{\includegraphics[width=0.48\textwidth]{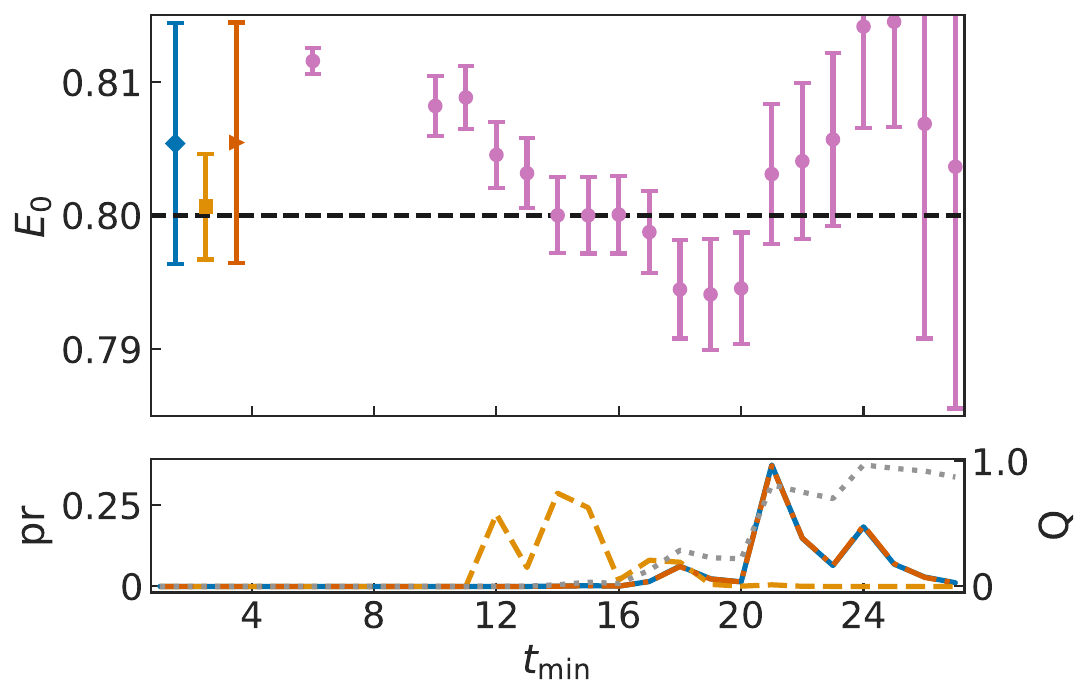}}
\hfill
\subfloat{\includegraphics[width=0.48\textwidth]{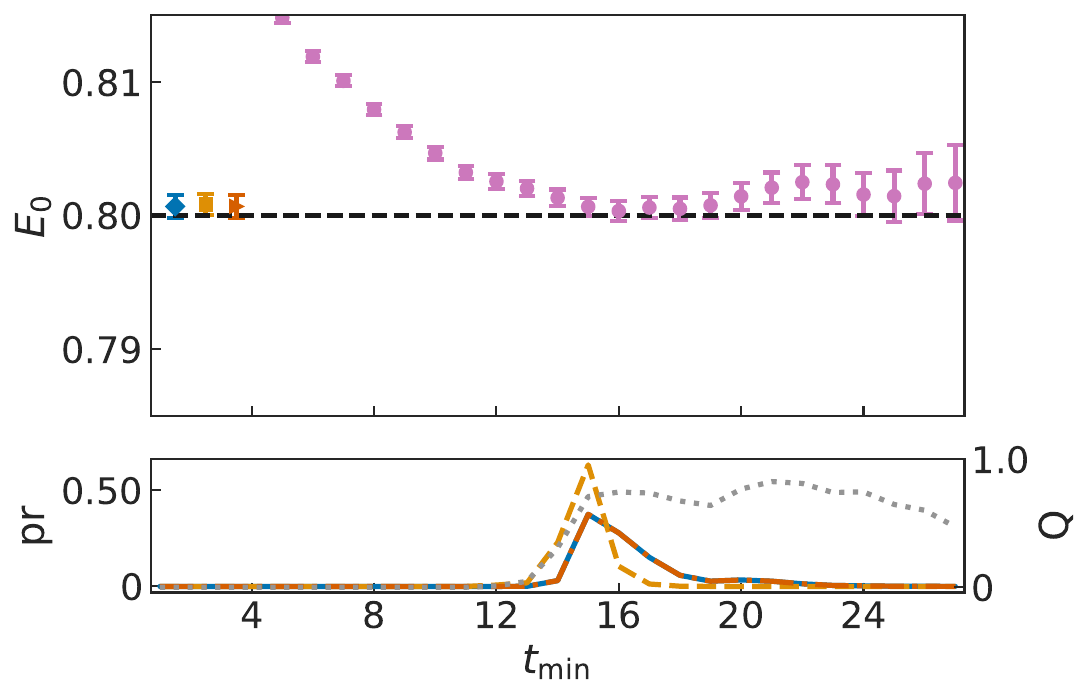}}
\hfill
\subfloat{\includegraphics[width=0.48\textwidth]{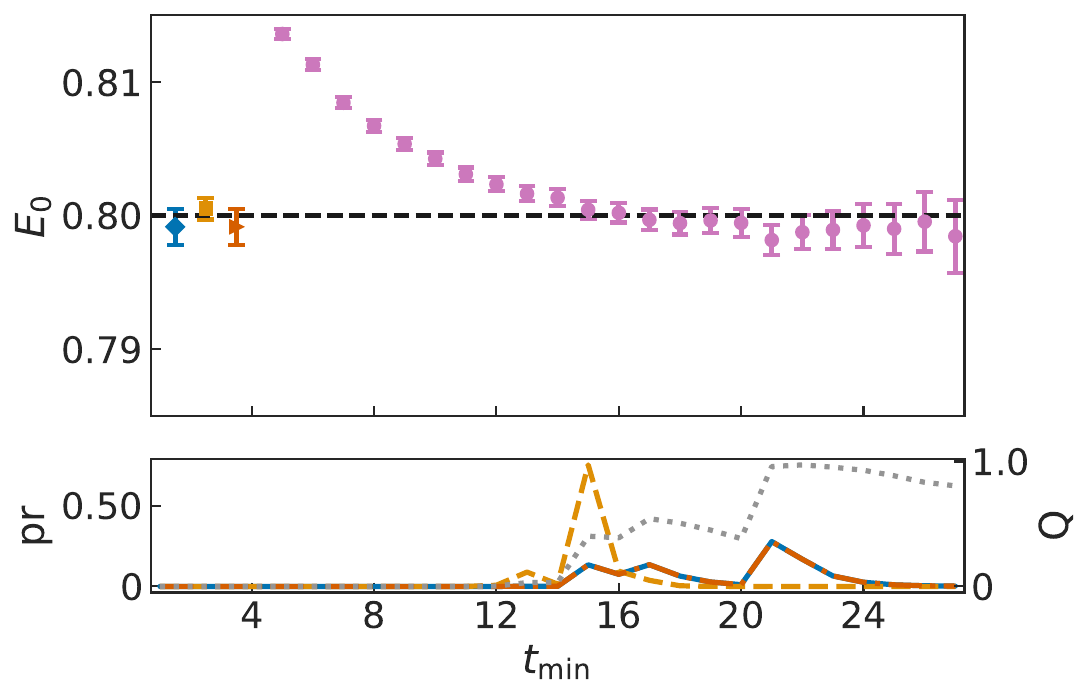}}
\hfill
\subfloat{\includegraphics[width=0.48\textwidth]{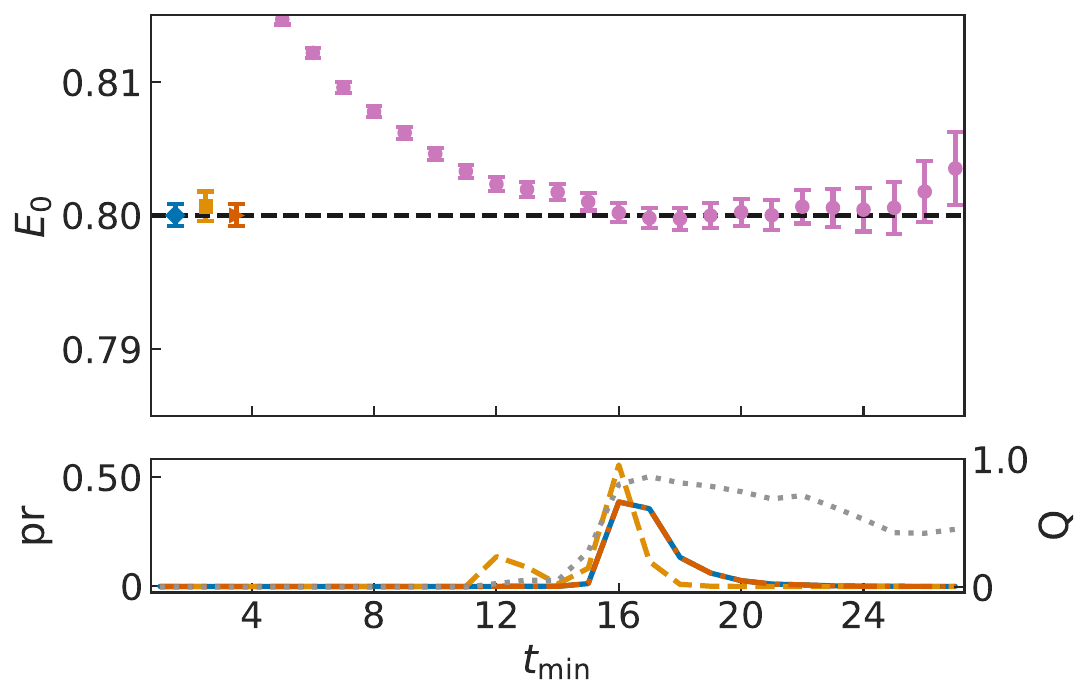}}
\caption{Fit results for the ground-state energy with true value $E_0=0.80$ (black dashed line) for $N=30$, $\sigma_{\eta}=0.3$, and $\sigma_{\theta}=0$, for data subset $t\in[t_{\min},31]$ (purple {\color{indv}\LARGE\textbullet}). Model-averaged results with the BAIC (blue {\color{baic}\ding{117}}), BPIC (yellow {\color{bpic}$\blacksquare$}), and PPIC (red {\color{ppic}$\blacktriangleright$}) agree well with model truth and each other in any case. The lower inset shows the standard $Q$-value (grey dotted line) and model weights $\pr(M_{\mu}|\yd)$ corresponding to the BAIC (blue solid curve), BPIC (yellow dashed curve), and PPIC (red dash-dotted curve). The four separate figures represent four random draws of correlated Gaussian fractional noise, but are otherwise identical.}
\label{fig:exp-fixed-n-no-floor}
\end{figure}

\begin{figure}[!htbp]
\subfloat{\includegraphics[width=0.48\textwidth]{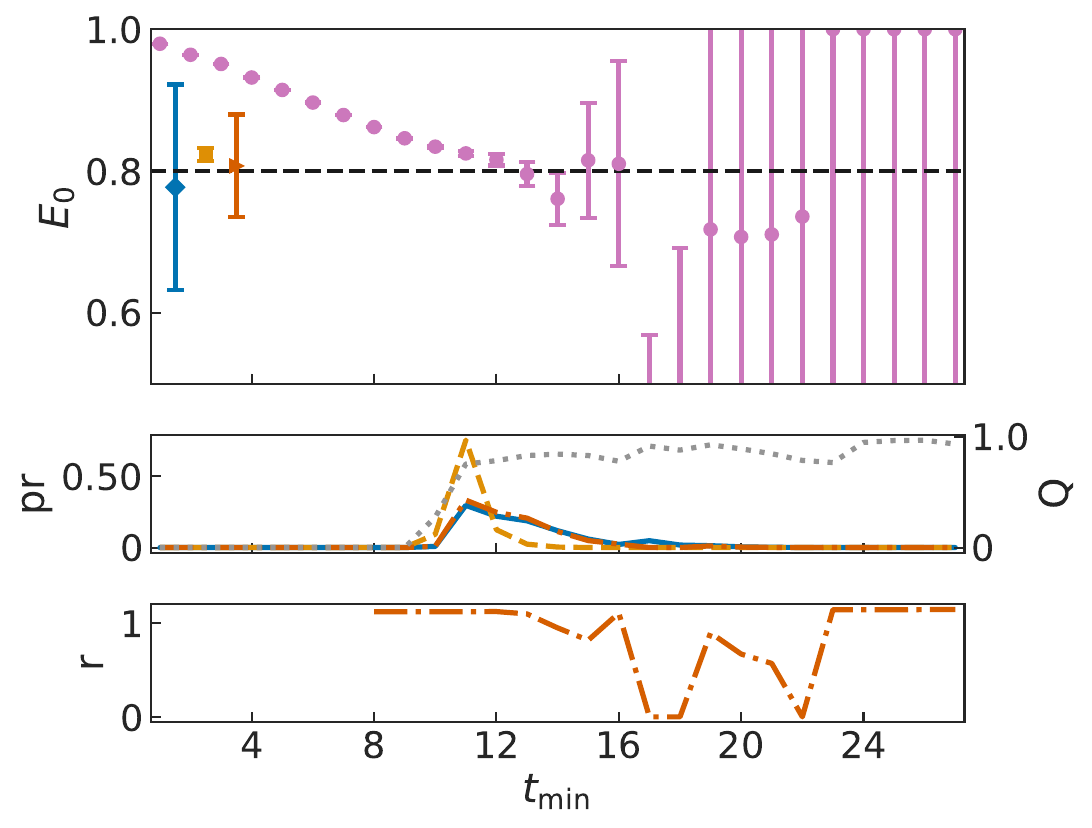}}
\hfill
\subfloat{\includegraphics[width=0.48\textwidth]{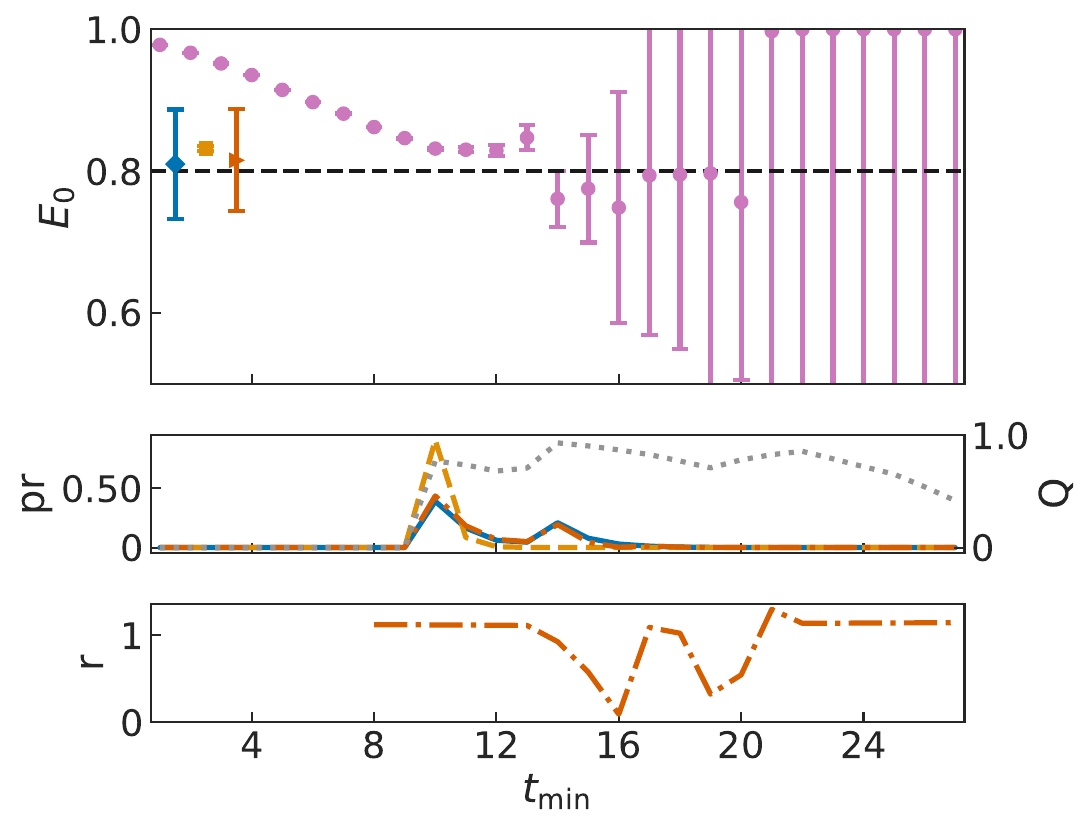}}
\hfill
\subfloat{\includegraphics[width=0.48\textwidth]{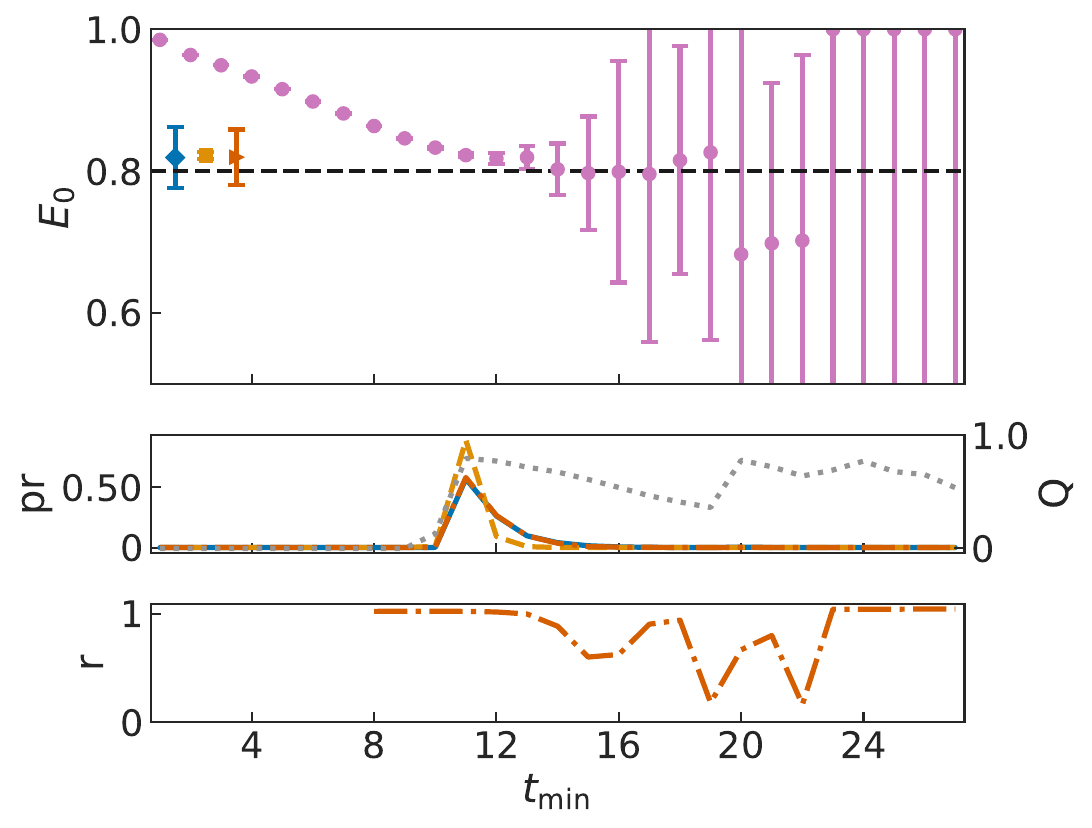}}
\hfill
\subfloat{\includegraphics[width=0.48\textwidth]{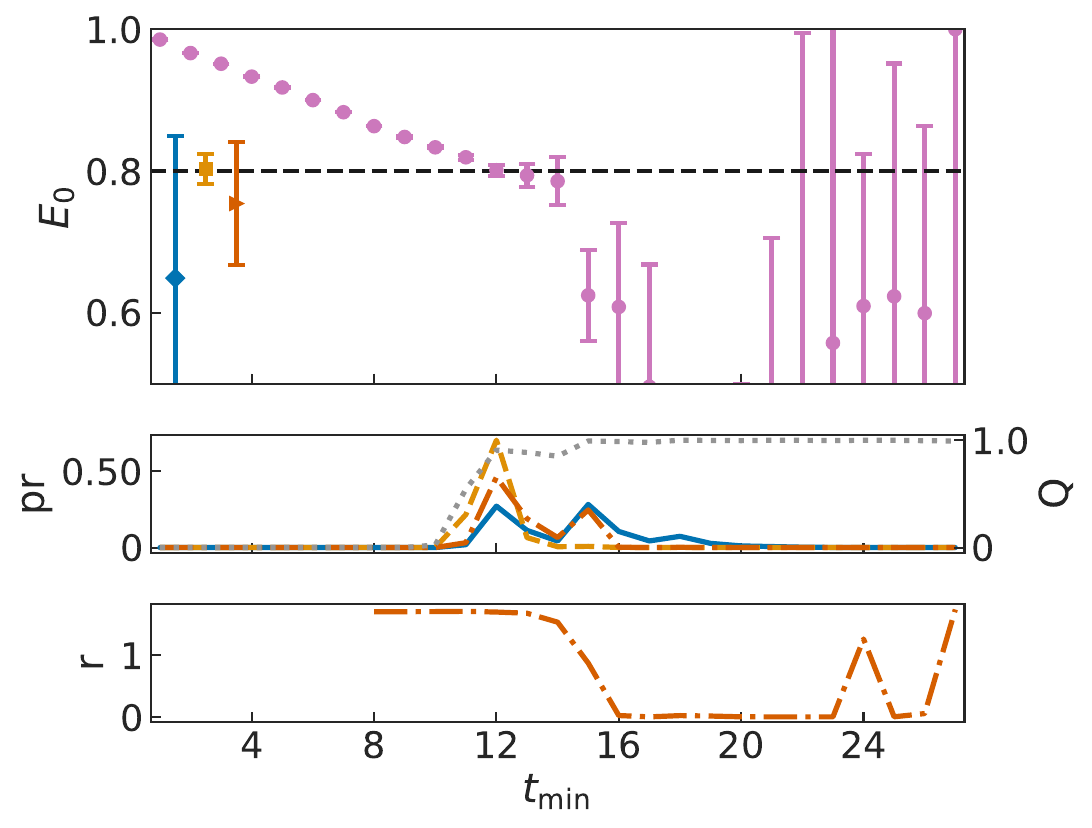}}
\caption{Fit results for the ground-state energy with true value $E_0=0.80$ (black dashed line) for $N=200$, $\sigma_{\eta}=0.003$, and $\sigma_{\theta}=10^{-5}$, for data subset $t\in[t_{\min},31]$ (purple {\color{indv}\LARGE\textbullet}). Model-averaged results with the BAIC (blue {\color{baic}\ding{117}}), BPIC (yellow {\color{bpic}$\blacksquare$}), and PPIC (red {\color{ppic}$\blacktriangleright$}) agree well with model truth and each other in any case. The middle inset shows the standard $Q$-value (grey dotted line) and model weights $\pr(M_{\mu}|\yd)$ corresponding to the BAIC (blue solid curve), BPIC (yellow dashed curve), and PPIC (red dash-dotted curve).  The lower inset shows the ratios of the
PPIC model weights to the BAIC model weights $\text{r}\equiv\pr(M_{\mu}|\yd)_{\PPIC}/\pr(M_{\mu}|\yd)_{\BAIC}$. The four separate figures represent four random draws of correlated Gaussian fractional noise and uncorrelated Gaussian additive noise, but are otherwise identical.}
\label{fig:exp-fixed-n-floor}
\end{figure}

The results of four independent trials of the above procedure with $(N,\sigma_{\eta},\sigma_{\theta})=(30,0.3,0)$ and $(N,\sigma_{\eta},\sigma_{\theta})=(200,0.003,10^{-5})$ are shown in \cref{fig:exp-fixed-n-no-floor} and \cref{fig:exp-fixed-n-floor}, respectively. Excited-state contamination is clear at low $t_{\min}$ as the second exponential state has influence over the fit results before it has decayed away. Model-averaged results agree well with model truth for all information criteria considered. Like the polynomial example, the model-averaged results favor parsimonious models; in the present context, parsimony corresponds to fits that cut away as little data as possible without compromising fit quality.

In \cref{fig:exp-fixed-n-no-floor}, where there is no noise floor, the BAIC and PPIC perform identically. The BPIC shows a preference for smaller $t_{\min}$ cuts due to the extra factor of $d_{\C}$ in its subset selection formula.  This leads to generally smaller uncertainty but potential for finite-sample bias, which will be more strongly evident below in the presence of additional noise.  

\cref{fig:exp-fixed-n-floor} represents a more challenging noisy-data case study, as the noise floor $\theta(t)$ is typically larger than the last several data points, which is reflected in the large $E_0$ error for fits at large $t_{\rm min}$. In this case, the PPIC consistently outperforms the BAIC in its estimation of the parameter mean and the error of this estimate. This results from the phenomenon discussed in \cref{subsec:ppic} where, in its use of information from each individual observation, the PPIC is able to penalize models that fail to predict the data and hence give poor parameter estimates. The improved error of the PPIC $E_0$ estimate is coupled to this effect as poor models tend to give larger parameter uncertainties.  The relatively low weight for noisy models at high $t_{\min}$ using PPIC versus BAIC can be seen from the lowest panel in each subfigure, which shows the ratio $\rm{r} \equiv \pr(M_\mu | \yd)_{\PPIC} / \pr(M_\mu | \yd)_{\BAIC}$ of the model weights using PPIC to BAIC. The PPIC also outperforms the BPIC in terms of finite-$N$ bias; while the BPIC often gives much smaller error estimates than the other information criteria, this is due to its overly aggressive penalty for $d_{\C}$ which heavily weights a single fit with small $t_{\min}$ as seen in the figure.  As a result, the BPIC often disagrees strongly with the true asymptotic value for $E_0$.

\begin{figure}[!htbp]
\subfloat{\includegraphics[width=0.8\textwidth]{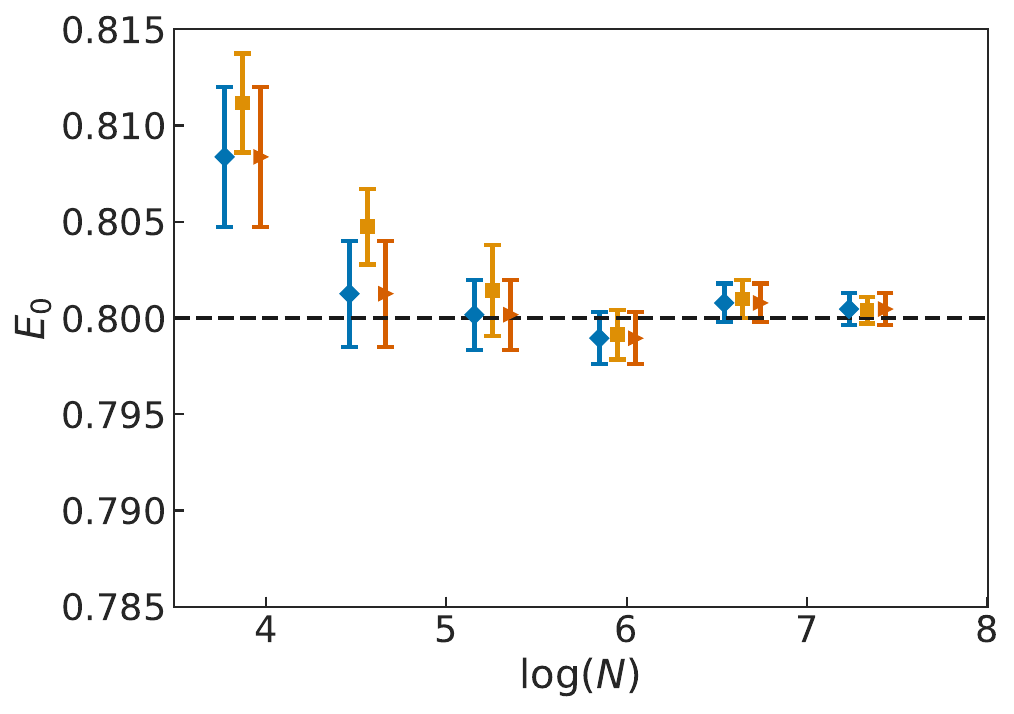}}
\hfill
\subfloat{\includegraphics[width=0.8\textwidth]{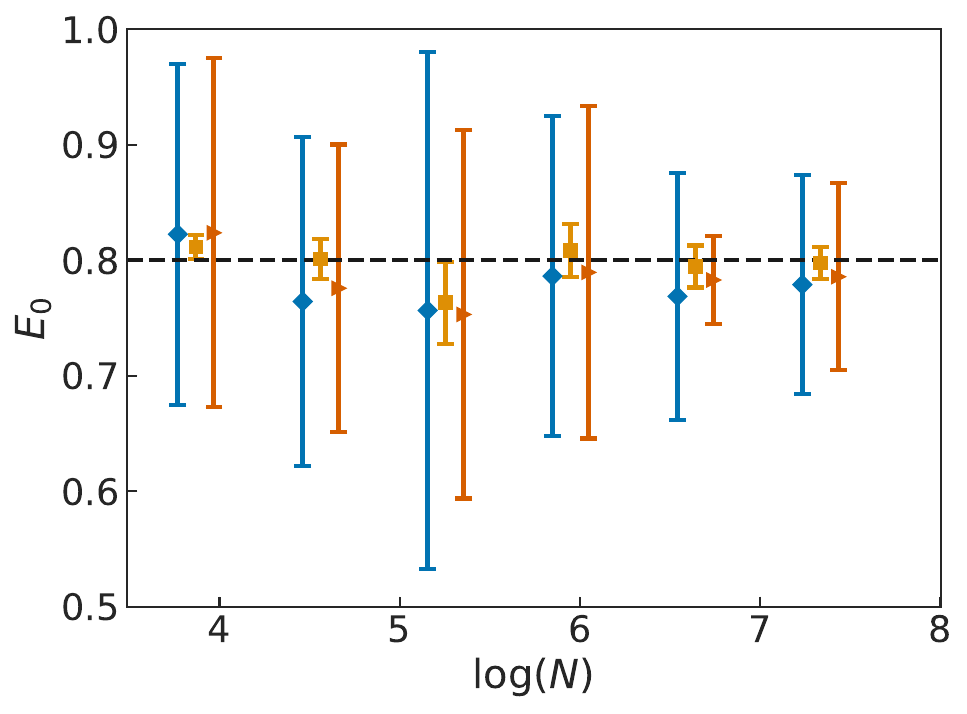}}
\caption{$N$-dependent scaling of the various estimates of the ground-state energy $E_0$ for $\sigma_{\eta}=0.3$ and $\sigma_{\theta}=0$ (top) and $\sigma_{\eta}=0.003$ and $\sigma_{\theta}=10^{-5}$ (bottom). The true value (black dashed line) is $E_0=0.80$. The model-averaged results using the BAIC (blue {\color{baic}\ding{117}}), BPIC (yellow {\color{bpic}$\blacksquare$}), and PPIC (red {\color{ppic}$\blacktriangleright$}) are consistent with model truth in all cases.} 
\label{fig:exp-n-dep}
\end{figure}

In \cref{fig:exp-n-dep}, we repeat the model averaging test with varying sample size $N$, again over $N=40, 80, 160, 320, 640, 1280$, with $\sigma_{\eta}=0.3$ and $\sigma_{\theta}=0$ (top panel) and with $\sigma_{\eta}=0.003$ and $\sigma_{\theta}=10^{-5}$ (bottom panel). All model-averaged results based on the tested information criteria agree well with model truth. As in \cref{fig:poly-n-dep}, model averaging leads to larger uncertainties compared to the result of fixing a single fit at fixed $t_{\min}$.  However, the fixed-model approach does not account for systematic error due to model truncation, as $t_{\min}$ must be adjusted as $N\rightarrow\infty$ for the estimate of $E_0$ to remain uncontaminated by higher-energy states.

\begin{figure}[!htbp]
\subfloat{\includegraphics[width=0.48\textwidth]{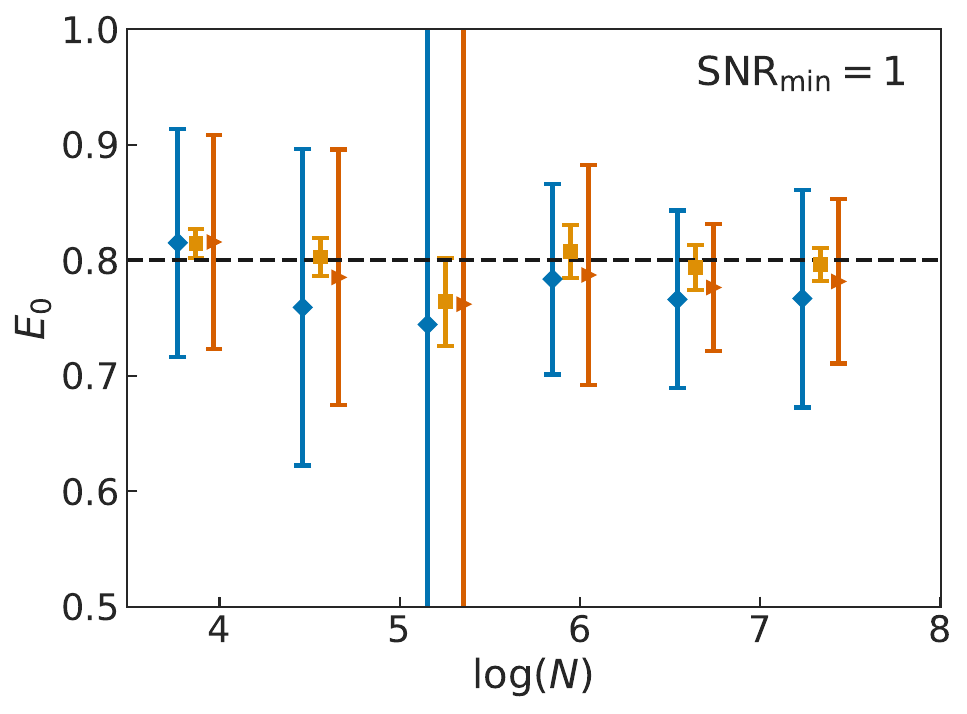}}\hfill
\subfloat{\includegraphics[width=0.48\textwidth]{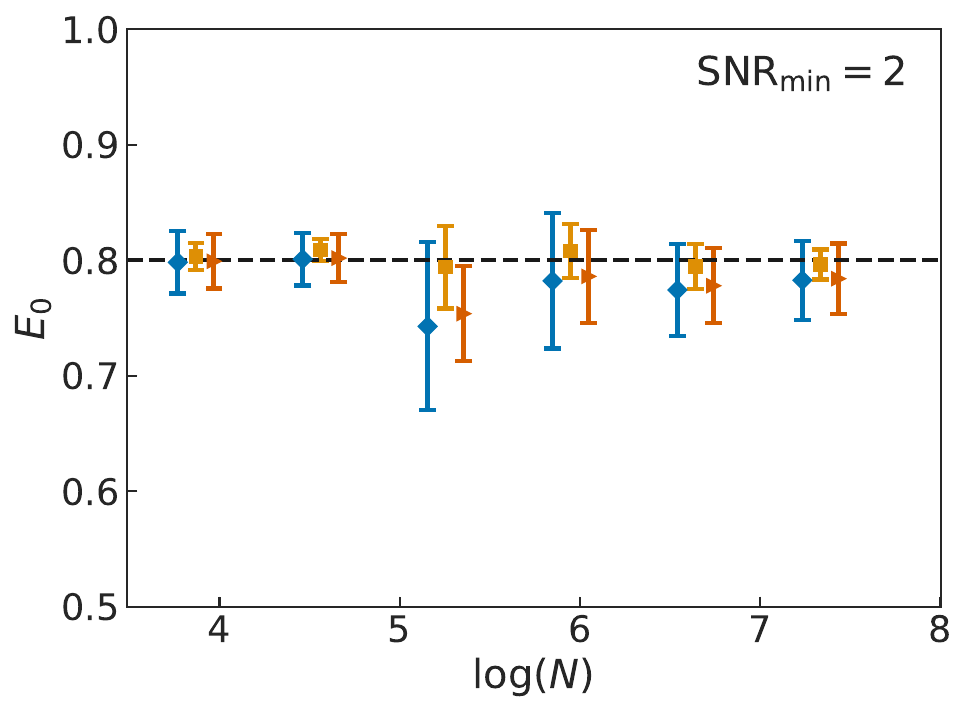}}\\
\subfloat{\includegraphics[width=0.48\textwidth]{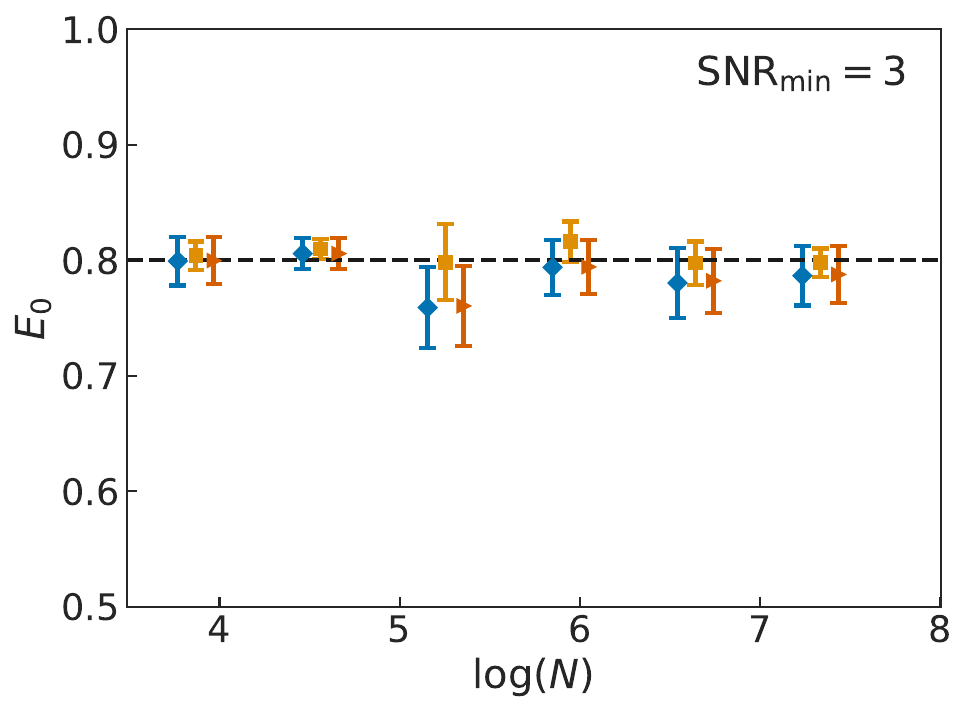}}\hfill
\subfloat{\includegraphics[width=0.48\textwidth]{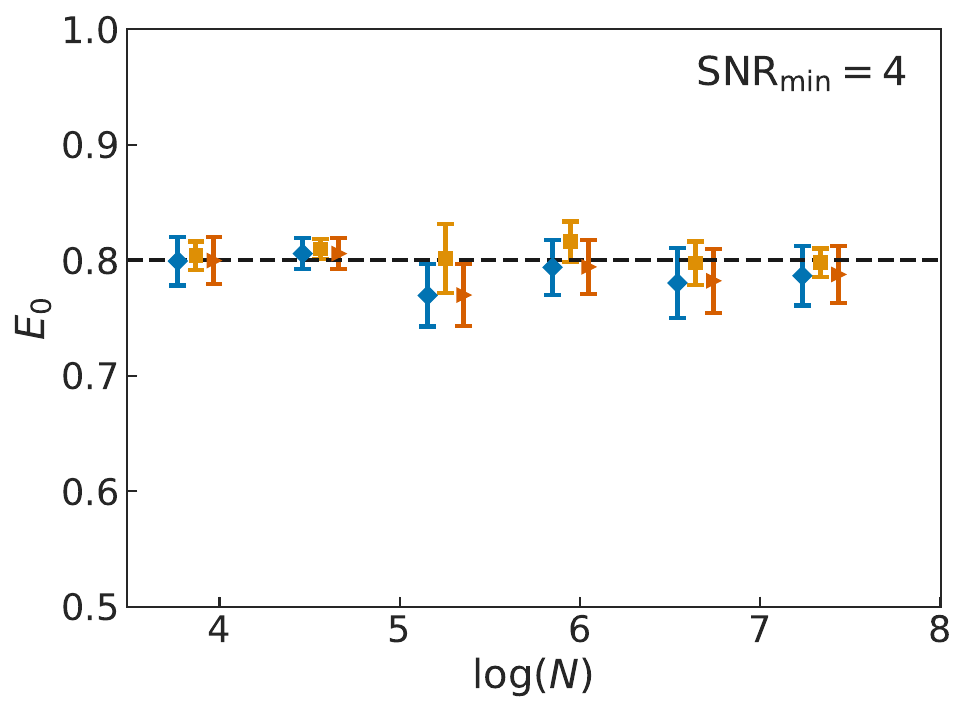}}
\caption{Model-averaging results versus the number $N$ of data samples included in the analysis.  The data is the same as the bottom panel of \cref{fig:exp-n-dep} and the subset sizes shown are $N=40, 80, 160, 320, 640, 1280$. The black dashed line shows the value of the model truth.  The model-averaged results shown use the BAIC (blue {\color{baic}\ding{117}}), BPIC (yellow {\color{bpic}$\blacksquare$}), and PPIC (red {\color{ppic}$\blacktriangleright$}). Each panel corresponds to a different value of minimum signal-to-noise ratio imposed on the data denoted by ${\rm SNR}_{\min}=1,2,3,4$ in the top right of each panel.  For ${\rm SNR}_{\rm min}=0$, see \cref{fig:exp-n-dep}.}
\label{fig:exp-snr-scaling}
\end{figure}

Though model averaging seems to protect the final results from excessively noisy data, the exponential signal-to-noise problem when the noise floor is present may cause concern about the accuracy of Laplace's method in these cases. In general, one should take caution in using data subset averaging procedure when a region of the data is effectively pure noise.  (Pure noise is, in some sense, drawn from the incorrect distribution, one centered at zero instead of around the desired signal.)  A straightforward approach to mitigating this effect is to impose a minimum signal-to-noise cut on the data before implementing the model averaging procedure. \cref{fig:exp-snr-scaling} shows the results of different minimum signal-to-noise ratios for the same $N$-scaling test data used in the bottom panel of \cref{fig:exp-n-dep}. Because the fit model decreases monotonically, we cut away the data for $t$ greater than the first time where the minimum signal-to-noise ratio is exceeded. 

We emphasize that the signal-to-noise ratio cut in these tests is distinct from the data subset selection procedure outlined in \cref{sec:data_subset_select}.  Unlike the choice of a $t_{\rm min}$ cut, a signal-to-noise cut removes data with minimal information content and therefore the choice of a specific threshold by the analyst will not be a significant source of systematic error.  Since a signal-to-noise cut discards the data before any further analysis, no information penalty need be assessed when it is used, i.e., $d_{\C}$ does not include the signal-to-noise cut data.\footnote{While this distinction is moot in the examples above (where the resulting penalty would change each IC by only an additive constant), it could be important if the BPIC or PPIC with long-range correlations are used (see \cref{sec:data_subset_select}).}

\begin{figure}[!htbp]
\includegraphics[width=0.7\textwidth]{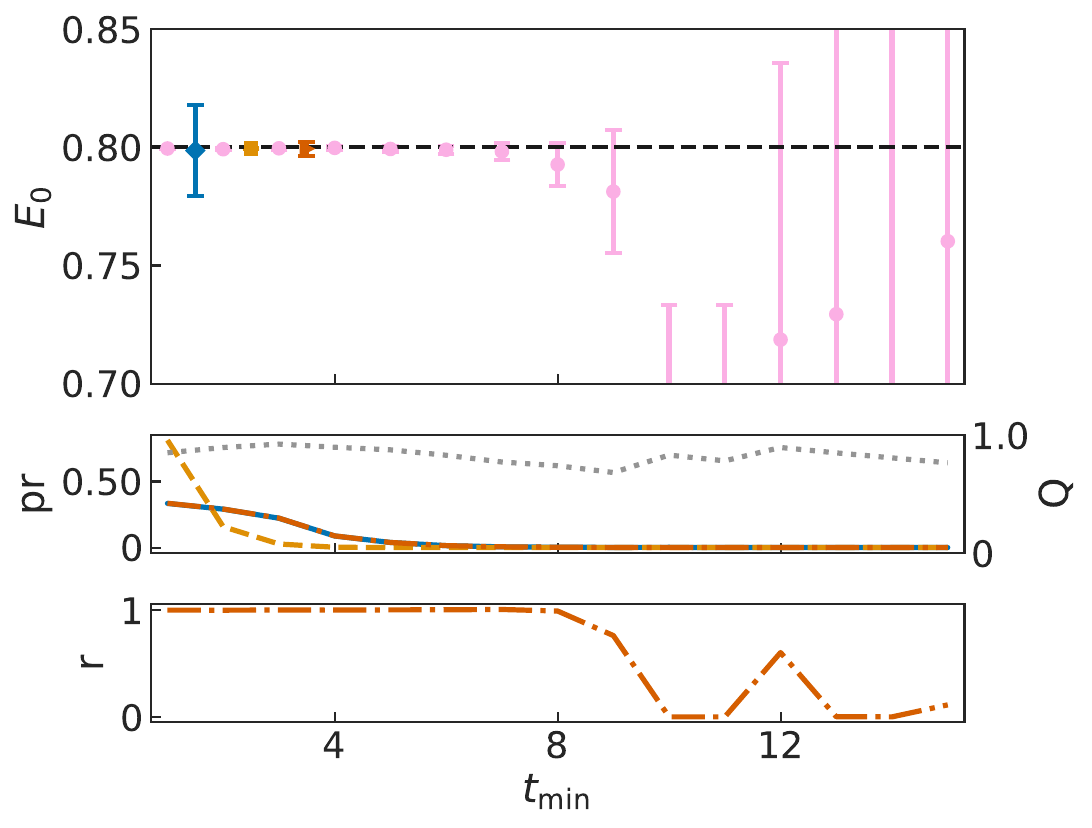}
\caption{Two-state fit results for the ground-state energy with true value $E_0=0.80$ (black dashed line) for $N=640$, $\sigma_{\eta}=0.003$, and $\sigma_{\theta}=10^{-5}$, for data subset $t\in[t_{\min},31]$ (pink {\color{indv_2state}\LARGE\textbullet}); a minimum signal to noise ratio of ${\rm SNR}_{\min}=4$ has been imposed on the data. Model-averaged results with the BAIC (blue {\color{baic}\ding{117}}), BPIC (yellow {\color{bpic}$\blacksquare$}), and PPIC (red {\color{ppic}$\blacktriangleright$}) agree well with model truth and each other in any case. The middle inset shows the standard $Q$-value (grey dotted line) and model weights $\pr(M_{\mu}|\yd)$ corresponding to the BAIC (blue solid curve), BPIC (yellow dashed curve), and PPIC (red dash-dotted curve). The lower inset shows the ratios of the
PPIC model weights to the BAIC model weights $\text{r}\equiv\pr(M_{\mu}|\yd)_{\PPIC}/\pr(M_{\mu}|\yd)_{\BAIC}$.}
\label{fig:exp-fixed-2state}
\end{figure}

\begin{figure}[!htbp]
\subfloat{\includegraphics[width=0.8\textwidth]{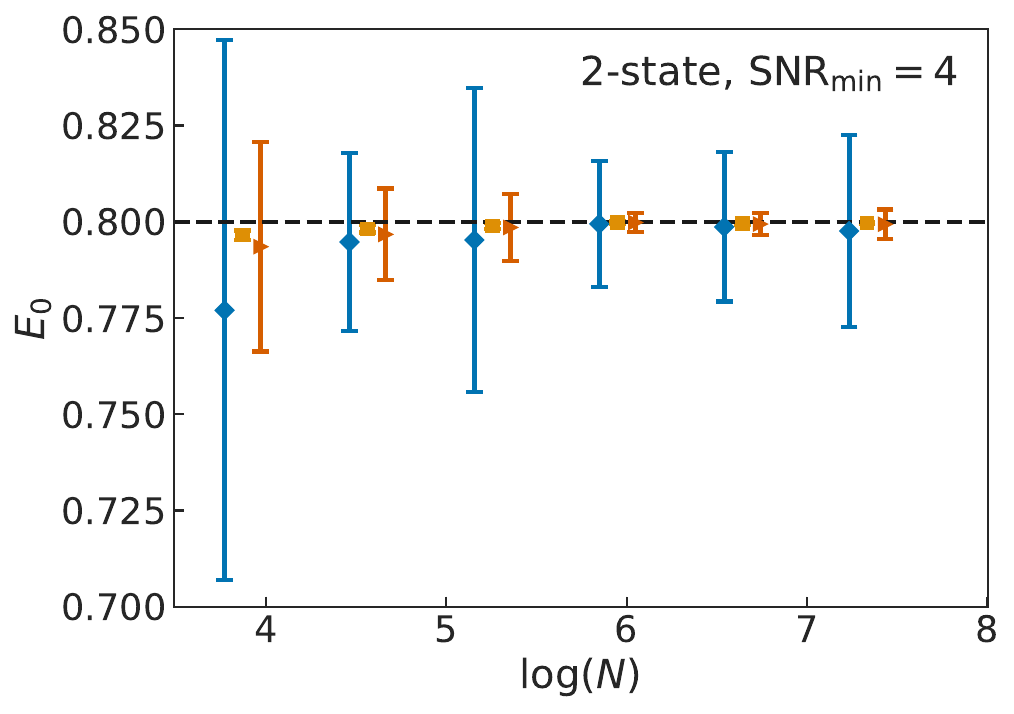}}
\caption{$N$-dependent scaling of the various estimates of the ground-state energy $E_0$ for $\sigma_{\eta}=0.003$ and $\sigma_{\theta}=10^{-5}$ using a two-state fit; a minimum signal to noise ratio of ${\rm SNR}_{\min}=4$ has been imposed on the data. The true value (black dashed line) is $E_0=0.80$. The model-averaged results using the BAIC (blue {\color{baic}\ding{117}}), BPIC (yellow {\color{bpic}$\blacksquare$}), and PPIC (red {\color{ppic}$\blacktriangleright$}) are consistent with model truth.}
\label{fig:exp-n-dep-2state}
\end{figure}

Even with the minimum signal-to-noise ratio imposed on the data in \cref{fig:exp-snr-scaling}, the variance on the model averaged results behaves counterintuitively. Namely, the uncertainty does not decrease monotonically with $N$. This is a symptom of the systematic error due to model truncation discussed in the context of the fixed-model in \cref{fig:exp-n-dep}. As $N$ increases, small values of $t$ are not well approximated by one-state exponential for the fit model \cref{eqn:exp-mod-func}.  For this reason, small $t_{\min}$ values contaminate the model averaged results, inflating parameter uncertainties.  Eventually, as $N$ increases, the data will be precise enough that there will be no region in which the one-state model is sufficient to describe the data.  Continuing to use model averaging over only one-state fits in this limit would lead to incorrect results due to model misspecification.  Instead, the model space should be expanded; a two-state fit to
\begin{align}\label{eqn:2state-exp-mod-func}
f_2(t)=A_0e^{-E_0t}+A_1e^{-E_1t},
\end{align}
can be performed to account for the excited state contamination. In contrast to correlator fits to real lattice data (cf. \cref{subsec:ex3}), \cref{eqn:2state-exp-mod-func} is of the same form as \cref{eqn:toy-model-truth} so there should be no further contamination from higher excited states. To improve numerical stability and ensure ordering of energy levels, the excited-state energy is fit in practice using the fit parameter $ldE_1 \equiv \log (E_1 - E_0)$ to replace $E_1$. \cref{eqn:2state-exp-mod-func} is fit to construct \cref{fig:exp-fixed-2state,fig:exp-n-dep-2state} for a minimum signal-to-noise ratio of ${\rm SNR}_{\min}=4$, which is sufficient to stabilize the results as shown in \cref{fig:exp-snr-scaling}. \cref{fig:exp-fixed-2state} shows the expected improved accuracy of the fits at small $t_{\min}$ using the two-state model.  Here the BPIC shows no bias with respect to the true value of $E_0$, since the two-state model is exactly the true model from which the data are drawn.

While \cref{fig:exp-n-dep-2state} shows improved behavior of the model-averaged uncertainty at large $N$ compared to \cref{fig:exp-snr-scaling} (particularly for the PPIC), the uncertainty still does not decrease monotonically as a function of $N$, as we would expect for, e.g., a parameter estimate from a single model.  This behavior may be concerning at face value since it indicates situations where adding more data does not result in reduction of error.  This behavior cannot persist indefinitely; since the error on individual sample estimates of $E_0$ scales as $1/\sqrt{N}$, and the model average is obtained from these estimates, the model-averaged uncertainty must decrease as $1/\sqrt{N}$ asymptotically.  The enhanced fluctuations in the model-averaged uncertainty versus sample size compared to single-model results are an inherent trade-off for robustness against incorrect model specification; see the discussion of bias-variance tradeoff in \cref{sec:conclusion}.

\begin{figure}[!htbp]
\subfloat{\includegraphics[width=0.8\textwidth]{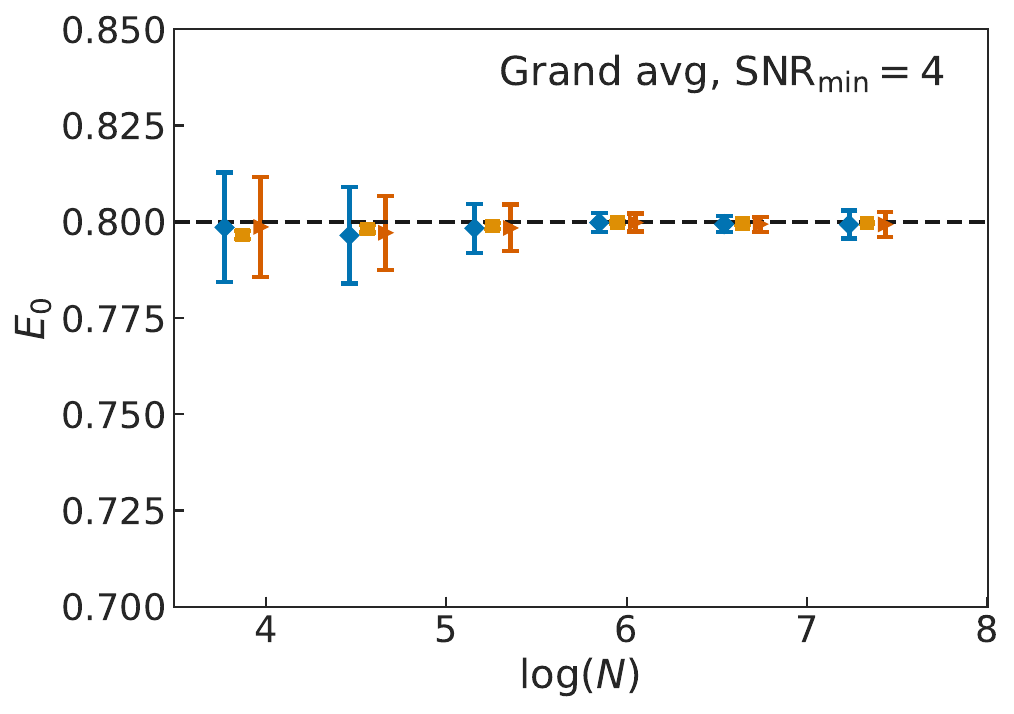}}
\caption{$N$-dependent scaling of the various estimates of the ground-state energy $E_0$ for $\sigma_{\eta}=0.003$ and $\sigma_{\theta}=10^{-5}$ averaging over both data subsets and the one- and two-state models; a minimum signal to noise ratio of ${\rm SNR}_{\min}=4$ has been imposed on the data. The true value (black dashed line) is $E_0=0.80$. The model-averaged results using the BAIC (blue {\color{baic}\ding{117}}), BPIC (yellow {\color{bpic}$\blacksquare$}), and PPIC (red {\color{ppic}$\blacktriangleright$}) are consistent with model truth.}
\label{fig:exp-n-dep-ga}
\end{figure}

\cref{fig:exp-n-dep-ga} shows the results of a ``grand average'' procedure, i.e., a model average including both one-state and two-state fits in addition to data subset selection.  If we attempt to compare these results to both \cref{fig:exp-snr-scaling,fig:exp-n-dep-2state}, we see that the grand average results actually give more precise estimates than either individual set of fits.  This shows concretely that expansion of the model space does not always result in increased uncertainty; in this case, the relative weight of individual fits with larger errors seems to be reduced in the grand average.

Briefly summarizing what we have found in our numerical results above, our main finding is that the PPIC shows the best overall performance in all tests.  Its precision is generally the same as or better than the precision of the BAIC, but without any signs of statistically significant bias versus the true value of $E_0$.  The PPIC is especially robust in cases where very noisy model estimates are part of the average; when signal-to-noise cuts are used, the PPIC's advantage tends to decrease relative to BAIC as the latter is improved more significantly, although there are counterexamples, e.g., \cref{fig:exp-n-dep-2state}.  The BAIC is the simplest criterion and also shows good performance with respect to the absence of statistically significant finite-$N$ bias.  On the other hand, the BPIC is overly aggressive in penalizing data cuts, resulting in the highest precision in many cases but together with a significant bias, which would be unacceptable in many applications in lattice field theory.

As a final remark on this example, we note that accuracy of the numerical approximations developed in \cref{subsec:laplace,subsec:super} have been corroborated in several test cases using the VEGAS algorithm, an importance-sampling-based Monte Carlo integration scheme \cite{Lepage:1977sw,Lepage:2020tgj,VegasGitHub}.  We do not show additional tables or figures with VEGAS evaluation of the full integrals, as in all cases tested these results are essentially indistinguishable from the approximate formulas.

\FloatBarrier
\subsection{Example 3: Lattice QCD correlation functions}
\label{subsec:ex3}

To further test our methodology on a more realistic example, we apply it to a real lattice QCD data set, specifically a nucleon two-point correlation function. The data consists of measurements on 615 configurations from the JLab/W\&M/MIT/LANL ensemble \texttt{a091m170} (see \cite{Mondal:2020ela} for details). On each configuration, correlators were measured on an even grid of 512 sources, projected to zero momentum, and averaged over sources. Gauge-invariant Gaussian smearing to radius 4.5 was applied at both the source and sink. We find no evidence of residual thermalization or autocorrelation effects, so we take these samples to be independent. All numerical values below are provided in implicit lattice units, i.e., $a=1$.

We carry out fits using two different model functions: a simple one-state model \cref{eqn:exp-mod-func},
and a two-state model \cref{eqn:2state-exp-mod-func}.
In practice, the parameter $ldE_1$ defined in \cref{subsec:ex2} is fit in lieu of $E_1$ for the same reasons discussed in \cref{subsec:ex2}.  The fits are done with ground-state priors $A_0 = 3(3000) \times 10^{-8}$ and $E_0 = 0.4(4)$; these were chosen primarily to ensure fit convergence.  For two-state fits, the priors are $A_1 = 3(10000) \times 10^{-8}$ and $ldE_1 = -0.5(1.0)$.  The choice of parameter priors does not affect the results qualitatively.

\begin{figure}[!htbp]
\includegraphics[width=0.7\textwidth]{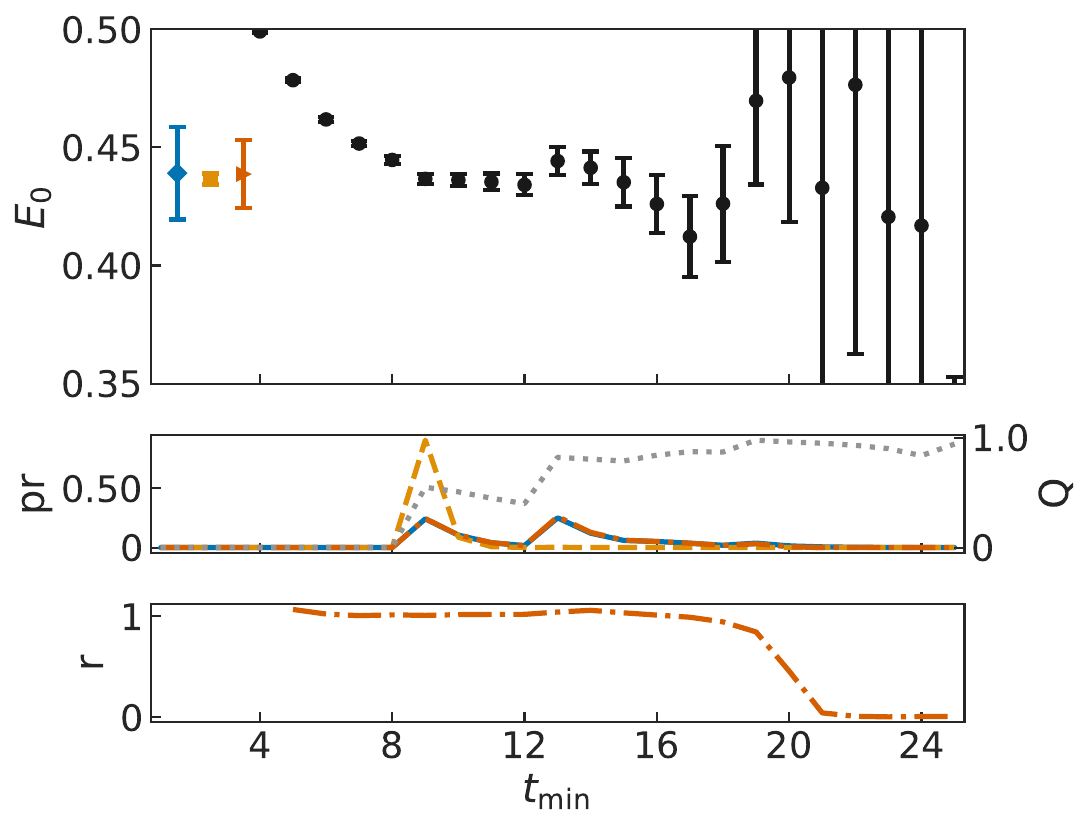}
\includegraphics[width=0.7\textwidth]{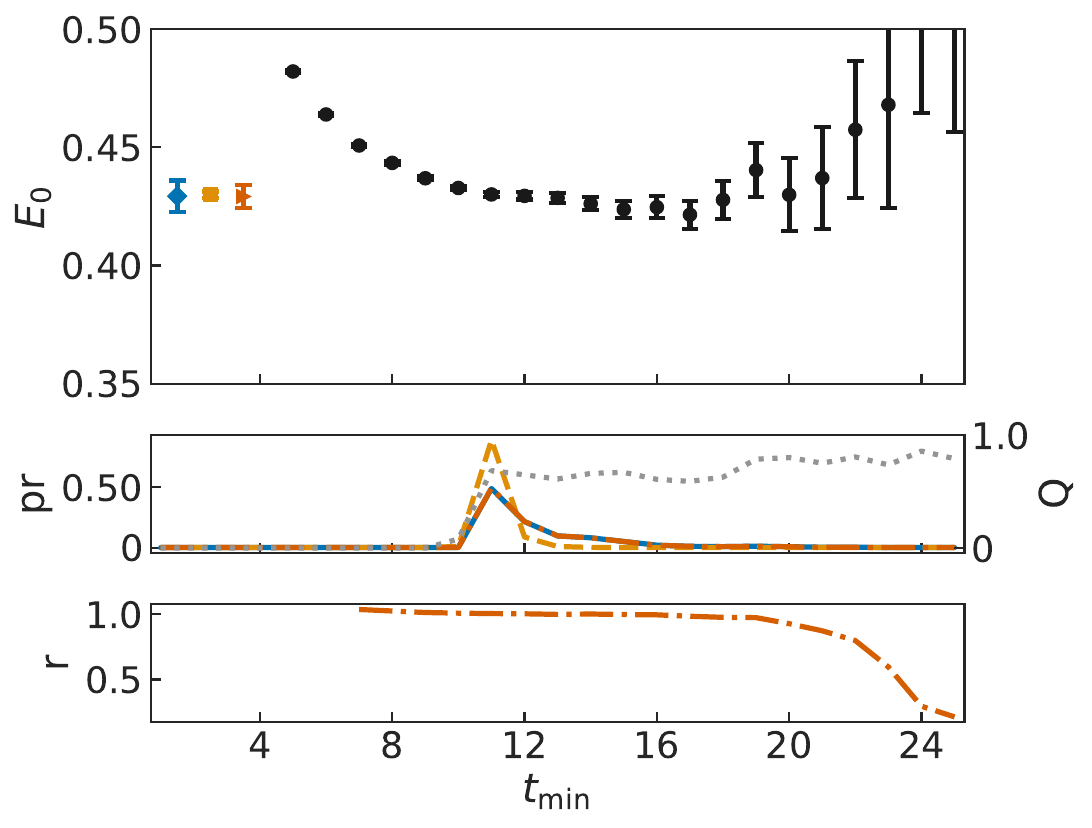}
\caption{Fit and model-averaging results for the lattice QCD nucleon data using the one-state model for $N=40$ (top panel) and $N=615$ (bottom panel).  Individual fit results to data subsets $t\in[t_{\min},30]$ are shown as black {\color{black}\LARGE\textbullet}.  The model-averaged results shown use the BAIC (blue {\color{baic}\ding{117}}), BPIC (yellow {\color{bpic}$\blacksquare$}), and PPIC (red {\color{ppic}$\blacktriangleright$}).  The lower inset shows the standard $Q$-value (grey dotted line) and model weights $\pr(M_{\mu}|\yd)$ corresponding to the BAIC (blue solid curve), BPIC (yellow dashed curve), and PPIC (red dash-dotted curve). The lower inset shows the ratios of the
PPIC model weights to the BAIC model weights $\text{r}\equiv\pr(M_{\mu}|\yd)_{\PPIC}/\pr(M_{\mu}|\yd)_{\BAIC}$.}
\label{fig:nuc-1state}
\end{figure}

\begin{figure}[!htbp]
\includegraphics[width=0.7\textwidth]{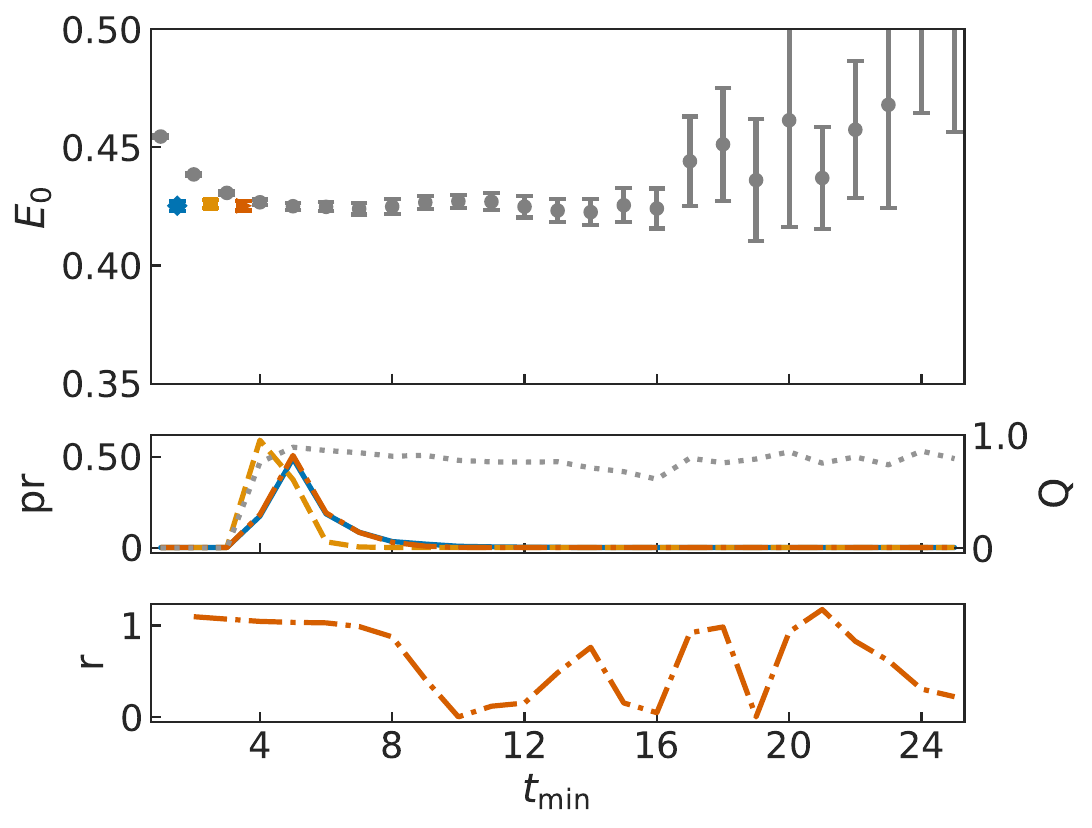}
\caption{Fit and model-averaging results for the lattice QCD nucleon data ($N = 615$) using the two-state model.  Individual fit results to data subsets $t\in[t_{\min},30]$ are shown as gray {\color{indv2}\LARGE\textbullet}. Other colors, symbols, and the lower sub-panel are all defined as in \cref{fig:nuc-1state}.}
\label{fig:nuc-2state}
\end{figure}

Individual fit results versus $t_{\rm min}$, as well as the corresponding model averaging results, are shown in \cref{fig:nuc-1state,fig:nuc-2state}.  Here we initially impose no signal-to-noise cut, although we would certainly advocate for doing so in a serious analysis of this data and will explore the effect of such a cut below.  Qualitatively, we see that the results are very similar to those obtained for the synthetic two-state exponential example above: overall the model-averaged results for the PPIC and BAIC are very similar, but the PPIC is slightly more precise.  The BPIC results tend to be much more precise, but suffer from potential bias, giving estimates at smaller $N$ which significantly disagree with the full data-set estimates.

Within \cref{fig:nuc-1state}, we also show the difference in results between the full dataset $N = 615$ and a much smaller sub-sample of $N = 40$.  Within the smaller dataset, increased model uncertainty is apparent in the plot of $\pr(M_\mu | \yd)$, with multiple local peaks appearing for the BAIC and PPIC.  This uncertainty is reflected in relatively large model-averaged uncertainty relative to any of the individual fits in the ``plateau'' region for these ICs.  With the full $N = 615$ dataset, the model probability becomes sharply peaked at the lowest value of $t_{\min}$ for which the one-state model is a good description of the data, decaying exponentially as $t_{\min}$ increases; this is precisely the behavior we expect at relatively large $N$.  Note that despite their close agreement for most fits, the PPIC once again results in a smaller uncertainty than the BAIC due to the rejection of noisy fit results at the largest $t_{\min}$, as visible from the ratio $r$ of PPIC to BAIC model weights which goes to zero at large $t_{\min}$.

\begin{figure}[!htbp]
\subfloat{\includegraphics[width=0.48\textwidth]{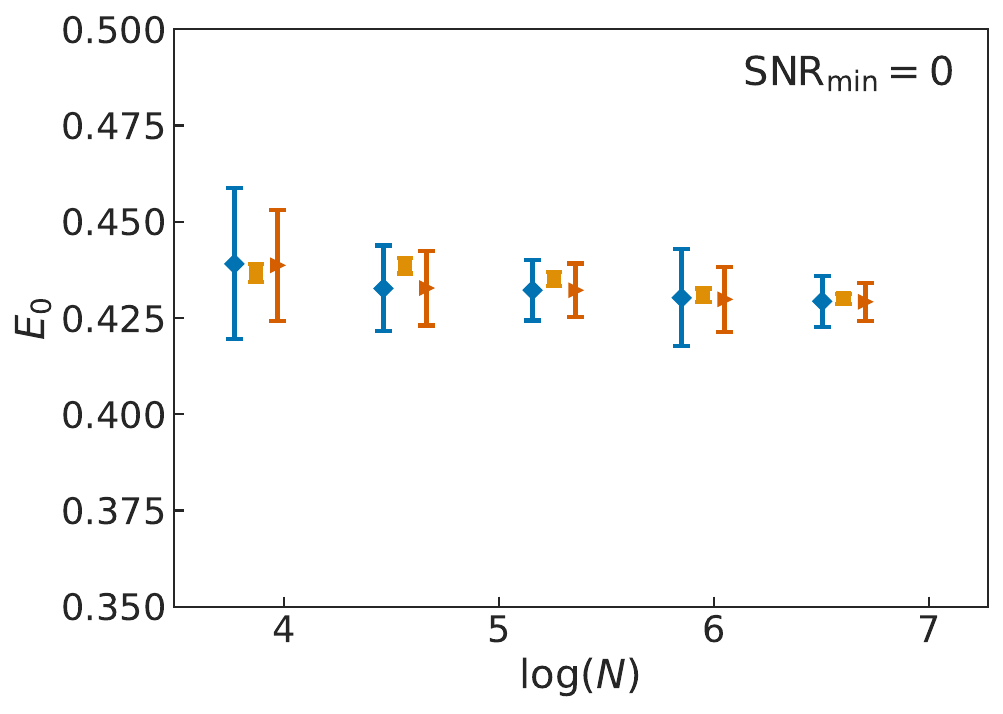}}
\subfloat{\includegraphics[width=0.48\textwidth]{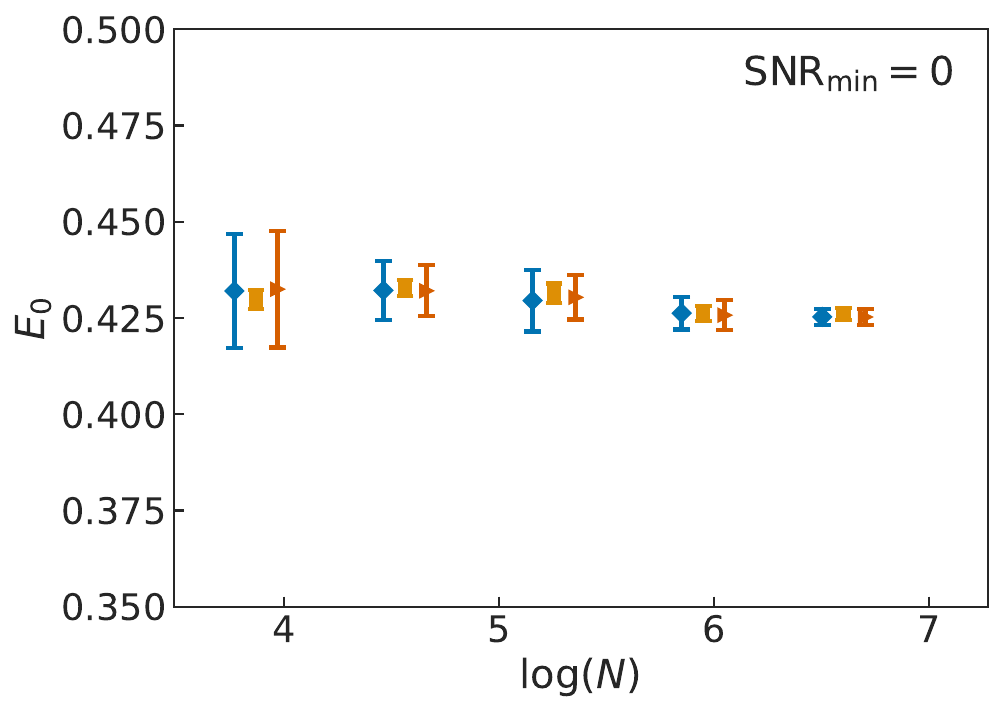}} \hfill
\subfloat{\includegraphics[width=0.48\textwidth]{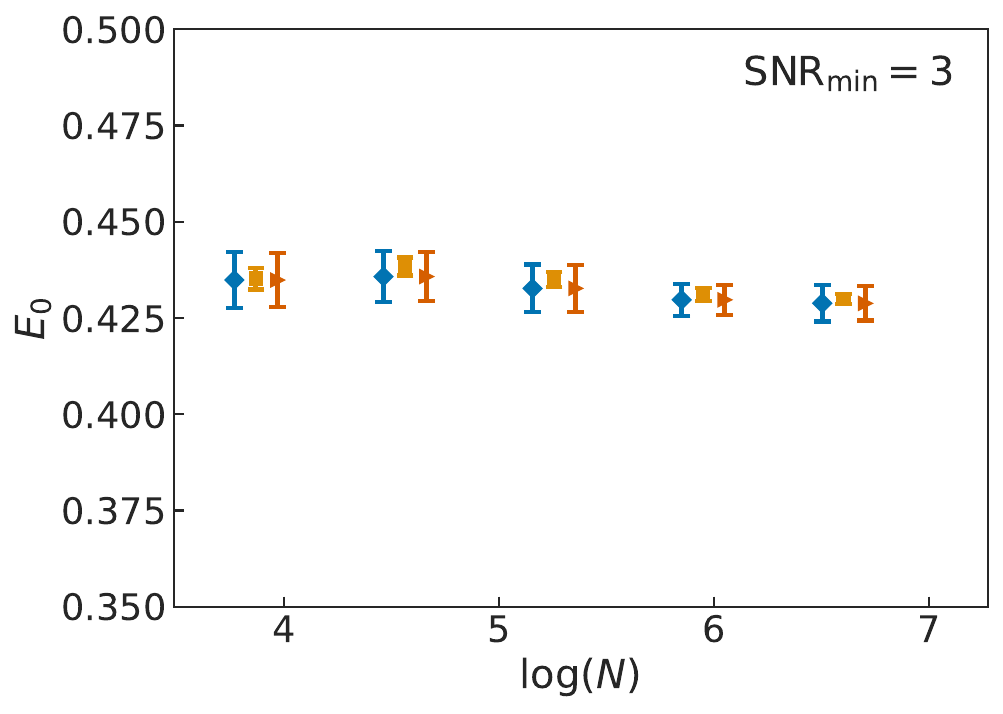}}
\subfloat{\includegraphics[width=0.48\textwidth]{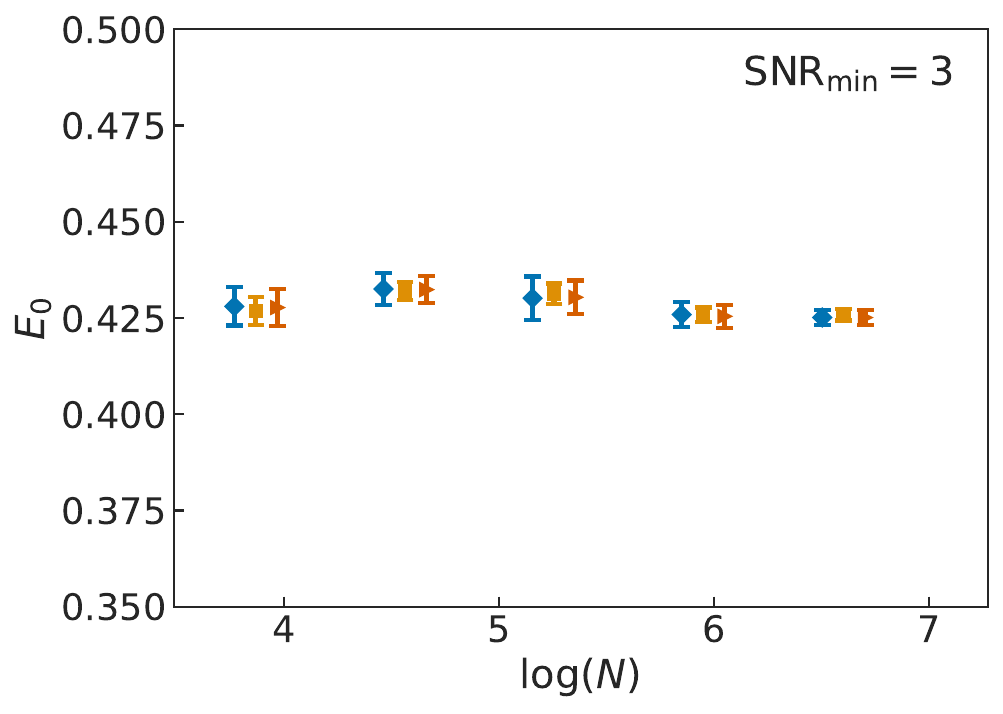}} \hfill
\subfloat{\includegraphics[width=0.48\textwidth]{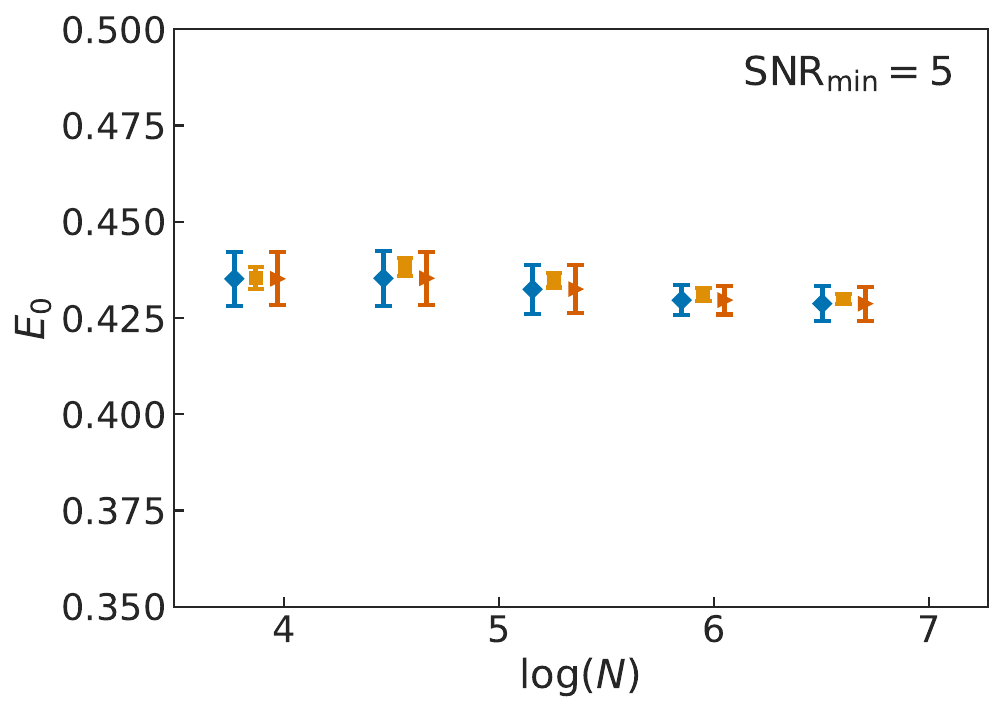}}
\subfloat{\includegraphics[width=0.48\textwidth]{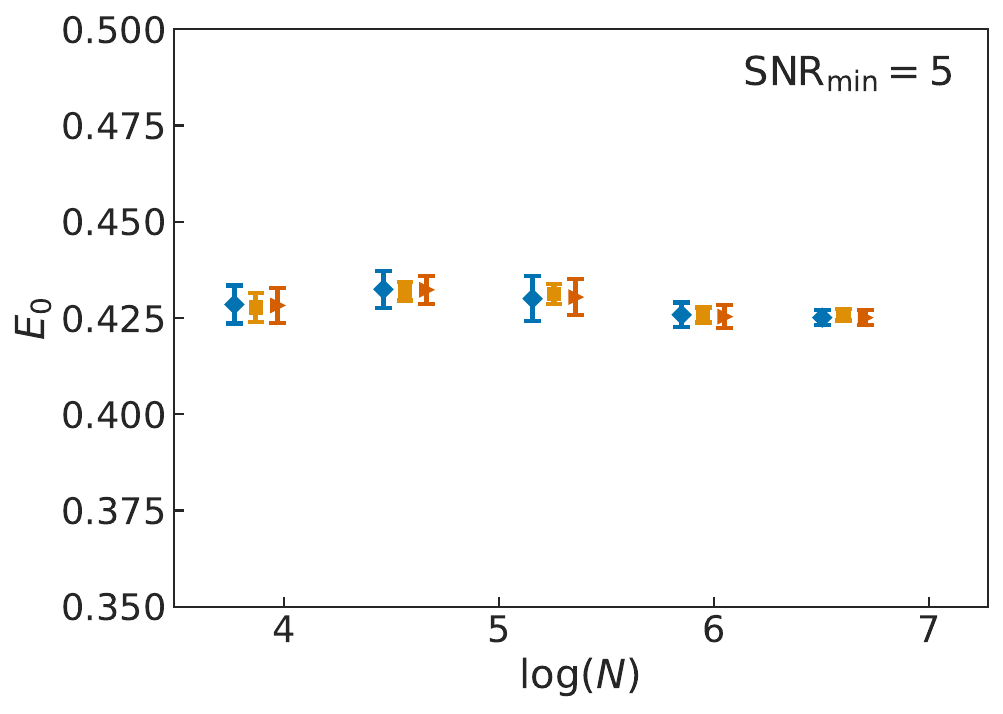}}
\caption{For the lattice QCD nucleon data, model-averaging results versus the number $N$ of data samples included in the analysis.  The data subset sizes shown are $N=40, 80, 160, 320, 615$.   The model-averaged results shown use the BAIC (blue {\color{baic}\ding{117}}), BPIC (yellow {\color{bpic}$\blacksquare$}), and PPIC (red {\color{ppic}$\blacktriangleright$}). The left column shows results obtained with the one-state fit model; the right column shows results using the two-state fit model.  Each row corresponds to a different value of minimum signal-to-noise cut imposed on the data: from the top row, the values used are ${\rm SNR}_{\min} = 0, 3, 5$.}
\label{fig:nuc-snr-scaling}
\end{figure}

\begin{figure}[!htbp]
\includegraphics[width=0.7\textwidth]{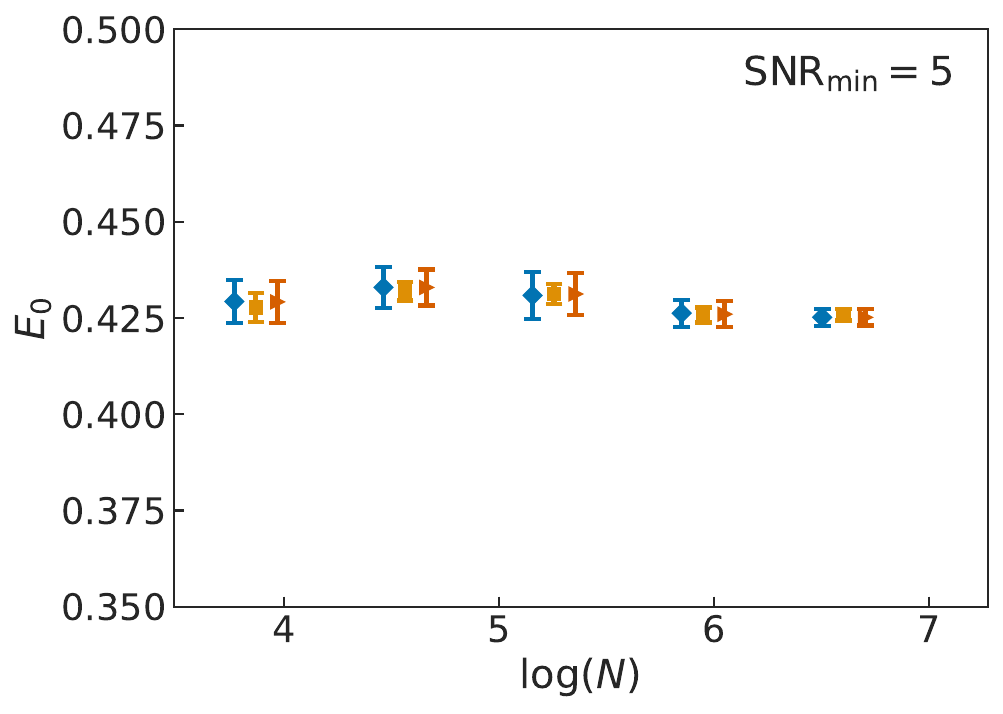}
\caption{For the lattice QCD nucleon data, combined results averaged across both data subsets and the single and double exponential model for ${\rm SNR}_{\min} = 5$ versus the number $N$ of data samples included in the analysis. The data subset sizes shown are $N= 40, 80, 160, 320, 615$.   The model-averaged results shown use the BAIC (blue {\color{baic}\ding{117}}), BPIC (yellow {\color{bpic}$\blacksquare$}), and PPIC (red {\color{ppic}$\blacktriangleright$}).}
\label{fig:nuc-ga}
\end{figure}

We further explore dependence on sample size using the nucleon data by cutting down to the first $N$ out of 615 measurements and repeating the model-averaging analysis, for $N$ taking on the values $\{40, 80, 160, 320, 615\}$.  The results of this procedure are shown in \cref{fig:nuc-snr-scaling}, imposing various levels of minimum signal-to-noise cut and for both one-state and two-state fits; the grand average over both models is shown in \cref{fig:nuc-ga}.  Most of the qualitative conclusions are very similar to those drawn from the synthetic data example; imposition of a signal-to-noise cut generally improves both the total uncertainty at fixed $N$ and the scaling of errors as a function of $N$.  For the one-state fits (left column), we can see a clear saturation of the error as $N$ increases, with no decrease in uncertainty from $N=320$ to $N=615$.  This effect can be explained by the effect discussed in the previous sub-section and clearly visible in \cref{fig:nuc-1state}, namely that as $N$ increases and data errors decrease, the ``plateau'' region in which the one-state model adequately describes the data shrinks.  The saturation of model-averaged uncertainty is thus an indicator that our model space is incomplete.  Indeed, we see that going instead to two-state fits (either exclusively or in the grand average) allows smaller error estimates to be obtained from the full $N=615$ dataset.

An important difference between this real-world nucleon data and the controlled example shown above is that for the nucleon data, the ``true model'' is not accessible---in principle, it is a sum over an infinite number of excited states.  This means that our true model is never contained in the model space we consider, strictly speaking, no matter how many excited states we include.  However, at any given $N$ and $t_{\rm min}$ the difference between the true model and a truncated model will be exponentially small as long as enough states are included.  For a given data sample, the Bayesian model averaging philosophy suggests to include all possible numbers of states and perform a ``grand average'' to determine the relative weights.  In practice, one should typically truncate the number of states once the number required to give stable descriptions of the data and saturation of error estimates, e.g. as done in \cite{Lepage2002} without model averaging.

Overall, the performance of the three information criteria tested against this real-world data mirrors what we saw in the toy-model case of \cref{subsec:ex2}.  The BAIC and PPIC are consistent with one another, with the PPIC often having smaller uncertainties particularly in the presence of noisy estimates for $E_0$.  The BPIC shows the smallest uncertainty but often also shows a consistent offset relative to the other ICs, and based on our other numerical results we would be concerned that its results are biased significantly.

\section{Conclusion }
\label{sec:conclusion}

We have adapted several information criteria from the model selection literature for use in Bayesian model averaging. The information criteria described give asymptotically unbiased estimates of the Kullback-Leibler divergence, which is sufficient to remove asymptotic bias from model averaged results when the regression procedure gives consistent model parameter estimates. By connecting these ideas to the K-L divergence, we are able to present a very general and rigorous statistical theory. We also provide specialized discussion of least squares, which is a consistent regression procedure. In the case of least squares, we derived numerically efficient and accurate asymptotic and superasymptotic approximations of the information criteria using Laplace's method and optimal truncation, respectively; these approximations are in fact exact for linear fit functions. For each information criteria, we extend the model averaging framework to data subset averaging.

The information criteria studied are the BAIC, BPIC, and PPIC, all of which are asymptotically unbiased with $o(N^{-1})$ finite-$N$ bias, where $N$ is the data sample size \cite{ZhouThesis,zhou2020posterior,Ando2007}. The approximate formulas provided for the BPIC and PPIC are at least $O(N^{-1})$ accurate; higher orders of accuracy (potentially exponential) may be achieved in cases of optimal truncation. We have chosen not to study the PAIC in detail in the body due to the lower order of accuracy ($O(1)$) of the integral approximation required and its theoretical similarity to the BPIC; the relevant PAIC formulas are given in Appendix~\ref{sec:paic_formulas}.

Each of the ICs have various strengths and weaknesses. The simplest is the BAIC, which only requires the number of parameters (and the number of the excluded data points) and evaluation of the likelihood function (i.e., $\hat{\chi}^2$) at the posterior mode $\vec{a}_{\PM}^*$ (i.e., the (Bayesian) best-fit point.) The other information criteria studied can be thought of as finite sample size corrections to the BAIC as they all give equivalent model probabilities in the large $N$ limit.  On the other hand, the use of a plug-in estimator in the BAIC is closer to frequentist than to Bayesian statistical practice as it assumes that there is a true value as opposed to a true distribution. A more natural treatment of parametric models in Bayesian inference should integrate over an estimate of this true distribution. One approach is to average the log likelihood over the posterior giving rise to the BPIC.
However, due to Jensen's inequality, the BPIC is unable to give a smaller K-L divergence than the PPIC where the log is taken after doing the posterior averaging. This seemingly minor change gives PPIC the ability to sense to individual fluctuations in the data leading to better performance in practice, particularly in situations where signal-to-noise is poor.  As seen in our numerical tests, this allows the PPIC to in some cases outperform both the BAIC, giving smaller error estimates, and the BPIC, giving greatly reduced finite-sample bias.  For this reason, we advocate for the use of the PPIC in applications of Bayesian model averaging. We note in passing that there are instances in which the individual data are not accessible and only the average statistics are known; in these cases, the PPIC cannot be computed and one of the other ICs must be used.

In the context of data subset selection, the BPIC in particular leads to an especially aggressive penalty for data cuts, which may result in significant finite-sample bias as seen in the numerical results of \cref{subsec:ex2,subsec:ex3}. Based on these results, we recommend against the use of BPIC in particular for problems where data subset selection is an important part of the model variation problem.  Both PPIC and BAIC should be used for subset variation, with preference for the PPIC when possible.

While the BAIC, BPIC, and PPIC perform differently in practice, this is merely an artifact of finite sample sizes. Being asymptotically unbiased estimators of the K-L divergence, all three ICs will become equivalent in the limit of infinite data. For this reason, it is logical to consider the BPIC and PPIC as finite-$N$ corrections to the BAIC. To better understand the behavior at finite $N$, it would be interesting in future work to consider higher-order corrections to the ICs; however, any corrections beyond $o(N^{-1})$ should be accompanied by a more detailed study of the finite-$N$ bias corrections.  Higher order bias correction similar to that of the so-called ``corrected AIC" (abbreviated as $\AIC_{\rm C}$ in the literature) could be a fruitful direction for future improvements \cite{Hurvich1989, Hurvich1995,Sugiura1978,Konishi2008,Anderson2004}.  Studies of non-asymptotic bias corrections are rare in the model selection literature but could be theoretically interesting, albeit practically challenging.

It should be noted that in all of the above cases, we make use of the approximate result $\tr[J^{-1} I] \rightarrow k$.  This is an asymptotic result that only holds if the true distribution belongs to the model space, as discussed in \cref{subsec:kl-div-plug}.  Use of the more general formula requires estimation of the $I$ and $J$ matrices, which can be numerically unstable particularly at smaller values of $N$.  An interesting direction for future work could be to explore more reliable numerical methods for estimation of this trace, such as shrinkage \cite{ledoit2004well,ledoit2012nonlinear,ledoit2017direct,Rinaldi:2019thf,FermilabLattice:2022gku}.

In our numerical studies of scaling of model averaging with sample size $N$, we sometimes encounter situations where increasing $N$ results in larger model average uncertainty.  This effect is sharply counterintuitive, since most familiar statistical error estimators (e.g., the standard error on a parameter obtained from a single model) decrease monotonically with $N$.  We interpret this effect as a manifestation of bias-variance tradeoff (see \cite{Mehta2019} for a general discussion of this phenomenon, and our own discussion in \cref{subsec:asymptotic-bias}.)  Model averaging typically results in increased uncertainty compared to the use of a single, fixed model.  This is a feature, as the increased error reflects systematic error due to model uncertainty that is neglected in the fixed-model case.  Removing this potential source of bias with model averaging results in increased error.  This tradeoff, combined with the discreteness of the model space, sometimes leads to complicated and counterintuitive behavior of uncertainty versus sample size.  Of course, it remains true that all of the individual model parameter estimates that go into the model average do have errors which decrease monotonically with sample size $N$; therefore, any large enough increase in $N$ must always tend to reduce the model averaged error as well.

It would be very interesting to explore the use of resampling techniques such as bootstrap estimation with model averaging.  For example, bootstrap methods could be used to directly estimate IC bias at finite sample size, possibly leading to improved model averaging performance at finite $N$.  Improvements to bias estimation could give better control over the bias-variance tradeoff.  It may also be interesting to explore whether direct bootstrap estimation of the K-L divergence itself, as in \cite{shibata1997bootstrap} for example, might lead to useful insights or practical improvements for model averaging.

While we have provided insight into the underlying statistical theory at play (e.g., the K-L divergence), another objective has been to give results that are useful in practical applications.  While our results are very general and can be applied to a wide array of fields, we are primarily motivated by applications to lattice field theory. Bayesian model averaging is well-suited for lattice applications due to the physical motivations and relevant functional forms of lattice models. Furthermore, the practical need for model and data truncation in lattice analyses fits naturally into the Bayesian model averaging framework.

As a brief aside, we should mention that some believe that the discrete optimization problem of model averaging (or model selection) is not a proper usage of Bayseian inference. For instance, \cite{Gelman2013} advocates for continuous model expansion, i.e., forming a larger model that includes the successful candidate models as special cases. This procedure poses obvious practical limitations as it continues \textit{ad infinitum} and could hide interesting physical insight within a large and complicated model. Despite philosophical qualms with model comparison, \cite{Gelman2013} agrees that model averaging is still a useful technique given a finite amount of available information. Furthermore, Bayesian model averaging is a natural procedure when the  model space is truly discrete as is the case for data subset selection in the analysis of lattice simulation data. That being said, it could be interesting to incorporate the idea of continuity into lattice analyses such as with Bayesian mixture models \cite{Dose2003}.

We conclude with some practical disclaimers. A notable advantage of model averaging over model selection is the removal of subjectivity from the analysis (e.g., in the analyst's choice of data subset to fit to a two-point correlator). To this end, we recommend the use of uniform parameter priors $\pr(M_{\mu})$ rather than, say, weighting parsimonious models more heavily as this is built into the ICs.  On the other hand, the factor $\pr(M_{\mu})$ should not be ignored in cases where there is good reason to include it.  For example, in cases where there is a strong theoretical reason \emph{not} to include a particular model, i.e., $\pr(M_\mu) \approx 0$, this should be reflected in the model priors rather than relying on the data entirely.  The model prior can also be used to adjust for situations where a uniform prior would result in bias; for example, if a family of highly similar models are included in the model space, a uniform prior may overweight this class of models solely due to the number of variations in the family.

We emphasize finally that model averaging is not an alternative to a statistically correct treatment of the data. While we have seen the PPIC to be somewhat robust against these effects, including incorrect results (e.g., failed fits, excessively small signal-to-noise ratios, autocorrelation effects) in the average can invalidate statistical estimates.

\begin{acknowledgments}

We give special thanks to the MIT lattice group for allowing us to make use of their nucleon two-point function data, generated using computing resources provided by the National Energy Research Scientific Computing Center, a DOE Office of Science User Facility supported by the Office of Science of the U.S. Department of Energy under Contract No. DE-AC02-05CH11231; the Oak Ridge Leadership Computing Facility at the Oak Ridge National Laboratory, which is supported by the Office of Science of the U.S. Department of Energy under Contract No. DE-AC05- 00OR22725; the USQCD Collaboration, which are funded by the Office of Science of the U.S. Department of Energy; and the Texas Advanced Computing Center (TACC) at The University of Texas at Austin.  We thank Daniel Hackett, George Fleming, and Will Jay for helpful conversations.  This work was supported in part by the U.S. Department of Energy (DOE), Office of Science, Office of High Energy Physics, under Award Number DE-SC0010005. 

\end{acknowledgments}

\bibliography{alternative-ICs}

\clearpage
\pagebreak

\appendix

\section{Marginalized information criteria}
\label{sec:app_last_paper}

In this appendix, we connect this work to \cite{Jay:2020jkz} and discuss the ABIC$_{\CV}$ in more detail. As discussed in \cref{sec:kl-div}, there are many ways to analyze parametric models using the K-L divergence. While the simplest is the plug-in estimator introduced in \cref{subsec:kl-div-plug}, a natural alternative is to marginalize over the parameter space giving the marginalized K-L divergence $\KL_{\marg}$ defined in \cref{eqn:kl-div-marg}. The corresponding information criteria is the ABIC \cite{Akaike1980,Kitagawa1997}, defined as
\begin{align}\label{eqn:abic-general}
\ABIC_{\mu}=-2\log\pr(M_{\mu})-2\log E_{\vec{a}}[\pr(\yd|\vec{a},M_{\mu})]+2k,
\end{align}
where we have written the marginalization with $E_{\vec{a}}[\dots]$, the expectation with respect to the prior distribution as defined in \cref{eqn:prior-expectation}.

In \cite{Jay:2020jkz}, the ABIC is applied to the cases of least-squares regression (see \cref{sec:least_squares} for notation).  Computing the integral to leading order via Laplace's method gives
\begin{align}\label{eqn:abic-log-det}
\ABIC_{\mu}\approx-2\log\pr(M_{\mu})+\chi_{\aug}^2(\vec{a}^*)+2k-2\log\frac{\det\tilde{\Sigma}}{\det\Sigma^*},
\end{align}
where $\tilde{\Sigma}$ and $\Sigma^*$ are the prior and best-fit covariance matrices, respectively.  To argue that the determinant terms are subleading in the limit of large sample size, \cite{Jay:2020jkz} consider the case of cross-validation, where the prior probabilities scale with the size of the data sample. In this cases, the ABIC reduces to the ``ABIC$_{\CV}$" (referred to as simply the ``AIC" in \cite{Jay:2020jkz}) defined in \cref{eqn:abic_cv-least-sqr} for least squares or more generally in \cref{eqn:abic_cv}.

This issue with this derivation is that the determinant terms can be significant in some cases, such as when priors are held fixed as is the case for the examples in \cref{sec:num-tests}. In fact, a similar derivation \cite{Konishi2008} that retains the $N$-dependence of $\det\tilde{\Sigma}/\det\Sigma^*$ leads to Schwarz's BIC \cite{Schwarz1978}:
\begin{align}
\BIC_{\mu}=-2\log\pr(M_{\mu})-2\log\pr(\yd|\vec{a}^*,M_{\mu})+\log(N)k.
\end{align}
which can be thought of as an alternative simplification to the ABIC$_{\CV}$. The BIC will clearly behave much differently from the BAIC for large $N$.

Though we do not show this rigorously here, the ABIC does not share the asymptotic unbiasedness property of the information criteria studied in the body of the text. Therefore, it can give asymptotically biased model averaging results when some models in the space of candidate models poorly reflect the data. This is a result of the outsized dependence of the ABIC on the prior as opposed to the data.  The BAIC, BPIC, PAIC, and PPIC all depend directly on the data either through the plug-in estimator $\vec{a}^*$ or through the posterior average $E_{\vec{a}|\yd}[\dots]$. In contrast, the ABIC uses an expectation over the prior. As a result, the ABIC may behave counterintuitively even in the limit of infinite data. This behavior is related to the so-called ``Jeffreys-Lindley paradox" in which a Bayesian analysis under certain choices of prior distribution can give incorrect results compared to, say, the analogous frequentist analysis. For further discussion, see \cite{Lindley:1957aa,Jeffreys1998,Cousins2017}.

\section{Asymptotic equivalence of information criteria}
\label{sec:asymptotic-appendix}

In this appendix, we compare the asymptotic forms of the different K-L divergences defined in \cref{sec:kl-div}. In doing so, we will establish that each form of the K-L divergence is asymptotically equivalent in the sense of model choice (e.g., model selection, model averaging). This is weaker than asserting that the different forms of the K-L divergence approach each other numerically and only requires convergence of the leading order terms in respective asymptotic expansions in the inverse sample size $N^{-1}$.

Reproducing only the model-dependent terms here in a convenient form, and omitting the expectation value $E_z[...]$ and the model choice $M_\mu$ (we want to compare them for a fixed model), we have
\begin{align}
\KL_{\plug} &\supset \log \pr(z|\vec{a}^*), \\
\KL_{\post} &\supset E_{\vec{a} | \yd} [\log \pr(z | \vec{a})], \\
\KL_{\pred} &\supset \log E_{\vec{a} | \yd} [ \pr(z | \vec{a}) ],
\end{align}
where
\begin{align}
E_{\vec{a} | \yd} [ ... ] \equiv \frac{\int d\vec{a}\ \pr(\vec{a} | \yd) (...)}{\int d\vec{a}\ \pr(\vec{a} | \yd)} = \frac{\int d\vec{a}\ \pr(\yd | \vec{a}) \pr(\vec{a}) (...)}{\int d\vec{a}\ \pr(\yd | \vec{a}) \pr(\vec{a})},
\end{align}
and $\pr(\vec{a})$ is the prior probability distribution for the given model. Throughout this appendix, we will assume that the prior information grows sufficiently slowly with $N$ compared to the data; specifically, we assume that \cref{eqn:prior_assumption} holds (e.g., fixed priors without data dependence). 

To proceed, note that for any continuous $f$
\begin{align}
\lim_{N\rightarrow\infty}\frac{\int d\vec{a}\ \pr(\yd | \vec{a}) \pr(\vec{a}) f(\vec{a})}{(2\pi)^{k/2}\abs{J_N^{-1}(a^*)}^{1/2}\pr(\vec{a}^*) f(\vec{a}^*)}=1,
\end{align}
which follows from convergence of the leading-order Laplace approximation. This is a standard result proven in many texts (e.g., \cite{Kirwin2010}). A more detailed discussion of the Laplace approximation is given in Appendix~\ref{sec:app_laplace_method}. Therefore,
\begin{align}
E_{\vec{a} | \yd} [\log \pr(z | \vec{a})]&=\frac{\int d\vec{a}\ \pr(\yd | \vec{a}) \pr(\vec{a}) \log \pr(z | \vec{a})}{\int d\vec{a}\ \pr(\yd | \vec{a}) \pr(\vec{a})}\\
&\rightarrow\frac{(2\pi)^{k/2}\abs{J_N^{-1}(a^*)}^{1/2}\pr(\vec{a}^*) \log \pr(z | \vec{a}^*)}{(2\pi)^{k/2}\abs{J_N^{-1}(a^*)}^{1/2}\pr(\vec{a}^*)}=\log \pr(z | \vec{a}^*),\\
\log E_{\vec{a} | \yd} [\pr(z | \vec{a})]&=\frac{\int d\vec{a}\ \pr(\yd | \vec{a}) \pr(\vec{a}) \pr(z | \vec{a})}{\int d\vec{a}\ \pr(\yd | \vec{a}) \pr(\vec{a})}\\
&\rightarrow \log \frac{(2\pi)^{k/2}\abs{J_N^{-1}(a^*)}^{1/2}\pr(\vec{a}^*) \pr(z | \vec{a}^*)}{(2\pi)^{k/2}\abs{J_N^{-1}(a^*)}^{1/2}\pr(\vec{a}^*)}= \log \pr(z | \vec{a}^*),
\end{align}
in the $N\rightarrow\infty$ limit. This establishes the asymptotic equivalence of $\KL_{\post}$ and $\KL_{\pred}$ to $\KL_{\plug}$. Since the various ICs we considered follow directly from the K-L divergence definitions, this establishes that both the PPIC and the PAIC/BPIC converge asymptotically to the BTIC/BAIC.

Similarly to \cref{subsec:laplace}, this result is established using Laplace's method. However, we emphasize that this equivalence is much more generally applicable than to just the case of least-squares discussed in \cref{subsec:laplace}. See \cite{Kirwin2010,Barndorff1989} for the required regularity conditions.

In effect, the above argument indicates
\beq\label{eqn:propto-delta}
\lim_{N \rightarrow \infty} \pr(\vec{a} | \yd) \propto \delta(\vec{a} - \vec{a}^*).
\eeq
To understand this we note that both $\KL_{\post}$ and $\KL_{\pred}$ contain the posterior probability 
\beq
\log \pr(\vec{a} | \yd) \propto \sum_i \log \left[ \pr(y_i | \vec{a}) \pr(\vec{a})^{1/N} \right].
\eeq
Since we have assumed \cref{eqn:prior_assumption}, the influence of the prior is negligible compared to that of the data as $N\rightarrow\infty$, so that this simply approaches the log likelihood function $\sum_i \log \pr(y_i | \vec{a})$. In terms of the parameters, the likelihood function becomes Gaussian asymptotically with a width decreasing proportional to $N$. As a result, we have the proportionality in \cref{eqn:propto-delta}. This connection is not made rigorous here, but is expected to hold except in pathological cases.

\section{A bound on the asymptotic bias of model averaging}
\label{sec:bias-appendix}

Here we derive the bound on the model averaging asymptotic bias given in \cref{eqn:asymptotic_bias}. For a general discussion of asymptotic bias and the relevant notation, see \cref{subsec:asymptotic-bias}. First, it will be useful to introduce some new notation for this appendix. Specifically, dependencies on the sample size will be shown explicitly with a subscript $N$. When the subscript $N$ is absent, this denotes the asymptotic value (e.g., $A\equiv\lim_{N\rightarrow\infty}A_N$ is the asymptotic value for a sequence of sample estimators $\{A_N\}_{N\in\mathbb{N}}$). The one exception is the sample data $\yd$, for which the $N$-dependence is clear and the asymptotic value (a random variable drawn from $\pr_{\T}$) is denoted by $z$.  For simplicity, we do not distinguish between $b_y$ (finite-sample bias) and $b_z$ (asymptotic bias) as defined in \cref{eqn:bias-finite-sample,eqn:bias-asymptotic} in this appendix, as making the $N$-dependence explicit is sufficient.

The bound in \cref{eqn:asymptotic_bias} holds (with probability $1$) if the parameter estimation procedure is consistent. The sequence of sample estimators $\{X(\yd)\}_{N\in\mathbb{N}}$ of $\xi$ is consistent if it satisfies \cite{larsen2005introduction}
\begin{equation}\label{eqn:weak-consistency}
\lim_{N \rightarrow \infty} \pr\left(|X(\yd) - \xi| > \epsilon\right) = 0,
\end{equation}
for any $\epsilon > 0$. This form of consistency is also known as weak consistency, in contrast to strong consistency where $\{X(\yd)\}_{N\in\mathbb{N}}$ satisfies \cite{Takeshi1985}
\begin{equation}\label{eqn:strong-consistency}
\pr\para{\lim_{N \rightarrow \infty}X(\yd)=\xi}=1.
\end{equation}
Weak consistency is defined using convergence in probability whereas strong consistency is defined using almost sure convergence.  Since almost sure convergence implies convergence in probability, strong consistency implies weak consistency.  Another related notion is convergence in the sense of distributions, which is implied by convergence in probability; convergence in probability and convergence in the sense of distributions are equivalent if the limiting random variable $X(z)$ is a constant.

As discussed in \cref{subsec:asymptotic-bias}, our primary goal is to remove asymptotic bias from the model average parameter estimates.  For concreteness, consider the bias of a single parameter $a_0$ given by
\begin{align}
b_z[\avg{a_0}_N]=E_z[\avg{a_0}_N]-a_{0,\T}^*.
\end{align}
By \cref{eqn:model_avg}, we have
\begin{align}
\avg{a_0}_N&=a_{0,\T,N}^*\pr(M_{\T}|\yd)+\sum_{\mu\neq\T}a_{0,\mu,N}^*\pr(M_\mu|\yd),
\end{align}
where $\pr(M_\mu|\{y\})$ satisfy \cref{eqn:asymptotic_model_weights}.

We wish to derive a bound on $b_z[\avg{a_0}_N]$ in terms of the asymptotic bias in the model weights. To that end, observe that
\begin{align}
\abs{b_z[\avg{a_0}_N]}&=\abs{E_z[\avg{a_0}_N]-a_{0,\T}^*}\\
&=\abs{\sum_\mu\left\{E_z[a_{0,\mu,N}^*\pr(M_\mu|\yd)-a_{0,\mu}^*\pr(M_{\mu}|z)]\right\}}\\
&=\abs{\sum_\mu\left\{E_z[(a_{0,\mu,N}^*-a_{0,\mu}^*)\pr(M_\mu|\yd)]+a_{0,\mu}^*b_z[\pr(M_\mu|\yd)]\right\}} \\
\label{eqn:pre-consistency} &\leq\sum_\mu\left\{\abs{E_z[(a_{0,\mu,N}^*-a_{0,\mu}^*)\pr(M_\mu|\yd)]}+\abs{a_{0,\mu}^*}\abs{b_z[\pr(M_\mu|\yd)]}\right\}.
\end{align}
Let $\epsilon>0$ and observe that
\begin{align}
\abs{E_z[(a_{0,\mu,N}^*-a_{0,\mu}^*)\pr(M_\mu|\yd)]}=&\abs{\int dF_{M_{\T}}(z)(a_{0,\mu,N}^*-a_{0,\mu}^*)\pr(M_\mu|\yd)}\\
\leq&\int dF_{M_{\T}}(z)\abs{a_{0,\mu,N}^*-a_{0,\mu}^*}\pr(M_\mu|\yd)\\
\leq&\int dF_{M_{\T}}(z)\abs{a_{0,\mu,N}^*-a_{0,\mu}^*}\\
\nonumber =&\int_{\Omega_{N,\epsilon}} dF_{M_{\T}}(z)\abs{a_{0,\mu,N}^*-a_{0,\mu}^*}\\
&+\int_{\Omega_{N,\epsilon}^c} dF_{M_{\T}}(z)\abs{a_{0,\mu,N}^*-a_{0,\mu}^*}\\
\label{eqn:norm}\nonumber \leq&\norm{a_{0,\mu,N}^*-a_{0,\mu}^*}_{L^{\infty}(\Omega_{N,\epsilon})}F_{M_{\T}}(\Omega_{N,\epsilon})\\
&+\norm{a_{0,\mu,N}^*-a_{0,\mu}^*}_{L^{\infty}(\Omega_{N,\epsilon}^c)}F_{M_{\T}}(\Omega_{N,\epsilon}^c),
\end{align}
where $\Omega_{N,\epsilon}=\left\{z\in\mathbb{R}^d:\abs{a_{0,\mu,N}^*-a_{0,\mu}^*}\leq\epsilon\right\}$ and $\Omega_{N,\epsilon}^c=\left\{z\in\mathbb{R}^d:\abs{a_{0,\mu,N}^*-a_{0,\mu}^*}>\epsilon\right\}$ is its complement. In \cref{eqn:norm}, $\norm{\dots}_{L^{\infty}(\Omega)}$ denotes the $L^{\infty}$-norm with respect to the true probability measure $F_{M_{\T}}(z)$ (defined in \cref{eqn:true-prob-meas}) on the set $\Omega$, i.e.,
\begin{align}
\norm{f}_{L^{\infty}(\Omega)} &\equiv \inf\{M\in\mathbb{R}:\abs{f(\Omega)}\leq M \text{ almost surely with respect to }F_{M_{\T}}(z)\}\\
&=\esssup_{z\in\Omega}(\abs{f(z)}),
\end{align}
where ``$\esssup$" denotes the essential supremum with respect to $F_{M_{\T}}(z)$.  For simplicity, we have also adopted the notation that the measure of a set $\Omega$ is denoted by
\begin{align}
F_{M_{\T}}(\Omega)\equiv\int_{\Omega}dF_{M_{\T}}(z).
\end{align}

To proceed, we assume the parameter estimation procedure is weakly consistent, i.e.,
\begin{equation}
\lim_{N \rightarrow \infty} \pr(|a_{0,\mu,N}^*-a_{0,\mu}^*| > \epsilon) = 0.
\end{equation}
This will be true for, say, least-squares regression, which is in fact strongly consistent in some cases \cite{Jennrich1969,Wu1981}. It follows from this assumption that, in the large $N$ limit, $\Omega_{N,\epsilon}$ has unit measure and $\Omega_{N,\epsilon}^c$ has measure zero.  Therefore,
\begin{align}
\lim_{N\rightarrow\infty}\abs{E_z[(a_{0,\mu,N}^*-a_{0,\mu}^*)\pr(M_\mu|\yd)]}\leq&\lim_{N\rightarrow\infty}\norm{a_{0,\mu,N}^*-a_{0,\mu}^*}_{L^{\infty}(\Omega_{N,\epsilon})}\leq\epsilon.
\end{align}
Taking $\epsilon$ arbitrarily small, \cref{eqn:pre-consistency} gives the following bound on the asymptotic bias:
\begin{align}
\abs{b_z[\avg{a_0}]}&=\lim_{N \rightarrow \infty}\abs{b_z[\avg{a_0}_N]}\leq\sum_\mu\abs{a_{0,\mu}^*}\lim_{N \rightarrow \infty}\abs{b_z[\pr(M_\mu|\yd)]}\\
&=\sum_\mu\abs{a_{0,\mu}^*}\abs{b_z[\pr(M_\mu|z)]},
\end{align}
which holds with probability $1$. Using \cref{eqn:model_avg}, a similar argument holds for $\avg{f(\mathbf{a})}$ giving the bound in \cref{eqn:asymptotic_bias}.

While this derivation seems rather technical, it is related to H\"{o}lder's inequality \cite{Hunter2001}:
\begin{align}
\abs{\int_{\Omega}d\mu f g}\leq\norm{f}_{L^p(\Omega)}\norm{g}_{L^q(\Omega)},
\end{align}
where $(\Omega,\mathcal{S},\mu)$ is a measure space ($\mathcal{S}$ being some $\sigma$-algebra on $\Omega$), $fg\in L^1(\Omega)$, $f\in L^p(\Omega)$ and $g\in L^q(\Omega)$ with respect to $\mu$. We have assumed that $1\leq p,q\leq\infty$ satisfy $1/p+1/q=1$ (i.e., $p$ and $q$ are H\"{o}lder conjugates); ``$1/\infty$" is defined as zero in this context.  We are able to achieve a somewhat sharper bound on $\abs{E_z[(a_{0,\mu,N}^*-a_{0,\mu}^*)\pr(M_\mu|\yd)]}$ using the fact that $\pr(M_\mu|\yd)$ is bounded between zero and one and assuming consistency (along with partitioning $\Omega$ as above).

\section{Formulas for the least-squares PAIC}
\label{sec:paic_formulas}

In this appendix, we discuss approximations to \cref{eqn:paic_expectation}---the PAIC for least-squares regression---in the various cases of interest and the practical complications that arise.

Similarly to the BPIC in \cref{subsec:bpic} and the PPIC in \cref{subsec:ppic}, the PAIC can be approximated by
\begin{align}\label{eqn:paic}
\PAIC_{\mu}\approx&\hat{\chi}^2(\vec{a}^*)+\frac{1}{2}\hat{H}_{ba}(\Sigma^*)_{ab}-\frac{1}{2}\hat{g}_dT_{cba}(\Sigma_2^*)_{abcd}+2k,
\end{align}
where we have assumed that the correct model is in the family of candidates so that $\tr\left[J^{-1}(\vec{a}^*)I(\vec{a}^*)\right]\rightarrow k$ and
\begin{align}
\hat{g}_a\equiv&\left.\frac{\partial\hat{\chi}^2}{\partial a_a}\right|_{\vec{a}=\vec{a}^*}, &
\hat{H}_{ab}\equiv&\left.\frac{\partial^2\hat{\chi}^2}{\partial a_a\partial a_b}\right|_{\vec{a}=\vec{a}^*}.
\end{align}
One might suspect that \cref{eqn:paic} is an NLO asymptotic expansion in the inverse sample size $N^{-1}$. However, there are some subtleties in the power counting for nominally $O(1)$ terms $\frac{1}{2}\hat{H}_{ba}(\Sigma^*)_{ab}$ and $-\frac{1}{2}\hat{g}_dT_{cba}(\Sigma_2^*)_{abcd}$. First, note that
\begin{align}
\frac{1}{2}\tr\left[\hat{H}\Sigma^*\right]=\tr\left[\left(\Sigma^*{}^{-1}-\frac{1}{2}\tilde{H}\right)\Sigma^*\right]=k-\frac{1}{2}\tr\left[\tilde{H}\Sigma^*\right],
\end{align}
where the substitution for $\hat{H}$ follows from the definition $\chi_{\rm aug}^2 = \hat{\chi}^2 + \tilde{\chi}^2$ and from \cref{eqn:covariance}. So, we see that the second term in \cref{eqn:paic} includes an $O(N^{-1})$ contribution $-\frac{1}{2}\tilde{H}_{ba}(\Sigma^*)_{ab}$.

Second, $\vec{a}^*$ is chosen so that the gradient of $\chi_{\aug}^2(\vec{a})$ vanishes. It follows that $-\hat{g}=\tilde{g}$, and thus
\begin{align}
-\frac{1}{2}\hat{g}_dT_{cba}(\Sigma_2^*)_{abcd}=\frac{1}{2}\tilde{g}_dT_{cba}(\Sigma_2^*)_{abcd},
\end{align}
giving rise to another $O(N^{-1})$ term.

Therefore,
\begin{align}
\PAIC_{\mu}=&\hat{\chi}^2(\vec{a}^*)+3k+O(N^{-1}).
\end{align}
In applying Laplace's method to \cref{eqn:paic_expectation}---an integral with $O(N)$ integrand---we have already neglected some $O(N^{-1})$ terms (cf. \cref{eqn:neglect_nnlo}). So, we again must neglect the $O(N^{-1})$ terms in order to maintain a consistent $O(1)$ asymptotic expansion giving
\begin{align}
\PAIC_{\mu}\approx&\hat{\chi}^2(\vec{a}^*)+3k.
\end{align}
The corresponding superasymptotic expansion is
\begin{align}\label{eqn:paic_opt_trunc}
\PAIC_{\mu} \approx &\begin{cases}
\hat{\chi}^2(\vec{a}^*)+3k, & k < \hat{\chi}^2(\vec{a}^*), \\
\hat{\chi}^2(\vec{a}^*) + 2k, & {\rm otherwise,}
\end{cases}
\end{align}
which should be used for nonlinear least squares. For linear least squares, the NLO expansion is exact and \cref{eqn:paic} can be used (with $T=0$), which is also identical to the form of the BPIC in \cref{eqn:bpic_linear}.

For data subset selection, the derivation is similar to that of the BPIC in \cref{sec:data_subset_select}; the PAIC is modified simply by the addition of a factor of $3d_{\C}$ for the cut portion of the data, i.e.
\begin{equation} \label{eqn:paic_subset}
\PAIC_{\mu,P} = \PAIC_{\mu} + 3d_{\C}
\end{equation}
with $\PAIC_{\mu}$ given by the superasymptotic formula in \cref{eqn:paic_opt_trunc}.  For use in model averaging, \cref{eqn:IC_MA} applies.

\section{Laplace's method and Gaussian integral formulas}
\label{sec:app_laplace_method}

Here we summarize the derivation of the NLO Laplace approximation applied to the integrals needed for the various information criteria appearing in the body of the paper in the case of least squares.  A more rigorous treatment of Laplace method can be found in many texts on asymptotics \cite{Bender2013,Bleistein1975}, and generalizations appear in the asymptotics literature \cite{Kirwin2010}.

Here we consider integrals of the form
\begin{align}
\mathcal{I}[\psi]=\int d\vec{a}\exp\left[-\frac{1}{2}\chi_{\aug}^2(\vec{a})\right]\psi(\vec{a}).
\end{align}
In the limit of large sample size, $\chi_{\aug}^2(\vec{a})=O(N)$. In the $N\rightarrow\infty$ limit, the integrand becomes sharply peaked about the best-fit parameter $\vec{a}^*=\argmin_{\vec{a}}\chi_{\aug}^2(\vec{a})$. Assuming for simplicity that $\vec{a}^*$ is on the interior of the parameter space (as is the case for the examples considered above), the integral becomes localizes to $B_{\epsilon}(\vec{a}^*)=\{\vec{a}:\norm{\vec{a}-\vec{a}^*}<\epsilon\}$, the ball of radius $\epsilon>0$ centered at $\vec{a}^*$:
\begin{align}
\label{eqn:eps_int}
\mathcal{I}[\psi]\approx\int_{\vec{a}\in B_{\epsilon}(\vec{a}^*)}d\vec{a}\exp\left[-\frac{1}{2}\chi_{\aug}^2(\vec{a})\right]\psi(\vec{a}).
\end{align}
Assuming $\epsilon$ is small, we can use truncated Taylor expansions about $\norm{\vec{a}-\vec{a}^*}=0$. To obtain the leading-order integral expansion, we would Taylor expand $\chi_{\aug}^2$ and $\psi$ to $O(\norm{\vec{a}-\vec{a}^*}^2)$ and $O(\norm{\vec{a}-\vec{a}^*}^0)$, respectively (see \cite{Jay:2020jkz} for example). For the NLO integral expansion, we need to include two more terms in both Taylor expansions (cf. \cite{Bleistein1975,Bender2013}):
\begin{align}
\label{eqn:chi2aug_taylor_exp}
\chi_{\aug}^2(\vec{a})=&\chi_{\aug}^2(\vec{a}^*)+\left({\Sigma^*}^{-1}\right)_{ba}\vec{\delta}_a\vec{\delta}_b+T_{cba}\vec{\delta}_a\vec{\delta}_b\vec{\delta}_c+F_{dcba}\vec{\delta}_a\vec{\delta}_b\vec{\delta}_c\vec{\delta}_d+O(\norm{\vec{\delta}}^5),\\
\label{eqn:psi_taylor_exp}\psi(\vec{a})=&\psi(\vec{a}^*)+g_a\vec{\delta}_a+\frac{1}{2}H_{ba}\vec{\delta}_a\vec{\delta}_b+O(\norm{\vec{\delta}}^3),
\end{align}
where $\vec{\delta}\equiv\vec{a}-\vec{a}^*$, the inverse parameter covariance matrix $\Sigma^*$ is defined in \cref{eqn:covariance}, and the remaining tensors $T,F,g,H$ are defined in \cref{eqn:TF,eqn:gH}. Substituting \cref{eqn:chi2aug_taylor_exp,eqn:psi_taylor_exp} into \cref{eqn:eps_int} gives
\begin{align}
\nonumber \mathcal{I}[\psi]\approx\int_{\vec{\delta}\in B_{\epsilon}(0)}d\vec{\delta}&\exp\left[-\frac{1}{2}\left(\chi_{\aug}^2(\vec{a}^*)+\left({\Sigma^*}^{-1}\right)_{ba}\vec{\delta}_a\vec{\delta}_b+T_{cba}\vec{\delta}_a\vec{\delta}_b\vec{\delta}_c+F_{dcba}\vec{\delta}_a\vec{\delta}_b\vec{\delta}_c\vec{\delta}_d\right)\right]\\
&\cdot\left(\psi(\vec{a}^*)+g_a\vec{\delta}_a+\frac{1}{2}H_{ba}\vec{\delta}_a\vec{\delta}_b\right)\\
\nonumber \approx\int_{\vec{\delta}\in B_{\epsilon}(0)}d\vec{\delta}&\exp\left[-\frac{1}{2}\left(\chi_{\aug}^2(\vec{a}^*)+\left({\Sigma^*}^{-1}\right)_{ba}\vec{\delta}_a\vec{\delta}_b\right)\right]\\
\nonumber &\cdot\left(\psi(\vec{a}^*)+\frac{1}{2}H_{ba}\vec{\delta}_a\vec{\delta}_b-\frac{1}{2}g_dT_{cba}\vec{\delta}_a\vec{\delta}_b\vec{\delta}_c\vec{\delta}_d-\frac{1}{2}\psi(\vec{a}^*)F_{dcba}\vec{\delta}_a\vec{\delta}_b\vec{\delta}_c\vec{\delta}_d\right.\\
\label{eqn:neglect_nnlo} &\hspace{1em}\left.+\frac{1}{8}\psi(\vec{a}^*)T_{fed}T_{cba}\vec{\delta}_a\vec{\delta}_b\vec{\delta}_c\vec{\delta}_d\vec{\delta}_e\vec{\delta}_f\right),
\end{align}
where the second approximation is obtained by Taylor expanding the highest order terms of the exponential $\exp\left[-\frac{1}{2}\left(T_{cba}\vec{\delta}_a\vec{\delta}_b\vec{\delta}_c+F_{dcba}\vec{\delta}_a\vec{\delta}_b\vec{\delta}_c\vec{\delta}_d\right)\right]$ about $\vec{\delta}=\vec{0}$, neglecting terms that only contribute to NNLO, and omitting odd terms that do not contribute to the integral. Again using the fact that the integral is sharply peaked about $\vec{\delta}=\vec{0}$ in the $N\rightarrow\infty$ limit, expanding the domain of integral to all of $\mathbb{R}^k$ introduces only a small error.  After doing so, we are left with
\begin{align}
\nonumber \mathcal{I}[\psi]\approx\int_{\mathbb{R}^k}d\vec{\delta}&\exp\left[-\frac{1}{2}\left(\chi_{\aug}^2(\vec{a}^*)+\left({\Sigma^*}^{-1}\right)_{ba}\vec{\delta}_a\vec{\delta}_b\right)\right]\\
\nonumber &\cdot\left(\psi(\vec{a}^*)+\frac{1}{2}H_{ba}\vec{\delta}_a\vec{\delta}_b-\frac{1}{2}g_dT_{cba}\vec{\delta}_a\vec{\delta}_b\vec{\delta}_c\vec{\delta}_d-\frac{1}{2}\psi(\vec{a}^*)F_{dcba}\vec{\delta}_a\vec{\delta}_b\vec{\delta}_c\vec{\delta}_d\right.\\
&\hspace{1em}\left.+\frac{1}{8}\psi(\vec{a}^*)T_{fed}T_{cba}\vec{\delta}_a\vec{\delta}_b\vec{\delta}_c\vec{\delta}_d\vec{\delta}_e\vec{\delta}_f\right).
\end{align}
After extending the domain, each term of the integral is proportional to one of the following Gaussian integrals:
\begin{align}
\int_{\mathbb{R}^k}d\vec{\delta}\exp\left[-\frac{1}{2}\left({\Sigma^*}^{-1}\right)_{ba}\vec{\delta}_a\vec{\delta}_b\right]&=(2\pi)^{k/2}(\det\Sigma^*)^{1/2},\\
\int_{\mathbb{R}^k}d\vec{\delta}\exp\left[-\frac{1}{2}\left({\Sigma^*}^{-1}\right)_{ba}\vec{\delta}_a\vec{\delta}_b\right]\vec{\delta}_a\vec{\delta}_b&=(2\pi)^{k/2}(\det\Sigma^*)^{1/2}(\Sigma^*)_{ab},\\
\int_{\mathbb{R}^k}d\vec{\delta}\exp\left[-\frac{1}{2}\left({\Sigma^*}^{-1}\right)_{ba}\vec{\delta}_a\vec{\delta}_b\right]\vec{\delta}_a\vec{\delta}_b\vec{\delta}_c\vec{\delta}_d&=(2\pi)^{k/2}(\det\Sigma^*)^{1/2}(\Sigma_2^*)_{abcd},\\
\int_{\mathbb{R}^k}d\vec{\delta}\exp\left[-\frac{1}{2}\left({\Sigma^*}^{-1}\right)_{ba}\vec{\delta}_a\vec{\delta}_b\right]\vec{\delta}_a\vec{\delta}_b\vec{\delta}_c\vec{\delta}_d\vec{\delta}_e\vec{\delta}_f&=(2\pi)^{k/2}(\det\Sigma^*)^{1/2}(\Sigma_3^*)_{abcdef},
\end{align}
where the high-order contractions of the covariance matrix are defined in \cref{eqn:covariance_contractions}. Using these integral identities, we obtain \cref{eqn:nlo_lap_approx}:
\begin{align} \label{eqn:nlo_lap_approx_appendix}
\nonumber \mathcal{I}[\psi]\approx&(2\pi)^{k/2}|\Sigma^*|^{1/2}\exp\left[-\frac{1}{2}\chi_{\aug}^2(\vec{a}^*)\right]\\
\nonumber &\cdot\left(\psi(\vec{a}^*)+\frac{1}{2}H_{ba}(\Sigma^*)_{ab}-\frac{1}{2}g_dT_{cba}(\Sigma_2^*)_{abcd}-\frac{1}{2}\psi(\vec{a}^*)F_{dcba}(\Sigma_2^*)_{abcd}\right.\\
&\hspace{1em}\left.+\frac{1}{8}\psi(\vec{a}^*)T_{fed}T_{cba}(\Sigma_3^*)_{abcdef}\right).
\end{align}

We will also need to consider the case were $\mathcal{I}[\psi]$ is normalized by $\mathcal{I}[1]$:
\begin{align}
\frac{\mathcal{I}[\psi]}{\mathcal{I}[1]}=\frac{\int d\vec{a}\exp\left[-\frac{1}{2}\chi_{\aug}^2(\vec{a})\right]\psi(\vec{a})}{\int d\vec{a}\exp\left[-\frac{1}{2}\chi_{\aug}^2(\vec{a})\right]}.
\end{align}
To approximate ratios such as this, we first apply the NLO Laplace approximation \cref{eqn:nlo_lap_approx_appendix} to both the numerator and denominator to obtain
\begin{align}
\frac{\mathcal{I}[\psi]}{\mathcal{I}[1]}\approx\frac{\psi(\vec{a}^*)+\frac{1}{2}H\Sigma^*-\frac{1}{2}gT\Sigma_2^*-\frac{1}{2}\psi(\vec{a}^*)F\Sigma_2^*+\frac{1}{8}\psi(\vec{a}^*)TT\Sigma_3^*}{1-\frac{1}{2}F\Sigma_2^*+\frac{1}{8}TT\Sigma_3^*},
\end{align}
where tensor indices are suppressed for simplicity.  By the tensor power counting summarized at the end of \cref{subsec:laplace}, we can expand the denominator as a geometric series; this expansion will maintain the same order of accuracy and is known in the probability literature \cite{Tierney1986}.  Keeping enough terms to maintain an overall NLO approximation, we obtain
\begin{align}
\frac{\mathcal{I}[\psi]}{\mathcal{I}[1]}&\approx\para{\psi(\vec{a}^*)+\frac{1}{2}H\Sigma^*-\frac{1}{2}gT\Sigma_2^*-\frac{1}{2}\psi(\vec{a}^*)F\Sigma_2^*+\frac{1}{8}\psi(\vec{a}^*)TT\Sigma_3^*}\para{1+\frac{1}{2}F\Sigma_2^*-\frac{1}{8}TT\Sigma_3^*}\\
&\approx\psi(\vec{a}^*)\para{1+\frac{1}{2}F\Sigma_2^*-\frac{1}{8}TT\Sigma_3^*}+\frac{1}{2}H\Sigma^*-\frac{1}{2}gT\Sigma_2^*-\frac{1}{2}\psi(\vec{a}^*)F\Sigma_2^*+\frac{1}{8}\psi(\vec{a}^*)TT\Sigma_3^*\\
&=\psi(\vec{a}^*)+\frac{1}{2}H_{ba}(\Sigma^*)_{ab}-\frac{1}{2}g_dT_{cba}(\Sigma_2^*)_{abcd},
\end{align}
where we have restored the indices in the last line.

Outside the context of the Laplace method, we also make use of some additional Gaussian integrals in the exact treatment of perfect model K-L divergences in \cref{sec:data_subset_select}.  Consider an integral of the following form:

\begin{equation}
\mathcal{J}_1 \equiv \int_{\mathbb{R}^k} d\vec{a} \exp \left[ -\frac{1}{2} (\mu_0 - \vec{a})^T (\Sigma_0^{-1}) (\mu_0 - \vec{a}) \right] (\mu_1 - \vec{a})^T \Sigma_1^{-1} (\mu_1 - \vec{a}).
\end{equation}
Defining the change of variables
\begin{align}
\delta &\equiv \vec{a} - \mu_0, \\
\xi &\equiv \mu_0 - \mu_1,
\end{align}
we can rewrite the integral as
\begin{equation}
\mathcal{J}_1 = \int_{\mathbb{R}^k} d\vec{\delta} \exp \left[ -\frac{1}{2} (\Sigma_0^{-1})_{ba} \delta_a \delta_b \right] (\Sigma_1^{-1})_{ba} (\delta + \xi)_a (\delta + \xi)_b.
\end{equation}
Using the Gaussian integral formulas above to simplify gives the result:
\begin{equation} \label{eqn:gaussian_integral_I2}
\mathcal{J}_1 = (2\pi)^{k/2} (\det \Sigma_0)^{1/2} \left( \xi^T \Sigma_1^{-1} \xi  +\tr [\Sigma_0 \Sigma_1^{-1} ]\right).
\end{equation}
We need one more additional integral:
\begin{equation}
\mathcal{J}_2 \equiv \int_{\mathbb{R}^k} d\vec{a} \exp \left[ -\frac{1}{2} (\mu_0 - \vec{a})^T (\Sigma_0^{-1}) (\mu_0 - \vec{a}) - \frac{1}{2} (\mu_1 - \vec{a})^T (N \Sigma_0)^{-1} (\mu_1 - \vec{a}) \right].
\end{equation}
where $N$ is an integer.  Applying the same change of variables as above and gathering terms, we have:
\begin{equation}
\mathcal{J}_2 = \int_{\mathbb{R}^k} d\vec{\delta} \exp \left[ -\frac{1}{2} \left(\frac{N+1}{N} \Sigma_0^{-1}\right)_{ba} \delta_a \delta_b - \frac{1}{2N} (\Sigma_0^{-1})_{ba} (\xi_a \xi_b + \xi_a \delta_b + \delta_a \xi_b) \right].
\end{equation}
We change variables again to $\delta' = \sqrt{(N+1)/N} \delta$ to absorb the extra factor in the first term.  Pulling the $\delta$-independent exponential factor out front, we then have
\begin{align}
\mathcal{J}_2 &= e^{-\xi^T \Sigma_0^{-1} \xi / (2N)} \left( \frac{N}{N+1}\right)^{k/2} \int_{\mathbb{R}^k} d\vec{\delta}' \exp \left[ -\frac{1}{2} (\Sigma_0^{-1})_{ba} \delta'_a \delta'_b - \frac{1}{2} \sqrt{\frac{1}{N(N+1)}} (\Sigma_0^{-1})_{ba} (\xi_a \delta'_b + \delta'_a \xi_b) \right] \\
&=e^{-\xi^T \Sigma_0^{-1} \xi / (2N)} \left( \frac{N}{N+1}\right)^{k/2} (2\pi)^{k/2} (\det \Sigma_0)^{1/2} \exp \left[ \frac{1}{2N(N+1)} \xi^T \Sigma_0^{-1} \xi \right] 
\end{align}
or simplifying,
\begin{equation} \label{eqn:gaussian_integral_I3}
\mathcal{J}_2 =  \left( \frac{N}{N+1}\right)^{k/2} (2\pi)^{k/2} (\det \Sigma_0)^{1/2} \exp \left[ -\frac{1}{2(N+1)} \xi^T \Sigma_0^{-1} \xi \right].
\end{equation}
%

\section{Alternative derivations for data subset selection}
\label{sec:app-subset-alt}

In this appendix, we give an alternative derivation for the data subset selection formulas given in \cref{sec:data_subset_select} that uses the partition of data and the least-squares ICs rather than computing the K-L divergences directly.

In the case of least-squares regression with correct model specification, the BPIC and PPIC (before the integral approximations) are given by
\begin{align}
\BPIC_{\mu}=&\chi_{\aug}^2(\vec{a}^*)-\frac{\int d\vec{a}\exp\left[-\frac{1}{2}\chi_{\aug}^2(\vec{a})\right]\tilde{\chi}^2(\vec{a})}{\int d\vec{a}\exp\left[-\frac{1}{2}\chi_{\aug}^2(\vec{a})\right]}+3k,\\
\PPIC_{\mu}=&-2\sum_{i=1}^{N}\log\frac{\int d\vec{a}\exp\left[-\frac{1}{2}\left(\chi_{\aug}^2(\vec{a})+\chi_i^2(\vec{a})\right)\right]}{\int d\vec{a}\exp\left[-\frac{1}{2}\chi_{\aug}^2(\vec{a})\right]}+2k.
\end{align}
For data selection, $k\rightarrow k+d_{\C}$, to account for the additional $d_{\C}$ parameters for the ``perfect'' model as before. We now must understand how the chi-squared functions and integrals transform as well. 

As described in the main body in \cref{sec:data_subset_select}, we partition the $\chi^2$ functions into kept, cut, and off-block-diagonal pieces:
\begin{align}
\hat{\chi}^2(\vec{a})=&\left(\bar{y}-\phi_{M,P}(\vec{a})\right)^T\hat{\Sigma}^{-1}\left(\bar{y}-\phi_{M,P}(\vec{a})\right)\\
=&\begin{pmatrix}\bar{y}_{\C}-\vec{a}_{\C}\\\bar{y}_{\K}-f_M(\vec{a}_{\K})\end{pmatrix}^T\begin{pmatrix} (\hat{\Sigma}^{-1})_{\C} & (\hat{\Sigma}^{-1})_{\off} \\ (\hat{\Sigma}^{-1})_{\off}^T & (\hat{\Sigma}^{-1})_{\K}\end{pmatrix}\begin{pmatrix}\bar{y}_{\C}-\vec{a}_{\C}\\\bar{y}_{\K}-f_M(\vec{a}_{\K})\end{pmatrix}\\
\equiv&\hat{\chi}_{\C}^2(\vec{a}_{\C})+\hat{\chi}_{\K}^2(\vec{a}_{\K})+2\hat{\chi}_{\off}^2(\vec{a}_{\C},\vec{a}_{\K}),
\end{align}
where
\begin{align}
\hat{\chi}_{\C}^2(\vec{a}_{\C})&\equiv\para{\bar{y}_{\C}-\vec{a}_{\C}}^T(\hat{\Sigma}^{-1})_{\C}\para{\bar{y}_{\C}-\vec{a}_{\C}},\\
\hat{\chi}_{\K}^2(\vec{a}_{\K})&\equiv\para{\bar{y}_{\K}-f_M(\vec{a}_{\K})}^T(\hat{\Sigma}^{-1})_{\K}\para{\bar{y}_{\K}-f_M(\vec{a}_{\K})},\\
\hat{\chi}_{\off}^2(\vec{a}_{\C},\vec{a}_{\K})&\equiv\para{\bar{y}_{\C}-\vec{a}_{\C}}^T(\hat{\Sigma}^{-1})_{\off}\para{\bar{y}_{\K}-f_M(\vec{a}_{\K})}\\
&=\para{\bar{y}_{\K}-f_M(\vec{a}_{\K})}^T(\hat{\Sigma}^{-1})_{\off}^T\para{\bar{y}_{\C}-\vec{a}_{\C}}.
\end{align}
Similarly, we define
\begin{align}
\chi_{\K,i}^2(\vec{a}_{\K})\equiv&\left(y_{\K,i}-f_M(\vec{a}_{\K})\right)^T(\Sigma^{-1})_{\K}\left(y_{\K,i}-f_M(\vec{a}_{\K})\right),\\
\chi_{\C,i}^2(\vec{a}_{\C})\equiv&\left(y_{\C,i}-\vec{a}_{\C}\right)^T(\Sigma^{-1})_{\C}\left(y_{\C,i}-\vec{a}_{\C}\right),\\
\chi_{\off,i}^2(\vec{a}_{\C},\vec{a}_{\K})\equiv&\para{y_{\C,i}-\vec{a}_{\C}}^T(\Sigma^{-1})_{\off}\para{y_{\K,i}-f_M(\vec{a}_{\K})}.
\end{align}
Furthermore,
\begin{align}
\tilde{\chi}^2(\vec{a})=&(\vec{a}_{\C}-\bar{y}_{\C})^T\tilde{\Sigma}_{\C}^{-1}(\vec{a}_{\C}-\bar{y}_{\C})+(\vec{a}_{\K}-\tilde{\vec{a}})^T\tilde{\Sigma}_{\K}^{-1}(\vec{a}_{\K}-\tilde{\vec{a}})\equiv\tilde{\chi}_{\C}^2(\vec{a}_{\C})+\tilde{\chi}_{\K}^2(\vec{a}_{\K}),
\end{align}
where $\tilde{\Sigma}^{-1}=\diag\left(\tilde{\Sigma}_{\C}^{-1},\tilde{\Sigma}_{\K}^{-1}\right)$ by construction. It follows that
\begin{align}
\chi_{\aug}^2(\vec{a})=\chi_{\C,\aug}^2(\vec{a}_{\C})+\chi_{\K,\aug}^2(\vec{a}_{\K})+2\hat{\chi}_{\off}^2(\vec{a}_{\C},\vec{a}_{\K}),
\end{align}
where $\chi_{\C,\aug}^2$ and $\chi_{\K,\aug}^2$ are defined analogously to \cref{eqn:chi2_aug} with the cut and kept statistics, respectively. We note for later use that $\hat{\chi}_{\C}^2(\bar{y}_{\C})$, $\tilde{\chi}_{\C}^2(\bar{y}_{\C})$, $\chi_{\C,\aug}^2(\bar{y}_{\C})$, and $\hat{\chi}_{\off}^2(\bar{y}_{\C},\vec{a}_{\K})$ vanish identically (for all $\vec{a}_{\K}$).

With these definitions, the information criteria become
\begin{align}
\label{eqn:BPIC_data_subset_start}
\nonumber \BPIC_{\mu,P}=&-\frac{\int d\vec{a}_{\C}d\vec{a}_{\K}\exp\left[-\frac{1}{2}\para{\chi_{\K,\aug}^2+\chi_{\C,\aug}^2+2\hat{\chi}_{\off}^2}\right]\para{\tilde{\chi}_{\K}^2+\tilde{\chi}_{\C}^2}}{\int d\vec{a}_{\C}d\vec{a}_{\K}\exp\left[-\frac{1}{2}\para{\chi_{\K,\aug}^2+\chi_{\C,\aug}^2+2\hat{\chi}_{\off}^2}\right]}\\
&+\chi_{\K,\aug}^2(\vec{a}_{\K}^*)+3(k+d_{\C}),\\
\label{eqn:PPIC_data_subset_start}
\nonumber \PPIC_{\mu,P}=&-2\sum_{i=1}^{N}\log\frac{\int d\vec{a}_{\C}d\vec{a}_{\K}\exp\left[-\frac{1}{2}\left(\chi_{\K,\aug}^2+\chi_{\C,\aug}^2+2\hat{\chi}_{\off}^2+\chi_{\K,i}^2+\chi_{\C,i}^2+2\chi_{i,\off}^2\right)\right]}{\int d\vec{a}_{\C}d\vec{a}_{\K}\exp\left[-\frac{1}{2}\para{\chi_{\K,\aug}^2+\chi_{\C,\aug}^2+2\hat{\chi}_{\off}^2}\right]}\\
&+2(k+d_{\C}),
\end{align}
where we have suppressed the arguments of the $\chi^2$ functions in the integrands for simplicity.

As in the main body, here we assume that the off-block-diagonal elements of the sample covariance $\hat{\Sigma}_{\off}$ are small (in the sense of induced operator norm) relative to the on-block-diagonal elements $\hat{\Sigma}_{\C}$ and $\hat{\Sigma}_{\K}$.  It follows from this approximation that $\chi_{i,\C}^2,\chi_{i,\K}^2\gg\chi_{i,\off}^2$ and $\hat{\chi}_{\C}^2,\hat{\chi}_{\K}^2\gg\hat{\chi}_{\off}^2$.  Therefore,
\begin{align}
\nonumber \BPIC_{\mu,P}\approx&\chi_{\K,\aug}^2(\vec{a}_{\K}^*)-\frac{\int d\vec{a}_{\K}\exp\left[-\frac{1}{2}\chi_{\K,\aug}^2\right]\tilde{\chi}_{\K}^2}{\int d\vec{a}_{\K}\exp\left[-\frac{1}{2}\chi_{\K,\aug}^2\right]}+3k\\
&-\frac{\int d\vec{a}_{\C}\exp\left[-\frac{1}{2}\chi_{\C,\aug}^2\right]\tilde{\chi}_{\C}^2}{\int d\vec{a}_{\C}\exp\left[-\frac{1}{2}\chi_{\C,\aug}^2\right]}+3d_{\C},\\
\nonumber \PPIC_{\mu,P}\approx&-2\sum_{i=1}^{N}\log\frac{\int d\vec{a}_{\K}\exp\left[-\frac{1}{2}\para{\chi_{\K,\aug}^2+\chi_{\K,i}^2}\right]}{\int d\vec{a}_{\K}\exp\left[-\frac{1}{2}\chi_{\K,\aug}^2\right]}+2k\\
&-2\sum_{i=1}^{N}\log\frac{\int d\vec{a}_{\C}\exp\left[-\frac{1}{2}\left(\chi_{\C,\aug}^2+\chi_{\C,i}^2\right)\right]}{\int d\vec{a}_{\C}\exp\left[-\frac{1}{2}\chi_{\C,\aug}^2\right]}+2d_{\C}.
\end{align}

The ``$\K$" integrals can be approximated using the NLO Laplace approximation as in the previous sections, leading to the same IC formulas that we found previously over the kept data, but with a subtle difference: the inverse covariance matrix appearing in $\hat{\chi}_{\K}^2$ is the sub-block of the full inverse matrix $(\hat{\Sigma}^{-1})_{\K}$, as opposed to the inverse of the sub-block covariance matrix $(\hat{\Sigma}_{\K})^{-1}$.  Under our block-diagonal assumption $\Sigma^{-1}\approx\diag\left(\Sigma_{\K}^{-1},\Sigma_{\C}^{-1}\right)$, these matrices are identical; even when off-diagonal correlations are present, the inverse of the sub-block is often used in practice to define $\hat{\chi}_{\K}^2$.  For further discussion of this point, see \cref{sec:data_subset_select}.

We will also compute the ``$\C$" integrals using our Laplace approximation formulas, but since we can take $M_{\perf,\mu}$ to be linear (e.g., a polynomial of degree $d_{\C}-1$), the BPIC formula will be exact; the PPIC is not exact here, but the result in \cref{sec:data_subset_select} is exact.  Beginning with the BPIC, we have the result
\begin{equation}
\label{eqn:bpic-cut-int} \frac{\int d\vec{a}_{\C}\exp\left[-\frac{1}{2}\chi_{\aug,\C}^2\right]\tilde{\chi}_{\C}^2}{\int d\vec{a}_{\C}\exp\left[-\frac{1}{2}\chi_{\aug,\C}^2\right]}=\tr\left[\tilde{\Sigma}_{\C}^{-1}\Sigma_{\C}^*\right],
\end{equation}
taking $\Sigma^*\approx\diag\left(\Sigma_{\K}^*,\Sigma_{\C}^*\right)$ which follows from our assumption that $\Sigma^{-1}\approx\diag\left(\Sigma_{\K}^{-1},\Sigma_{\C}^{-1}\right)$.  Since we are working in the limit of infinitely diffuse priors over the cut data, we have $(\Sigma_{\C}^*)^{-1} = \hat{\Sigma}_{\C}^{-1} + \tilde{\Sigma}_{\C}^{-1} \rightarrow \hat{\Sigma}_{\C}^{-1}$, and thus
\begin{equation}
\tr\left[\tilde{\Sigma}_{\C}^{-1}\Sigma_{\C}^*\right] \rightarrow \tr\left[\tilde{\Sigma}_{\C}^{-1} \hat{\Sigma}_{\C}\right] \rightarrow 0.
\end{equation}
So the only additional contribution to the BPIC for data subset selection is a penalty term of $+3d_{\C}$.  We turn next to the PPIC, where the Laplace approximation formula gives us the result:
\begin{equation}
\label{eqn:ppic-cut-int} \sum_{i=1}^{N}\log\frac{\int d\vec{a}_{\C}\exp\left[-\frac{1}{2}\left(\chi_{\aug,\C}^2+\chi_{\C,i}^2\right)\right]}{\int d\vec{a}_{\C}\exp\left[-\frac{1}{2}\chi_{\aug,\C}^2\right]}=\sum_{i=1}^{N}\log\left[1+\frac{1}{2}\tr\left[\left(\frac{1}{4}(g_{\C,i})(g_{\C,i})^T-\frac{1}{2}H_{\C,i}\right)\Sigma_{\C}^*\right]\right].
\end{equation}

In summary, the ICs (up to constant terms) for data subset selection are
\begin{align}
\BAIC_{\mu,P}=&\hat{\chi}^2(\vec{a}^*)+2k+2d_{\C},\\
\BPIC_{\mu,P}\approx&\hat{\chi}^2(\vec{a}^*)-\frac{1}{2}\tilde{H}_{ba}(\Sigma^*)_{ab}+\frac{1}{2}\tilde{g}_dT_{cba}(\Sigma_2^*)_{abcd}+3k+3d_{\C},\\
\nonumber \label{eqn:ppic_subset_alt} \PPIC_{\mu,P}\approx&\hat{\chi}^2(\vec{a}^*)+2k+2\left(1 + \frac{1}{2N} \right) d_{\C}\\
&-2\sum_{i=1}^{N}\log\left[1+\frac{1}{2}\left(\frac{1}{4}(g_i)_b(g_i)_a-\frac{1}{2}(H_i)_{ba}\right)(\Sigma^*)_{ab}+\frac{1}{4}(g_i)_dT_{cba}(\Sigma_2^*)_{abcd}\right],
\end{align}
where we have dropped all ``$\K$" subscripts.

It is worth noting that the PPIC in \cref{eqn:ppic_subset_alt} disagrees with \cref{eqn:ppic_subset} even to $O(N^{-1})$.  This disagreement comes from higher-order terms in the bias correction of the PPIC on the perfect model.  By computing the $\KL_{\C}$ exactly, the result in \cref{eqn:ppic_subset} accounts for these corrections where as \cref{eqn:ppic_subset_alt} does not.  Studying higher-order corrections to IC bias could be an interesting direction for future work.

\section{Relevant derivatives}
\label{sec:derivatives}

In this appendix, we give expressions for the relevant derivative tensors used in \cref{sec:num-tests}. We write these in terms of derivatives of the model function $f(\vec{a})$ (with the model index $\mu$ suppressed), which we calculated using auto-differentiation (specifically using the Python package \texttt{jax} \cite{jax2018github}) to obtain the results in \cref{sec:num-tests}. For brevity, we only give expressions for derivatives for $\chi_{\aug}^2$, as the other relevant derivatives (i.e., those of $\tilde{\chi}^2$, $\hat{\chi}^2$, and $\chi_i^2$) can be deduced from
\begin{align}
\chi_{\aug}^2(\vec{a})=\tilde{\chi}^2(\vec{a})+\hat{\chi}^2(\vec{a})=\tilde{\chi}^2(\vec{a})+\sum_{i=1}^{N}\chi_i^2(\vec{a})-(N-1)d.
\end{align}
The derivative tensors are written in index summation notation where indices at the beginning of the (roman) alphabet (i.e., $a,b,c,d$) denote parameter dimensions and at the end of the alphabet (i.e., $x,y$) data dimensions. The expressions are as follows:
\begin{align}
\chi_{\aug}^2=&(\vec{a}-\tilde{\vec{a}})_a(\tilde{\Sigma}^{-1})_{ab}(\vec{a}-\tilde{\vec{a}})_b+\para{\bar{y}-f}_x(\hat{\Sigma}^{-1})_{xy}\para{\bar{y}-f}_y,\\
\frac{\partial\chi_{\aug}^2}{\partial a_a}=&2\bracket{(\tilde{\Sigma}^{-1})_{ab}(\vec{a}-\tilde{\vec{a}})_b-\para{\frac{\partial f}{\partial a_a}}_x(\hat{\Sigma}^{-1})_{xy}\para{\bar{y}-f}_y},\\
\frac{\partial^2\chi_{\aug}^2}{\partial a_a\partial a_b}=&2\bracket{(\tilde{\Sigma}^{-1})_{ab}-\para{\frac{\partial^2f}{\partial a_a\partial a_b}}_x(\hat{\Sigma}^{-1})_{xy}\para{\bar{y}-f}_y+\para{\frac{\partial f}{\partial a_a}}_x(\hat{\Sigma}^{-1})_{xy}\para{\frac{\partial f}{\partial a_b}}_y},\\
\nonumber \frac{\partial^3\chi_{\aug}^2}{\partial a_a\partial a_b\partial a_c}=&2\bracket{-\para{\frac{\partial^3f}{\partial a_a\partial a_b\partial a_c}}_x(\hat{\Sigma}^{-1})_{xy}\para{\bar{y}-f}_y
+\para{\frac{\partial^2f}{\partial a_a\partial a_b}}_x(\hat{\Sigma}^{-1})_{xy}\para{\frac{\partial f}{\partial a_c}}_y\right.\\
&\hspace{1em}\left.+\para{\frac{\partial^2f}{\partial a_a\partial a_c}}_x(\hat{\Sigma}^{-1})_{xy}\para{\frac{\partial f}{\partial a_b}}_y
+\para{\frac{\partial^2f}{\partial a_b\partial a_c}}_x(\hat{\Sigma}^{-1})_{xy}\para{\frac{\partial f}{\partial a_a}}_y}.
\end{align}
Note that we omit $(\partial^4\chi_{\aug}^2)/(\partial a_a\partial a_b\partial a_c\partial a_d)$ as it cancels to NLO in any case (see \cref{subsec:bpic} for details).

\end{document}